\documentclass{lmcs} %%% last changed 2026-09-14
\pdfoutput=1
\usepackage[utf8]{inputenc}

\usepackage{lastpage}
\lmcsdoi{22}{4}{4}
\lmcsheading{}{\pageref{LastPage}}{}{}%
{Jun.~20,~2025}{Oct.~09,~2026}{}

\keywords{Petri net, conservative net, structural liveness.}

\usepackage{macros}
\usepackage{tikz}
\usetikzlibrary{arrows,shapes,decorations,automata,backgrounds,petri,calc,positioning}
\usepackage{amssymb,amsmath}
\usepackage{bm}
\usepackage{xspace}
\usepackage{stmaryrd}
\let\origtextsc\textsc
\renewcommand{\textsc}[1]{\textnormal{\origtextsc{#1}}}
\SetSymbolFont{stmry}{bold}{U}{stmry}{m}{n}

\usepackage{hyperref}
\def\eg{{\em e.g.}}

\begin{document}

% If the title is longer than 55 characters, then specify a shorter running title as the optional argument to \title. The running title should be roughyl at most 55 characters:
\title{Structural Liveness of Conservative Petri Nets}
\titlecomment{{\lsuper*}
This is a revised and extended version of the
conference paper~\cite{DBLP:conf/fossacs/JancarLV25}.}
%OPTIONAL comment concerning the title, \eg,
%if a variant or an extended abstract of the paper has appeared elsewhere.}
\thanks{Ji\v{r}\'{i} Val\r{u}\v{s}ek was supported by
Grants No.\ IGA\_PrF\_2024\_024 and IGA\_PrF\_2025\_018 of IGA of Palack\'y University Olomouc.
We also thank the reviewers for their helpful comments.}	%optional
\specialissue{Selected Papers of the 28th International Conference on Foundations of Software Science and Computation Structures (FoSSaCS 2025)}
% affiliations are numbered automatically with a, b, c (see below)
% use the optional argument to indicate the affiliation(s) of each author
% omit the argument if there is only one author, or only one affiliation
\author[P.~Jan\v{c}ar]{Petr Jan\v{c}ar\lmcsorcid{0000-0002-8738-9850}}[a]
\author[J.~Leroux]{J{\'{e}}r{\^{o}}me Leroux\lmcsorcid{0000-0002-7214-9467}}[b]
\author[J.~Val\r{u}\v{s}ek]{Ji\v{r}{\'{i}} Val\r{u}\v{s}ek\lmcsorcid{0000-0003-1345-1321}}[a]

% affiliation 1 (automatically numbered a)
\address{Dept of Comp. Sci., Faculty of Science, Palack\'y
University, Olomouc, Czechia}	%optional
% write emails for all authors having that affiliation
%\email{name1@email1, name2@email1, name3@email1}  %optional

% affiliation 2 (automatically numbered b)
\address{LaBRI, CNRS, Univ. Bordeaux, France}	%optional
%\email{name2@email2}  %optional

%% etc.

%% required for running head on odd and even pages, use suitable
%% abbreviations in case of long titles and many authors:

%%%%%%%%%%%%%%%%%%%%%%%%%%%%%%%%%%%%%%%%%%%%%%%%%%%%%%%%%%%%%%%%%%%%%%%%%%%

%% the abstract has to PRECEDE the command \maketitle:
%% be sure not to issue the \maketitle command twice!

\begin{abstract}
  \noindent We show that the EXPSPACE-hardness result for
	structural liveness of Petri nets [Jan\v{c}ar and
	Purser, 2019] holds even for a~simple subclass of
	conservative nets. As our main result, we prove that
	for structurally live conservative nets, the values of the
	minimal live markings are at most doubly exponential in the
	size of the net. This implies the EXPSPACE-completeness
	of structural liveness for conservative Petri nets. 
	The result also applies to structurally bounded Petri nets,
	whereas
the complexity of the general case remains open. 

As a proof ingredient of independent interest, we present an extension of known results on the bounds of minimal integer solutions to Boolean combinations of linear equalities, inequalities, and divisibility constraints.
\end{abstract}

\maketitle

\section{Introduction}

Petri nets are a well-known model of a class of distributed systems;
for an introduction, we refer, \eg, to the monographs~\cite{reisig2013} and~\cite{BestDevillers2024}.

The \emph{reachability} problem asks whether a given target marking is reachable from a given initial marking. Here, a \emph{marking} is the standard term for a Petri net configuration, assigning a number of \emph{tokens} to each \emph{place} (local state). The reachability problem for Petri nets is well known for its computational complexity: its Ackermann-completeness has only recently been established (see~\cite{DBLP:conf/focs/Leroux21,DBLP:conf/focs/CzerwinskiO21} for the lower bound and~\cite{DBLP:conf/lics/LerouxS19} for the upper bound).

Given a Petri net and an initial marking, the \emph{boundedness} problem asks whether the reachability set is finite, and the \emph{liveness} problem asks whether all \emph{transitions} (system actions) remain live forever. Both are among the standard analysis problems. While boundedness is known to be EXPSPACE-complete~\cite{DBLP:conf/stoc/CardozaLM76,DBLP:journals/tcs/Rackoff78}, liveness is tightly related to reachability~\cite{DBLP:conf/focs/Hack74} and is thus now known to be Ackermann-complete.

There are also natural structural versions of boundedness and liveness.
The \emph{structural boundedness} problem asks, given a Petri net, whether the net is bounded for every initial marking, while the \emph{structural liveness} problem asks, given a Petri net, whether there exists an initial marking for which the net is live.

While structural boundedness is easily shown to be in PTIME (and is thus substantially easier than boundedness), structural liveness is only known to be EXPSPACE-hard and decidable~\cite{DBLP:journals/acta/JancarP19}. In fact, the decidability result can be strengthened by recent results on the home-space problem~\cite{DBLP:conf/concur/JancarL23,DBLP:journals/lmcs/JancarL24}, which readily imply an Ackermannian upper bound for structural liveness as well; nevertheless, the large complexity gap still calls for clarification.

\emph{Our contribution.}
As a step towards clarifying the complexity of \emph{structural
liveness} for general Petri nets, we establish
\emph{EXPSPACE-completeness} for the class of \emph{conservative
nets}, which preserve a weighted sum of tokens throughout their
executions. We recall that deciding whether a given net is
conservative is also in PTIME, similarly to structural boundedness
(see, \eg,~\cite{DBLP:conf/apn/MayrW14}).

A crucial notion in our proof is \emph{structural reversibility},
referred to simply as \emph{reversibility} in this paper. 
A net is reversible if there exists a sequence of transitions in which
each transition occurs at least once and whose overall effect is zero,
that is, executing the sequence does not change the marking.
Equivalently, the effect of any transition can be ``undone'' by some transition sequence.
Reversibility is also in PTIME, and it can be easily shown to be a~necessary condition for structural liveness in the case of structurally bounded nets.

Moreover, it is immediate that every conservative net is structurally bounded, and a straightforward application of Farkas’ lemma shows that every reversible structurally bounded net is conservative. Hence, our EXPSPACE-completeness result can equivalently be stated for the class of \emph{structurally bounded nets}.

As a natural first step in our future research, we plan to address several subtle issues that would allow us to extend the result to the full class of reversible nets.

The lower bound, EXPSPACE-hardness, is achieved by adapting the
construction of~\cite{DBLP:journals/acta/JancarP19}, which shows a
reduction from the EXPSPACE-complete word problem for commutative
semigroups. This problem can also be phrased as a coverability problem
for reversible Petri nets~\cite{DBLP:conf/stoc/CardozaLM76,
mayr1982complexity, DBLP:journals/jc/Mayr97}. We recall that
coverability is a weaker form of reachability: it asks whether there
exists a reachable marking that is component-wise at least as large as the target. Our adaptation demonstrates EXPSPACE-hardness of structural liveness even for nets where each transition has precisely two input and 
two output places, that is, for nets that naturally correspond to population protocols~\cite{DBLP:journals/dc/AngluinADFP06}.

Our main result is an EXPSPACE upper bound. The crucial step
establishes
that for every structurally live conservative net, there exists a live
marking with at most a doubly exponential (2-exp) number of tokens.
In fact, we establish this 2-exp bound in a more general setting: for nets
that possess live bottom strongly connected components in their
reachability graphs. Conservative nets constitute a
special case, for which such a 2-exp upper bound readily implies an
EXPSPACE upper bound for the structural liveness problem.

The aforementioned 2-exp bound is achieved by showing that for any
such net, there exists a specific quantifier-free
Presburger formula that admits a ``small'' (2-exp) solution and whose
solutions are live markings. More precisely, a solution to this
formula represents a set of at most exponentially many mutually
reachable markings, which together demonstrate that all of them are live.

A key aspect of our approach is the replacement of
reachability with a conceptually simpler \emph{virtual reachability},
which allows for negative numbers of tokens in (virtual) markings. To
express virtual reachability in reversible nets, we employ linear
systems---that  is, Boolean combinations of linear equalities,
inequalities, and
divisibility constraints---which constitute the aforementioned
special
Presburger formulas. As a technical contribution of independent
interest, we provide exponential bounds on the minimal solutions of
such linear systems, extending known results such as those
in~\cite{DBLP:conf/rta/Pottier91} (see also the survey~\cite{DBLP:journals/siglog/Haase18}).

\emph{Related research.}
This paper can be viewed as a continuation of the research line
initiated by \cite{DBLP:journals/ipl/BestE16}, which explicitly
indicated that even the decidability of structural liveness for Petri
nets remained open at that time. As already mentioned, decidability
and EXPSPACE-hardness are now established
\cite{DBLP:journals/acta/JancarP19}; furthermore, the decidability
result was recently strengthened by \cite{DBLP:conf/concur/JancarL23,
DBLP:journals/lmcs/JancarL24}, implying an Ackermannian upper bound.
Another result in this research line is the PSPACE-completeness of
structural liveness for IO-nets (Immediate Observation Petri Nets),
which were introduced in \cite{DBLP:conf/apn/EsparzaRW19} and inspired
by a subclass of population protocols in
\cite{DBLP:journals/dc/AngluinAER07}. While
\cite{DBLP:conf/apn/EsparzaRW19} does not consider structural liveness
explicitly, a~PSPACE upper bound follows directly from its results on liveness; an explicit, self-contained proof of PSPACE-completeness is provided in \cite{DBLP:journals/fuin/JancarV22}.

Numerous studies have investigated liveness for various subclasses of
Petri nets, often exploring related structural properties (see
the monographs~\cite{reisig2013, BestDevillers2024,
desel_esparza_1995}, or~\cite{DBLP:journals/fuin/HujsaD18} as 
an example of a paper explicitly studying structural liveness 
for a specific subclass). In particular, we use established results
regarding liveness for conservative nets, for which we refer to~\cite{DBLP:conf/apn/MayrW14}.

\emph{Organization of the paper.} 
Section~\ref{sec:basicdef} introduces basic notions and notation,
defines linear systems, 
and states the main results; it also establishes the equivalence of conservativeness and structural boundedness for reversible nets. Section~\ref{sec:upperbound} establishes the aforementioned 2-exp upper bound, drawing on results proven in Sections~\ref{sec:smallsolutions} and~\ref{sec:virtual}. Finally, Section~\ref{sec:hardness} addresses the EXPSPACE lower bound.

We note that throughout this paper, we adopt a convention where lemmas are considered more significant than propositions.

\section{Basic Definitions, and Results}\label{sec:basicdef}
By $\setN$, $\setN_+$, $\setZ$, and $\setR$ we denote the sets of nonnegative
integers, positive integers, integers, and reals, respectively.
For $i,j\in\setZ$ we put $[i,j]=\{i,i{+}1,\dots,j\}$.
The unary operation $|.|$ denotes the absolute value for numbers, the
cardinality for sets, and the length for sequences.
For a~vector $x\in\setZ^d$ ($d\in\setN$), we denote by $x(i)$ the value of its
component $i\in[1,d]$, 
hence $x=(x(1),x(2),\ldots,x(d))$. 
We use the \emph{component-wise (partial) order} $\leq$ on $\setZ^d$.

\paragraph{Vectors, restrictions.}
It will be clear from context when a vector is understood as a column vector; for example, if $B$ is an $m\times n$ matrix, then in the equation $Bx=b$ the vectors $x$ and $b$ are viewed as column vectors of dimensions $n$ and $m$, respectively. By $\mathbf{0}$ we denote the zero vector, whose dimension is clear from context.
Sometimes we consider vectors as elements of $\setZ^J$, where $J$ is a finite subset of $\setN$; for example, $\setN^d$ is identified with $\setN^{[1,d]}$.
For $x\in \setZ^d$ and $J\subseteq[1,d]$, we denote by $x\restr{J}$
the restriction of $x$ to $J$, that is, the vector in $\setZ^J$
satisfying $x\restr{J}(i)=x(i)$ for each $i\in J$. For
$X\subseteq\setZ^d$, we put $X\restr{J}=\{x\restr{J}\mid x\in X\}$.

\paragraph{Multiplication, dot product, rank.}
We use “$\cdot$” for standard multiplication (including multiplication of a vector or a matrix by a scalar), but for $x,y\in\setZ^d$ we use $\scalar{x}{y}$ to denote the dot product $\sum_{i=1}^d x(i)\cdot y(i)$.
The \emph{rank} of an $m\times n$ matrix $B$ is denoted by $\rank{B}$;
in particular, $\rank{B}\leq\min\{m,n\}$.

\paragraph{Norms for vectors, sets, and matrices.}
We use the following \emph{norms} for vectors $x\in\setZ^J$ and finite sets $X\subseteq\setZ^J$ (where $\setZ^J$ may in particular be $\setZ^d$):
\[
	\nnorm{x}=\max_{i\in J}|x(i)|, \ \ 
	\norm{1}{x}=\sum_{i\in J}|x(i)|, \ \ 
\nnorm{X}=\max_{x\in X}\nnorm{x}, \ \ 
	\norm{1}{X}=\max_{x\in X}\norm{1}{x};
\]
we stipulate that $\max\emptyset=0$.
Hence,
	\[0\leq\nnorm{x}\leq\norm{1}{x}\leq |J|\cdot\nnorm{x},
	{\textnormal{ and }}
0\leq\nnorm{X}\leq\norm{1}{X}\leq |J|\cdot\nnorm{X}.
	\]
We also define $\nnorm{B}$ and $\norm{1}{B}$ \emph{for matrices over
$\setZ$}, viewing a matrix as the set of its rows.
For an $m\times n$ matrix $B$, we have  
$\norm{}{B}=\max_{i,j}|B_{ij}|$ and
 $\norm{1}{B}=\max_{i\in[1,m]}\sum_{j=1}^n|B_{ij}|$.

\paragraph{Submononoid \bm{$X^*$}.} 
For $X\subseteq \setZ^d$, we denote by $X^*$ the \emph{submonoid of $(\setZ^d,+)$ generated by $X$}, that is, the set of all finite sums of elements of $X$.

\begin{figure}[t]
\begin{center}
\begin{tikzpicture}[node distance=1.5cm,>=stealth',bend angle=20,auto]
    \tikzstyle{place}=[circle,thick,draw=black!75,fill=blue!20,minimum size=4mm]
    \tikzstyle{transition}=[rectangle,thick,draw=black!75,
    fill=black!20,minimum size=3mm]
    \tikzstyle{dots}=[circle,thick,draw=none,minimum size=6mm]

    \tikzstyle{every label}=[black]

    \begin{scope}[
      % move tokens apart such they do not intersect with the border of the place
      token distance=5.2pt
      ]
        \node [transition] (a1) [] {$a_1$};
        \node [place,tokens=1] (p3) [label=right:$p_3$,below right of=a1] {}
        edge [post,bend right] (a1)
        edge [pre,bend left] node {2} (a1);
        \node [transition] (a4) [below right of=p3] {$a_4$}
        edge [post] (p3);
        \node [place,tokens=2] (p5) [label=right:$p_5$,above right of=p3] {}
        edge [post] (a1)
        edge [post,bend right] (a4)
        edge [pre,bend left] node {2} (a4);
        \node [place] (p4) [label=right:$p_4$,below left of=a4] {}
        edge [post] (a4);
        \node [transition] (a3) [left of=p4] {$a_3$}
        edge [pre] node[swap] {2} (p3)
        edge [post] (p4);
        \node [place] (p2) [label=left:$p_2$,left of=a3] {}
        edge [pre] (a1);
        \node [transition] (a2) [above left of=p2] {$a_2$}
        edge [pre] (p2);
        \draw[shorten >=0.3mm,shorten <=0.3mm,<->] (a2) to[in=150,out=0] ($(-1cm,-2.5cm)$) to[in=135,out=330] (p4);
        \node [place,tokens=1] (p1) [label=left:$p_1$,above right of=a2] {}
        edge [post] (a1)
        edge [pre] (a2);
        \draw[shorten >=0.3mm,shorten <=0.3mm,<->] (p1) to (a4);
    \end{scope}
    \begin{scope}[xshift=6.5cm]
        \node [transition] (a1) [] {$a_1$};
        \node [place] (p3) [label=right:$p_3$,below right of=a1] {}
        edge [post,bend right] (a1)
        edge [pre,bend left] node {2} (a1);
        \node [transition] (a4) [below right of=p3] {$a_4$}
        edge [post] (p3);
        \node [place] (p4) [label=right:$p_4$,below left of=a4] {}
        edge [post] (a4);
        \node [transition] (a3) [left of=p4] {$a_3$}
        edge [pre] node[swap] {2} (p3)
        edge [post] (p4);
        \node [place] (p2) [label=left:$p_2$,left of=a3] {}
        edge [pre] (a1);
        \node [transition] (a2) [above left of=p2] {$a_2$}
        edge [pre] (p2);
        \draw[shorten >=0.3mm,shorten <=0.3mm,<->] (a2) to[in=150,out=0] ($(-1cm,-2.5cm)$) to[in=135,out=330] (p4);
        \node [place] (p1) [label=left:$p_1$,above right of=a2] {}
        edge [post] (a1)
        edge [pre] (a2);
        \draw[shorten >=0.3mm,shorten <=0.3mm,<->] (p1) to (a4);
    \end{scope}
\end{tikzpicture}
\end{center}
\caption{A 5-dim net $A_1$ (left), and a 4-dim net $A_2$
	(right), a restriction of $A_1$.}
    \label{fig:example}
\end{figure}

\paragraph{Petri nets.}
A common definition presents a Petri net as a tuple $(P,T,W)$ where $P$ and
$T$ are finite disjoint sets of \emph{places} and \emph{transitions},
respectively, and $W\colon (P\times T)\cup(T\times P)\to \setN$
is viewed as a set of \emph{weighted arcs}.
Figure~\ref{fig:example} 
depicts places as circles, transitions as boxes, and nonzero-weight
arcs, while leaving weights equal to $1$ implicit; an arc with arrows
at both ends denotes two arcs, one in each direction. 
Nevertheless, we adopt an equivalent definition that we find notationally more convenient in our context.

\paragraph{Petri nets as sets of actions.}
A \emph{Petri net} $A$ \emph{of dimension} $d\in\setN$, a
\emph{$d$-dim net $A$} for short, is a finite set of pairs
$a=(a_-,a_+) \in \Nat^d \times \Nat^d$ 
which are referred to as 
 \emph{actions} (a synonym for \emph{transitions}).
 Each action $a=(a_-,a_+)$ has an associated \emph{displacement},
defined as $\Delta(a)=(a_+-a_-)\in\setZ^d$. 
We set 
	\[
	\nnorm{A}=\max_{a\in A}\nnorm{a},
	\textnormal{ where }
\nnorm{a}=\nnorm{\{a_+,a_-\}}.
\]
We note the close relation to \emph{vector addition
	systems} (VASs), where a Petri net action
$a=(a_-,a_+)$ is replaced by the plain vector $\Delta(a)$.

\paragraph{Markings, and relations  \bm{$\step{a}$},
\bm{$\xdashrightarrow{a}$}.}
A \emph{marking} of $A$ is a vector $x\in\Nat^d$ assigning the value $x(i)$, that is, the number of \emph{tokens}, to each \emph{component} (a synonym for \emph{place}) $i\in[1,d]$.
An \emph{action} $a$ is
\emph{enabled at a marking} $x$ if $x\geq a_-$.
The ``one-step'' \emph{relation} $\step{a}$
 is defined on $\setN^d$  by setting ${x}\step{a}{y}$
if  $x\geq a_-$ and $y=x+\Delta(a)=x-a_-+a_+$.
The \emph{virtual} ``one-step'' \emph{relation} $\virtual{a}$
 is defined on $\setZ^d$  by setting ${x}\virtual{a}{y}$
if  $y-x=\Delta(a)$.
Hence, $x \step{a} y$ implies $x \virtual{a} y$, whereas the converse, even for $x,y \in \setN^d$, does not necessarily hold.

\paragraph{Conservative nets.}
A $d$-dim \emph{net} $A$ is \emph{conservative} if there exists
$w\in(\setN_+)^d$ such that $\scalar{w}{\Delta(a)}=0$ for all $a\in A$;
in this case  ${x}\step{a}{y}$ implies $\scalar{w}{x}=\scalar{w}{y}$.
If, moreover, $w(i)=1$ for all $i\in[1,d]$, then $A$ is \emph{$1$-conservative}.
Note that in a $1$-conservative net, performing actions preserves the
total number of tokens, whereas in a conservative net, only a \emph{weighted} sum of tokens is maintained.

\begin{exa}\label{exa:basicnet}
Figure~\ref{fig:example}(left) shows
a $5$-dim net $A_1$ with $4$ actions. 
For instance,
$a_1=
\left((a_1)_-,(a_1)_+\right)=
\left((1,0,1,0,1),(0,1,2,0,0)\right)$,
$\Delta(a_1)=({-}1,1,1,0,{-}1)$,
	$\Delta(a_2)=(1,{-}1,0,0,0)$,
$\Delta(a_3)=(0,0,{-}2,1,0)$.
We have 
$(0,2,0,1,0)\step{a_2}(1,1,0,1,0)$, and
	$(0,2,0,0,0)\virtual{a_2}(1,1,0,0,0)$,
	but \emph{not}
$(0,2,0,0,0)\step{a_2}(1,1,0,0,0)$.
The net $A_1$ is conservative, as demonstrated by the vector 
$w=(1,1,1,2,1)$.
The net $A_2$ in Figure~\ref{fig:example}(right)
	is obtained from $A_1$ by removing the fifth component, that
	is, 
	place $p_5$. Note that $A_2$ is 
	not
	conservative; indeed, $\Delta(a_1a_2)=(0,0,1,0)$ precludes the
	existence of a preserved 
 weighted sum with positive component weights.
\end{exa}

\paragraph{The relation \bm{$\step{\sigma}$}, 
reachability set \bm{$\Reach(x)$}, virtual reachability relation
\bm{$\xdashrightarrow{\sigma}$}.}
Given a $d$-dim net $A$,
we define the relation
$\step{\sigma}\, \subseteq \setN^d\times\setN^d$
for any action sequence $\sigma=a_1a_2\ldots a_k$
as the
composition $\step{a_1}\circ\step{a_2}\cdots\circ\step{a_k}$.
The \emph{displacement of} $\sigma$ is given by 
$\Delta(\sigma)=\sum_{j=1}^k \Delta(a_j).$
We also refer to $x\step{\sigma}y$ as an
\emph{execution in} $A$, \emph{from} a marking $x$ \emph{to} a marking $y$.

The \emph{reachability relation in} $A$ 
is $\reach\,\subseteq\setN^d\times\setN^d$, where
$x\xrightarrow{*}y$ holds if $x\xrightarrow{\sigma}y$ for some
action sequence $\sigma$. The \emph{reachability set of} a marking $x$
is defined as $\Reach(x)=\{y\mid x\reach y\}$.
We also define the relation $\virtual{\sigma}\, \subseteq
\setZ^d\times\setZ^d$ by setting $x\virtual{\sigma}y$ if
$y-x=\Delta(\sigma)$,
and we refer to  $x\virtual{\sigma}y$ as a \emph{virtual execution in $A$}.
The \emph{virtual reachability relation in} $A$ 
is $\vreach\,\subseteq\setZ^d\times\setZ^d$, where
$x\vreach y$ holds if $x\virtual{\sigma}y$ for some
action sequence $\sigma$.
 We set
\[A\D=\{\Delta(a)\mid a\in A\},\]
and note that 
 $x\virtual{*} y$ if and only if $(y-x)$ belongs to the submonoid $(A\D)^*$ of
 $(\setZ^d,+)$.
We further write $A\D^*$ instead of $(A\D)^*$.

\paragraph{Reversible nets.}
A~net $A$ is 
\emph{structurally reversible}, referred to as \emph{reversible} in this
paper, if the monoid $A\D^*$ is a~subgroup of $(\setZ^d,+)$, that is, if for every
$a\in A$ we have $-\Delta(a)\in A\D^*$.
Note that $A$ is reversible if and only if
 virtual reachability is symmetric ($x\virtual{*}y$ iff
$y\virtual{*}x$).

\paragraph{Restricted nets and their executions.}
For a $d$-dim net $A$ and $I\subseteq [1,d]$, 
\begin{equation*}
	\textnormal{the $|I|$-dim net $A\restr{I}$}
\end{equation*}
	is the \emph{restriction of} $A$ \emph{to} the
	\emph{components} (with indices)  in $I$.
	Each action $a=(a_-,a_+)$ of $A$
	gives rise to an action
	$a\restr{I}=({a_-}\restr{I},{a_+}\restr{I})$ of $A\restr{I}$.
We refer to 
executions 
$x\step{\sigma\restr{I}}y$ in $A\restr{I}$,  implicitly assuming 
that $x,y\in\setN^I$, and
adopt the convention that
\begin{equation*}
	\textnormal{$x\restr{I}\step{\sigma}y\restr{I}$ stands for
	$x\restr{I}\step{\sigma\restr{I}}y\restr{I}$.} 
\end{equation*}
Analogous conventions apply to virtual executions
$x\virtual{\sigma\restr{I}}y$, 
$x\restr{I}\virtual{\sigma}y\restr{I}$.

\begin{rem}\label{rem:dimzero}
Our definition does not exclude the nets of dimension $0$. Though these
trivial nets are of no interest, they may arise in our context as
$A\restr{I}$ when $I=\emptyset$. Formally, the empty tuple
	$()$ is the unique marking of the net $A\restr{\emptyset}$,
	and we have $()\step{a}()$ for all actions $a\in A$.
\end{rem}	

\begin{prop}[\textbf{Restrictions preserve reversibility}]\label{prop:restrpresrev}
\hfill\\
If a $d$-dim net $A$ is reversible and $I\subseteq[1,d]$, then 
$A\restr{I}$ is reversible.
\end{prop}

% Fix Indentation
\noindent
Note that this implication does not hold for
conservativeness, as demonstrated by Example~\ref{exa:basicnet}, where
$A_1$ is conservative but its restriction $A_2=(A_1)\restr{[1,4]}$ is not.

\paragraph{Liveness, structural liveness, dead markings.}
A \emph{marking} $x\in\setN^d$ of a $d$-dim net $A$ is \emph{live} if
for every action $a\in A$ and every $x'\in \Reach(x)$, 
 there exists $y\in \Reach(x')$ such that
 $y\geq a_-$ (that is, $a$ is enabled at $y$, allowing the step
 $y\step{a}y+\Delta(a)$).  A~ $d$-dim \emph{net} $A$ is \emph{structurally
 live} if there exists a marking $x\in\setN^d$ of $A$ that is live.
 Note the following standard fact:

\begin{prop}[\textbf{Liveness is closed under
	reachability}]\label{prop:liveunderreach}
\hfill\\	
If a marking $x$ of a net $A$ is live, then all
 markings $y\in\Reach(x)$ are also live.
\end{prop}

% Fix Indentation
\noindent
An \emph{action} $a\in A$ is \emph{dead at} $x$ if 
no marking in $\Reach(x)$ enables
$a$; that is, for each $y\in
\Reach(x)$ we have $y\ngeq a_-$.
For technical convenience, we say that
a \emph{marking} $x$ 
is \emph{dead} if some \emph{action} $a\in A$ is \emph{dead at} $x$.
(While in other contexts a marking is typically considered dead only if
\emph{all} actions are
dead, we adopt this weaker notion throughout the paper.) 
With this terminology, we observe the following simple fact:

\begin{prop}[\textbf{Nonliveness as reachability of dead markings}]\label{prop:nonlivereachdead}
\hfill\\
	A marking $x$ of a net $A$ is nonlive if and only if 
	$\Reach(x)$ contains a dead marking.
\end{prop}

\begin{exa}\label{exa:basiclive}
The net $A_1$ in Figure~\ref{fig:example} 
is structurally live. Indeed, $x_0=(1,0,1,0,2)$ is a
    live marking, which is easily verifiable since there is only one (infinite)
    execution from $x_0$: $(1,0,1,0,2) \step{a_1} (0,1,2,0,1) \step{a_3}
    (0,1,0,1,1) \step{a_2} (1,0,0,1,1) \step{a_4}
    (1,0,1,0,2)\step{a_1}\cdots$.

   By adding a token in $p_4$, we obtain a nonlive marking
	$x'_0=(1,0,1,1,2)$. Specifically,
	$x'_0\step{a_1a_2a_1}y=(0,1,3,1,0)$,
where $y$ is dead because actions $a_1$ and $a_4$ are dead at $y$
	(due to
	$y(5)=0$).
We could also verify that a necessary condition for $x$ to be live is
that 
$x(i)< 3$ for all $i\in[1,4]$; however, there is no such bound on $x(5)$.
(If $x(i)\geq 3$ for some $i\in[1,4]$, then $x$ can reach a~dead
	marking; on the other hand, there exist live markings $x$ with arbitrarily
	large $x(5)$.)
\end{exa}

\begin{rem}
Our example also demonstrates that liveness is not monotonic, even
in conservative nets; that is, $x\leq x'$ and $x$ being live does not
	necessarily imply that $x'$
is live. In general, the set of live markings of
	a $1$-conservative net is not even semilinear (that is, not expressible
	in Presburger arithmetic); the example
	from~\cite{DBLP:journals/acta/JancarP19} can be adapted to
	$1$-conservative nets.
\end{rem}

\paragraph{Linear Systems \bm{$S$}, and their size-parameters
\bm{${\nnorm{S}}$},
\bm{${\mlcm{S}}$}.} 
We now define systems that are, in essence, 
quantifier-free formulas of Presburger arithmetic with divisibility 
constraints. Such a formula is $d$-dimensional if it contains $d$ 
integer variables.

Let $\bvar{x}$ be a variable vector ranging over $\setZ^d$, a \emph{$d$-dim
variable} for short, viewed as a tuple
$\bvar{x}=(\bvar{x}_1,\bvar{x}_2,\dots,\bvar{x}_d)$ of $d$ integer
variables.
A constraint of
the form $\scalar{\alpha}{\bvar{x}}\sim c$ where $\alpha\in \setZ^d$,
$c\in\setZ$, and $\sim\mathop{\in}\,\{=,\geq\}$ is an \emph{equality
constraint} if $\sim$ is $=$, and an \emph{inequality constraint} if
$\sim$ is $\geq$. It is a~\emph{homogeneous constraint} if $c=0$.
For $m\in\setN_+$, a~constraint 
$\scalar{\alpha}{\bvar{x}}\equiv c\,(\bmod\, m)$, where
$\alpha\in[0,m{-}1]^d$ and $c\in [0,m{-}1]$, is a \emph{divisibility constraint}.

A \emph{$d$-dim linear system} $S$
is a propositional formula whose atomic propositions are 
equality, inequality, and divisibility constraints for a fixed
$d$-dim variable.
The \emph{set of solutions} $\sem{S}$ 
 consists of the vectors $x\in\setZ^d$
satisfying $S$; if $\sem{S}\neq\emptyset$, then 
$S$ is \emph{satisfiable}. 

By the \emph{norm}
$\nnorm{S}$ we mean the least $s\in\setN$ such that 
$\max\{\nnorm{\alpha},|c|\}\leq s$
for all equality and inequality constraints $\scalar{\alpha}{\bvar{x}}\sim c$ in
$S$.
Moreover, $\mlcm{S}$ is the \emph{least common multiple} of all $m$
occurring in $(\bmod\, m)$ in 
divisibility constraints 
in $S$, with $\mlcm{S}=1$ if no such
constraints exist.

\paragraph{Small solution theorems.}
When dealing with solutions of linear systems, \emph{small-solution
theorems} provide bounds on the size of minimal solutions. 
Given a set $M\subseteq \setN^d$, 
by $\min_\leq(M)$ we denote the set of minimal vectors in $M$ with
respect to
the component-wise order $\leq$. Since $\leq$ is 
a~well-partial-order on $\setN^d$, the set $\min_\leq(M)$ is
always finite.
In~\cite{DBLP:conf/rta/Pottier91}, 
Pottier provided several
bounds for minimal
nonnegative solutions of a conjunction of homogeneous equality constraints. 
We now recall a bound that is particularly useful for our purposes.

\begin{lemC}[\cite{DBLP:conf/rta/Pottier91} (\textbf{Homogeneous
	equality constraints})]\label{lem:smallsolution}
\hfill\\	
  Let $M=\{x\in\setN^n \mid Bx=\mathbf{0}\}$ where $B$ is an $m\times n$
	matrix over $\setZ$. Then
	$X=\min_{\leq}(M\smallsetminus\{\mathbf{0}\})$ is a finite set
	such that $M=X^*$. Moreover the following bound holds, where
	$r=\rank{B}$:
\begin{equation*}
\norm{1}{X}\leq (1+\norm{1}{B})^r.
\end{equation*}
\end{lemC}

% Fix Indentation
\noindent
Note that the rank bound $r$ in
Lemma~\ref{lem:smallsolution} satisfies $r\leq\min\{m,n\}$.
There is also a~bound on the
solutions of a conjunction of inequality constraints, 
$Bx\geq b$, in~\cite{DBLP:conf/rta/Pottier91}, but it involves an exponent
of $m$. This is not convenient for us when $m$, the number of
constraints, is much larger than $n=d$, the dimension of our problem.
Therefore in Section~\ref{sec:smallsolutions}, we provide a proof of
the following theorem:

\begin{thm}[\textbf{Bounds on solutions of linear systems}]\label{thm:smallsolutions}
\hfill\\
	Every satisfiable $d$-dim linear system $S$ 
 has a solution $x\in\setZ^d$ such that
	\[\nnorm{x}\leq
		\mlcm{S} \cdot \left(1+ (d+1)\,!\cdot (d\cdot
		\nnorm{S}+1)^{d}\right).\]
\end{thm}

\begin{rem}
Later, in
the proof of Theorem~\ref{th:upper}, we employ a $d'$-dim linear system $S$
with $d'=d\cdot2^d$. 
It is crucial for our purposes that Theorem~\ref{thm:smallsolutions} then yields a solution that is
	at most doubly exponential (2-exp) in $d$.  
In contrast, using~\cite[Theorem 3.12]{DBLP:journals/tocl/Klaedtke08},
	we could only derive
	that for $S$ (even without divisibility constraints),  
the size of the minimal automaton encoding $\sem{S}$ 
in binary is bounded by $(2+2\cdot\nnorm{S})^{|S|}$,
where $|S|$ is the number of constraints in $S$. In our case, $|S|\geq
2^d$, and thus the bounds on the minimal solutions of $S$ derived from the
shortest accepting paths in such an automaton would be at least
$2^{(2+2\cdot\nnorm{S})^{|S|}}$,
	which is triple-exponential (3-exp) in $d$.
\end{rem}

\paragraph{Linear Systems for Subgroups of \bm{$\setZ^d$}.}
In the sequel, by a~\emph{group} we mean 
a \emph{subgroup} of $(\setZ^d,+)$,
which is also called a \emph{lattice} in this context.
The group \emph{spanned} by a set $X\subseteq \setZ^d$ is 
the monoid $(X\cup-X)^*$, that is, the set of finite integer linear
combinations of elements of $X$. 
In Section~\ref{sec:virtual}, we prove the following theorem, which
provides a way to encode any such group by a linear system, with a
bound on its  size.

\begin{thm}[\textbf{Groups expressed by linear systems of bounded size}]\label{thm:virtual2system}
\hfill\\	
Let $L$ be the group spanned by a finite set $X\subseteq
\setZ^d$. There exists a~$d$-dim linear system $S$ such that
\begin{center}
$\sem{S}=L$ and $\max\{\nnorm{S},\mlcm{S}\}\leq d\,!\cdot
	\norm{}{X}^d$.
\end{center}
\end{thm}

% Fix Indentation
\noindent
We thus obtain a corollary characterizing the virtual reachability of
reversible nets, recalling that $A\D=\{\Delta(a)\mid a\in A\}$: 
\begin{cor}[\textbf{Reversible virtual reachability expressed by a linear system}]\label{cor:virtual2system}
\hfill\\
	For every reversible $d$-dim net $A$, there exists
	a~$2d$-dim linear system $S_A$ such that
		\begin{center}
		$\sem{S_A}=\{(x,y)\in\setZ^d\times\setZ^d \mid
			x\virtual{*} y\}$, and
	$\max\{\nnorm{S_A},\mlcm{S_A}\}\leq d\,!\cdot\nnorm{A\D}^d$.
\end{center}
\end{cor}
\begin{proof}
	We recall that $x\virtual{*} y$ iff $y-x\in A\D^*$, and that 
	$A\D^*$ is a subgroup of $(\setZ^d,+)$ since $A$ is
reversible.
	Let $S$ be the linear system guaranteed 
	by 
	Theorem~\ref{thm:virtual2system} for the group $L=A\D^*$
	(spanned by the set $A\D$)
	 with a~$d$-dim variable
	 $\bvar{x}'=(\bvar{x}'_1,\bvar{x}'_2,\dots,\bvar{x}'_d)$.

	We construct $S_A$ with a~$2d$-dim variable 
	$(\bvar{x}_1,\bvar{x}_2,\dots,\bvar{x}_d,\bvar{y}_1,\bvar{y}_2,\dots,\bvar{y}_d)$ 
	 by replacing every occurrence of
	$\bvar{x}'_i$ in $S$ with the expression
	$(\bvar{y}_i-\bvar{x}_i)$. It follows that 
	\[\sem{S_A}=\{(x,y)\in\setZ^d\times\setZ^d\mid
	y-x\in\sem{S}\},\]
	which is precisely the set of pairs $(x,y)$ such that $x\virtual{*} y$.
Since the transformation only involves substituting a difference of
	variables for a single variable, the norms and the lcm remain
	unchanged: $\nnorm{S_A}=\nnorm{S}$, $\mlcm{S_A}=\mlcm{S}$.
 The claimed bound $d\,!\cdot\nnorm{A\D}^d$
then follows directly from
	Theorem~\ref{thm:virtual2system}.
\end{proof}
% Fix Indentation
\noindent
We will employ Theorem~\ref{thm:smallsolutions} and
Corollary~\ref{cor:virtual2system} to prove our main result,
Theorem~\ref{th:upper}. Before stating the result, we introduce
some further relevant notions.

\paragraph{Live bottom SCCs in reachability graphs.}
Given a $d$-dim net $A$, we denote by $(\setN^d,\to)$ 
the \emph{reachability graph} of
$A$, where $x\to y$ if $x\step{a}y$
for some $a\in A$. A nonempty set $X\subseteq \setN^d$ such that
$\Reach(x)=X$ for all $x\in X$ corresponds to 
 a \emph{bottom strongly connected component}  
in the graph; we simply call $X$ a \emph{bottom SCC} of $A$ in this case. 
A bottom SCC $X$ is \emph{live} if some $x\in X$ is live (which
implies that all $x\in X$ are live, by
Proposition~\ref{prop:liveunderreach}).

\paragraph{1-exp and 2-exp functions.}
We call a function $f\colon\setN^2\to\setN$ 
a \emph{1-exp} function if 
\[f(m,i)\leq (2+m)^{\text{poly}(i)},\]
meaning that there exists a polynomial $p\colon\setN\to\setN$
such that 
$f(m,i)\leq (2+m)^{p(i)}$ for all $m,i\in\setN$.
It is a \emph{2-exp} function if 
\[f(m,i)\leq (2+m)^{2^{\text{poly}(i)}}.\]
Throughout the paper, we specifically use values of the form $f(\nnorm{A},d)$ or 
$f(\nnorm{A},|I|)$ for $I\subseteq[1,d]$, where $A$ is a $d$-dim net.

\begin{thm}[\textbf{Bounds on minimal live markings in 
	nets with live bottom SCCs}]\label{th:upper}
	There exists a 2-exp function $f$ such that 
	for every $d$-dim net $A$ possessing a live bottom SCC, 
	there exists a live marking
	$x\in\setN^d$
	satisfying $\nnorm{x}\leq f(\nnorm{A},d)$.
\end{thm}

\paragraph{Live bottom SCCs, reversibility, conservativeness,
	structural boundedness.}
 We say that a \emph{marking} $x\in\setN^d$ of a~$d$-dim net $A$ is
 \emph{bounded} if the set of reachable markings $\Reach(x)=\{y\mid
 x\reach y\}$ is finite; that is, 
 there exists $B_x\in\setN$ such that $\nnorm{y}\leq B_x$ for
 all $y\in\Reach(x)$. A \emph{net} $A$ is \emph{structurally
 bounded} if every marking $x$ of $A$ is bounded.

The following lemma shows that structurally live conservative nets coincide
	with structurally live structurally bounded nets, and that 
	Theorem~\ref{th:upper} applies to these nets.

\begin{lem}[\textbf{Conservativeness vs.~structural boundedness for
	live nets}]\label{lem:consstrbound}\hfill
	\begin{enumerate}
\item
Every conservative net is structurally bounded.
\item
Every structurally bounded net that is structurally live has a live
			bottom SCC.	
		\item
Every net possessing a live bottom SCC is reversible.			
		\item
Every structurally bounded net that is reversible is conservative.
\end{enumerate}
\end{lem}
\begin{proof}
	(1)
	If a $d$-dim net $A$ is conservative, there exists $w\in(\setN_+)^d$
	such that $\scalar{w}{\Delta(a)}=0$ for all $a\in A$. It
	follows that 
	$x\reach y$ implies $\scalar{w}{x}=\scalar{w}{y}$. Since  
 all components of $w$ are positive, the set $C_x=\{y\in\setN^d\mid
	\scalar{w}{y}=\scalar{w}{x}\}$ is finite for any $x\in\setN^d$. Since
	$\Reach(x)\subseteq C_x$, the net $A$ is structurally bounded.

(2)
Let $x$ be a live marking of a structurally bounded net $A$.
	Since $x$ is bounded, $\Reach(x)$ is finite, which implies
	that there must
	exist a (finite) bottom SCC $X\subseteq\Reach(x)$ in the
	reachability graph of $A$.
	Since $x$ is live, every $y\in\Reach(x)$ is also live (by
	Proposition~\ref{prop:liveunderreach});
	consequently, $X$ is a live bottom SCC of $A$.

	(3) Let $X$ be a (finite or infinite) live bottom SCC of a net
	$A$.  By the definition of 
liveness,
	for
 each action $a\in A$, there exists some $y\in X$ at which $a$ is enabled,
that is, 
 $y\step{a}y+\Delta(a)$. Since $X$ is a bottom SCC, we must have
	$y+\Delta(a)\reach y$.
	This implies that $y-(y+\Delta(a))=-\Delta(a)\in A\D^*$. 
	Therefore, $-\Delta(a)\in A\D^*$ for every $a\in A$, which
	shows that $A$ is
	reversible.

	(4)
	Let $A=\{a_1,a_2,\ldots,a_k\}$ be a $d$-dim net 
that is structurally bounded and reversible.
	For each $i\in[1,d]$, let $e_i\in\setN^d$
	be the $i$th unit vector (satisfying
	$e_i(i)=1$ and $e_i(j)=0$ for all $j\neq i$). We observe that
	there exists no
	$v\in \setN^k$ such that
	\begin{equation}\label{eq:biggerei}
		\sum_{j=1}^{k} v(j)\cdot\Delta(a_j) \;\geq\; e_i.
	\end{equation}
Indeed, if such $v$ existed, then for an action sequence 
$\sigma$ with Parikh vector $v$  
	(\eg, $\sigma=(a_1)^{v(1)}\ldots (a_k)^{v(k)}$)
	and a sufficiently large marking $x$, we would have
	 $x\step{\sigma}x+\Delta(\sigma)$ with $\Delta(\sigma)\geq
	 e_i$.
This would imply $x\step{\sigma^n}x+n\cdot\Delta(\sigma)\geq x+n\cdot
	e_i$ for all $n\in\setN$, contradicting structural
	boundedness.

Since $A$ is reversible, every $\Delta(a_j)$ can be ``undone'' by
	some action sequence; that is, $-\Delta(a_j)=\sum_{\ell=1}^k
	u(\ell)\cdot\Delta(a_\ell)$ for some $u\in\setN^k$.
Hence every integer combination of the vectors $\Delta(a_j)$ equals some
	nonnegative one, so \eqref{eq:biggerei} has no solution
	$v\in\setZ^k$ either.
	Finally, the $\Delta(a_j)$ are integer vectors, so if
	\eqref{eq:biggerei} has a real solution then it has a rational one,
	and hence an integer one after multiplying by a common denominator.
	Therefore \eqref{eq:biggerei} has no solution $v\in\setR^k$ either.

Now we recall the following variant of Farkas' lemma 
(see, \eg, Proposition 6.4.3(iii)
in~\cite{DBLP:books/daglib/0016926}):
\begin{quote}
	Let $M$ be a real matrix with $m$ rows and $n$ columns, and let
	$b \in \setR^m$ be a vector.
	The system $Mx \leq b$ has a solution $x \in \setR^n$ if and only if
	every nonnegative $y \in \setR^m$ with $y^{\transp} M =
	\mathbf{0}^{\transp}$ also satisfies $y^{\transp} b \geq 0$.
\end{quote}
Fix $i\in[1,d]$, let $M$ be the $d\times k$ matrix whose $j$th column
is $-\Delta(a_j)$, and let $b=-e_i$.  Then $Mv\leq b$ is
precisely~\eqref{eq:biggerei}, which has no solution $v\in\setR^k$.
By the lemma, there exists a nonnegative
$y\in\setR^d$ with $y^{\transp}M=\mathbf{0}^{\transp}$ and
$y^{\transp}b<0$.  The $j$th entry of $y^{\transp}M$ is
$\scalar{y}{-\Delta(a_j)}$, so the former condition amounts to
$\scalar{y}{\Delta(a_j)}=0$ for all $j\in[1,k]$; and
$y^{\transp}b=\scalar{y}{-e_i}=-y(i)$, so the latter amounts to
$y(i)>0$.
By the rationality argument above, $y$ may be chosen to be a rational
vector, and hence an integer vector after multiplying by a common
denominator; both conditions are preserved.
Writing $w_i$ for the resulting vector, we
obtain for each $i\in[1,d]$ some $w_i\in\setN^d$ with $w_i(i)>0$ and
$\scalar{w_i}{\Delta(a_j)}=0$ for all $j\in[1,k]$.

	Setting $w=\sum_{i=1}^d w_i$ yields a vector $w\in(\setN_+)^d$
	satisfying $\scalar{w}{\Delta(a_j)}=0$ for all $j\in[1,k]$.
	Consequently, the net $A$ is conservative.
\end{proof}

% Fix Indentation
\noindent
The EXPSPACE-hardness result  
from~\cite{DBLP:journals/acta/JancarP19} will be slightly
strengthened by Theorem~\ref{thm:lowerbound} in
Section~\ref{sec:hardness}; Theorem~\ref{th:upper} thus allows us to
obtain a matching
upper bound.

\begin{cor}[\textbf{EXPSPACE-completeness of structural liveness}]\label{cor:conservEScomplete}
\hfill
	\begin{enumerate}
		\item			
	Structural liveness for conservative nets is EXPSPACE-complete.
\item
	Structural liveness for structurally bounded nets is EXPSPACE-complete.
	\end{enumerate}
	\end{cor}	
\begin{proof}
	Since EXPSPACE$=$NEXPSPACE,
it suffices to provide a nondeterministic algorithm working in
	exponential space
	that verifies positive instances of the considered problems.
	By Lemma~\ref{lem:consstrbound}, the positive instances of
	both problems coincide; in particular, every structurally live
	structurally bounded net is conservative. 

	Given a structurally live conservative $d$-dim net $A$, an 
	algorithm can guess a live
	marking $x\in\setN^d$ as guaranteed by Theorem~\ref{th:upper}.
 Since  $\nnorm{x}\leq f(\nnorm{A},d)$ for the corresponding 2-exp function
	$f$
	satisfying	$f(m,i)\leq(2+m)^{2^{\text{poly}(i)}}$, the
	binary representation of $x$ has size at most
	$O(2^{\text{poly}(d)}\cdot\log (2+\nnorm{A}))$, which fits in exponential space.
Finally, verifying the liveness of $x$ for $A$ can be performed in
	polynomial space with respect to
	the binary encoding of $A$ and $x$ (see, \eg,~\cite[Lemma
	5]{DBLP:conf/apn/MayrW14}), which remains within exponential
	space relative to the original input size. 
\end{proof}

\section{Upper Bound (Proof of Theorem~\ref{th:upper})}\label{sec:upperbound}
In Section~\ref{sec:infproof}, a proof of Theorem~\ref{th:upper} is
presented; the proofs of some auxiliary results are deferred to
Sections~\ref{subsec:dead} and~\ref{subsec:virtrevers}. 
We begin with several
observations and conventions.

\paragraph{1-exp and 2-exp functions.}

We have defined 1-exp and 2-exp functions as functions $f\colon\setN^2\to\setN$
satisfying
\begin{equation*}
	f(m,i)\leq (2+m)^{\text{poly}(i)}\text{ and }
f(m,i)\leq (2+m)^{2^{\text{poly}(i)}},
\end{equation*}
	respectively.
For technical convenience, we implicitly assume that each such 
function $f$ that we consider is \emph{non-decreasing} in both arguments; that is, 
$m \leq m'$ and $i \leq i'$ imply $f(m, i) \leq f(m', i')$. 
Furthermore, we assume that $f$ is \emph{extensive}, meaning that $m \leq f(m,i)$.

It is straightforward to verify that the class of 1-exp functions, as
well as  the class of 2-exp functions, 
 is
closed under standard operations such as sum, product, and
composition, where the latter is
defined as \[(f\circ g)(m,i)=f(g(m,i),i).\]
For instance, if $f(m,i)\leq	(2+m)^{{p_1(i)}}$
and 
 $g(m,i)\leq	(2+m)^{{p_2(i)}}$ with $p_1(i)\geq 1$ and
 $p_2(i)\geq 1$, then  
\[(f\circ g)(m,i)\leq
\left(2+(2+m)^{{p_2(i)}}\right)^{{p_1(i)}}\leq 
(2+m)^{{2\cdot p_2(i)\cdot p_1(i)}}.\]
The functions $p_1, p_2$
can be, in particular, polynomials or exponential functions.

 We define the $j$-th iterate of $f$ recursively
as follows:
\begin{equation*}
	f^{(0)}(m,i)=m,\text{ and }
f^{(j+1)}(m,i) = f(f^{(j)}(m,i), i).
\end{equation*}
Note that 
if $f(m,i)$ is a 1-exp function or a 2-exp function, then $f'(m,i)=f^{(i)}(m,i)$ 
is a 2-exp function. Indeed,
if $f(m,i)\leq (2+m)^{{p(i)}}$, where $p$ is a polynomial or
exponential function, then
\[f'(m,i)=f^{(i)}(m,i)\leq (2+m)^{2^{i-1}\cdot
(p(i))^i}=(2+m)^{2^{(i-1)+ i\cdot \log
p(i)}}\leq(2+m)^{2^{\text{poly}(i)}}.\]
We use these closure properties
implicitly when deriving the existence of specific 1-exp or 2-exp functions 
throughout this section.

\paragraph{1-exp and 2-exp constants.}

For convenience, when considering a $d$-dim net $A$, we also
refer to \emph{1-exp} and \emph{2-exp constants}. We say that $C$ is a 1-exp (resp.~2-exp)
constant related to $A$ if
$C=f(\nnorm{A},d)$ for some 1-exp (resp.~2-exp) function that is independent of $A$.
For instance, we will use a 2-exp constant $C\dead^A$ associated
with a 2-exp function $f\dead$; the superscript $A$ clarifies that 
$C\dead^A=f\dead(\nnorm{A},d)$, where $d$ is the dimension of the net $A$.

\subsection{Proof of
Theorem~\ref{th:upper}.}\label{sec:infproof}
\hfill\\

We aim to show the existence of a 2-exp function $f$ such that for every $d$-dim
net $A$ possessing a live bottom SCC $X\subseteq\setN^d$ (where every
$x\in X$ is a live marking and $\Reach(x)=X$ for all
$x\in X$),
 there exists a live marking
$x\in\setN^d$ satisfying $\nnorm{x}\leq f(\nnorm{A},d)$.
 Recall that 
such a net $A$ is reversible by Lemma~\ref{lem:consstrbound}.

In this section, we provide the overall proof, deferring certain
technical parts 
to the subsequent sections.

\begin{rem}\label{rem:Lerresults}
Some technical issues are addressed by referring to results from~\cite{DBLP:conf/fsttcs/Leroux19}. Specifically, these concern
$d$-dimensional \emph{extractors} (used to distinguish between small
	and large components of markings, or of sets of markings),
a 1-exp \emph{virtual reachability} function $f\vr$ for reversible
	Petri nets with states (PNSs), and 
	a 2-exp \emph{mutual-reachability} function $f\mr$ for general
	Petri nets.
\end{rem}

\paragraph{Dead markings, 2-exp constant $\bm{C\bdead^A}$}
Recall that a marking $x\in\setN^d$ of a $d$-dim net $A$ is live if and only if there is no
dead marking in $\Reach(x)$
(Proposition~\ref{prop:nonlivereachdead}). By our convention, a
marking $y$ is dead if at least one action $a$ is dead at $y$, that
is, $y'\not\geq a_-$ for all $y'\in\Reach(y)$.
Our goal is to bound the minimal size of solutions $x\in\setN^d$ to the constraint
\begin{equation}\label{eq:firstconstr}
	\textnormal{``for each dead marking $y$, we have  $x\not\reach
	y$''.}
\end{equation}
In Section~\ref{subsec:dead} we recall, by
Lemma~\ref{lem:rackoffdead} using  Rackoff's
result in~\cite{DBLP:journals/tcs/Rackoff78}, that 
\begin{equation*}
	\textnormal{\emph{the set of dead markings} of a $d$-dim net
	$A$} 
\end{equation*}
	can be presented
as the set of solutions $y\in\setN^d$ 
of a~disjunction
\begin{equation}\label{eq:alldead}
	\bigvee_{I\subseteq [1,d]} \,\bigvee_{j=1}^{k_I}
	\left(\bvar{y}\restr{I}=\alpha_{(I,j)}\right)
\end{equation}
where 
each (restricted) marking $\alpha_{(I,j)}\in\setN^{I}$ is \emph{dead in the 
restricted net} $A\restr{I}$ and satisfies 
\[\nnorm{\alpha_{(I,j)}}< C\dead^A,\] for some 2-exp
constant $C\dead^A$.
We allow $k_I=0$ for some $I\subseteq [1,d]$; the corresponding
disjunction $\bigvee_{j=1}^{0}$ is empty, and thus false.  The
disjunction~\eqref{eq:alldead} can therefore be restricted to
$\bigvee_{I\in\mathcal{D}}\cdots$, where
$\mathcal{D}=\{I\subseteq[1,d] \mid k_I>0\}$.  
Note that $\emptyset\notin\mathcal{D}$: in the restricted net
$A\restr{\emptyset}$ no action is ever dead (recall
Remark~\ref{rem:dimzero}), hence no dead $\alpha_{(\emptyset,j)}$
exists.

\begin{rem}
By our convention, the expression ``some 2-exp constant $C\dead^A$''
	means that there exists a non-decreasing function $f\dead(m,i)\leq
	(2+m)^{2^{\text{poly}(i)}}$
such that $C\dead^A=f\dead(\nnorm{A},d)$ for
any $d$-dim net $A$.
A more detailed analysis would involve stating
that	$\nnorm{\alpha_{(I,j)}}< f\dead(\nnorm{A},|I|)$, where 
	$f\dead(\nnorm{A},1)=\nnorm{A}$, and discussing the tight
	relation to the coverability problem (for markings $a_-$). 
	However, for our  purposes, it suffices to simplify the
	analysis by referring only to
	$C\dead^A$, even when a restricted net $A\restr{I}$ is
	considered.

	Technically, we do not rely on the fact that the set of dead
	markings of $A$ is downward closed (that is, if $y$ is dead and $y'\leq y$,
	then $y'$ is also dead). We only use the property that,
	given a marking $y$ of a net $A$, 
if $y\restr{I}$ is dead in the
	restricted net $A\restr{I}$, then $y$ is dead in $A$
	(Proposition~\ref{prop:restrdead}).
\end{rem}

\paragraph{(Virtual) reachability, 1-exp
$\bm{f\bvr}$ and $\bm{C\bvr^A}$, 2-exp $\bm{C\bsuf^A=\max\{C\bdead^A,C\bvr^A\}}$.}
Our proof of Theorem~\ref{th:upper} is based on replacing
the non-reachability relation $\not\reach$ in 
constraint~(\ref{eq:firstconstr}) with the virtual reachability relation
$\vreach$, in a way that will be clarified later. 
Markings in which all components are sufficiently large will play a
crucial role.

\begin{defi}[\textbf{(Sub)markings with all components above a level,
	$\bm{x\in\upC{C}}$}]
\hfill\\
We say that a \emph{vector} $x\in\setN^d$ is \emph{above} a
\emph{level} $C\in\setN$, or simply
$x\in\setN^d$ is \emph{above} $C\in\setN$, denoted
 $x\in\upC{C}$,	if $x(i)\geq C$ for all $i\in[1,d]$.
	A restricted vector
	$x\restr{J}$ with 
	$J\subseteq[1,d]$ is above $C$,
	denoted $x\restr{J}\in\upC{C}$, if  $x(i)\geq C$ for all $i\in
	J$.
\end{defi}

% Fix Indentation
\noindent
The following lemma is a consequence of~\cite[Lemma~5]{DBLP:conf/fsttcs/Leroux19}.

\begin{lem}[\textbf{Virtual and standard reachability in
	reversible nets, $\bm{f\bvr}$,
	$\bm{C\bvr^A}$}]\label{lem:fvr}
\hfill\\	
	There exists a 1-exp function $f\vr$ such that 
for every reversible $d$-dim net $A$ we have:
	\begin{enumerate}
		\item
If $x\vreach y$, then  $x\virtual{\sigma}y$
for some $\sigma$ satisfying
			$|\sigma|\leq
			f\vr(\nnorm{A},d)\cdot\nnorm{y-x}=
		C\vr^A\cdot \nnorm{y-x}$.
\item
If $x\vreach y$
			and  $x,y\in\upC{C\vr^A}$, 
	then $x\step{\sigma}y$
	for some $\sigma$ satisfying
$|\sigma|\leq
		C\vr^A\cdot \nnorm{y-x}$.
	\end{enumerate}
	\end{lem}
% Fix Indentation
\noindent
Intuitively, this can be seen by recalling that $x\vreach y$ means
$y-x\in A\D^*$. 
We highlight the fact that, given a reversible $d$-dim net $A$,  virtual reachability coincides with
standard reachability above $C\vr^A$:
\begin{equation*}
	x\vreach y\iff x\reach y,\text{ for all
	}x,y\in\setN^d\cap \upC{C\vr^A}.
\end{equation*}
For later convenience, we define a sufficiently large 2-exp
constant:
\begin{equation}\label{eq:Csuf}
C\suf^A=\max\{C\dead^A,C\vr^A\}.
\end{equation}

\paragraph{Segment decomposition of virtual executions in reversible nets,
1-exp $\bm{C\bshift^A}$.}
\ \\
In Section~\ref{subsec:virtrevers}, 
we show that
if $x\virtual{*}y$ in a \emph{reversible} $d$-dim net $A$, then
there exists
a~virtual execution 
\begin{equation}\label{eq:virtsegment}
x=x_0\virtual{*}x_1\virtual{*}x_2\cdots\virtual{*}x_k=y
\end{equation}
consisting of \emph{1-exp segments} 
that \emph{stepwise approach the
target} $y$ from $x$.
The segments $x_j\vreach x_{j+1}$ are 1-exp bounded in the sense that
\begin{equation}\label{eq:Cshift}
0<\nnorm{x_{j+1}-x_j}\leq C\shift^A
\end{equation}
for some 1-exp constant $C\shift^A$ (satisfying
$C\shift^A=f\shift(\nnorm{A},d)\leq
(2+\nnorm{A})^{\text{poly}(d)}$).
By the term \emph{stepwise approaching},
	we mean that the sequence $x = x_0, x_1, \dots, x_k = y$ satisfies:
\begin{itemize}
    \item For each $i \in [1, d]$ with $x(i) \leq y(i)$, there exists
	    $j_i\in [0,k]$ such that
		\[ x(i) < x_1(i) < \cdots < x_{j_i}(i) = x_{{j_i}+1}(i) = \cdots = x_k(i)
		= y(i); \]
    \item For each $i \in [1, d]$ with $x(i) > y(i)$,  there exists
	    $j_i\in [0,k]$ such that
		\[ x(i) > x_1(i) > \cdots > x_{j_i}(i) = x_{{j_i}+1}(i) = \cdots = x_k(i)
		= y(i). \]
\end{itemize}

% Fix Indentation
\noindent
Note that Lemma~\ref{lem:fvr}(1) implies that for each segment $x_j\vreach x_{j+1}$
of~(\ref{eq:virtsegment}) there exists  $\sigma$ with
$|\sigma|\leq C\vr^A\cdot C\shift^A$ such that
$x_j\virtual{\sigma}x_{j+1}$,
and even $x_j\step{\sigma}x_{j+1}$ if $x_j,x_{j+1}\in\upC{C\vr^A}$.

\paragraph{Small and large components of vectors, relative to
extractors.}
We need to distinguish between relatively small and large components 
of vectors $x \in \setN^d$ (markings of a net $A$). 
To identify small components, we employ the notion of 
extractors, following~\cite{DBLP:conf/fsttcs/Leroux19}.

\begin{defi}[\textbf{Extractors $\bm{\lambda}$	of dimension $\bm{d}$}]\label{def:extract}
\hfill\\	
For $d\geq 1$, by a \emph{$d$-dim extractor} $\lambda$
we mean a tuple 
	\[(\lambda_1,\lambda_2,\dots,\lambda_d)\in(\setN_+)^d
	\text{ where }\lambda_1\leq\lambda_2 \leq \dots \leq
	\lambda_{d}.\]
For technical convenience, we also refer to the values $\lambda_0$
	and $\lambda_{d+1}$, by
implicitly setting $\lambda_0=1$ and $\lambda_{d+1}=\lambda_d$.
\end{defi}

% Fix Indentation
\noindent
The following claim is straightforward (for further details
see~\cite{DBLP:conf/fsttcs/Leroux19}).

\begin{prop}[\textbf{The set $\bm{I^\lambda_x\subseteq[1,d]}$ of
	(indices of) small
	components of $\bm{x\in\setN^d}$}]\label{prop:Ilambdax}
\hfill\\	
Given a $d$-dim extractor $\lambda$,
for each $x\in\setN^{d}$ there exists a unique maximal
	set $I_x^\lambda\subseteq [1,d]$ such that 
	\[\bigwedge_{i\in
	I^\lambda_x}\left(x(i)<\lambda_{|I^\lambda_x|}\right)\ \land\ 
	\bigwedge_{i\in[1,d]\smallsetminus
	I^\lambda_x}\left(x(i)\geq\lambda_{|I^\lambda_x|+1}\right).
	\]
	More generally, for each subset $J\subseteq [1,d]$, there is 	a unique maximal
	set $I_x^{(\lambda,J)}\subseteq J$ such that 
	\[\bigwedge_{i\in
	I^{(\lambda,J)}_x}\left(x(i)<\lambda_{|I^{(\lambda,J)}_x|}\right)\
	\land \
	\bigwedge_{i\in J\smallsetminus
	I^{(\lambda,J)}_x}\left(x(i)\geq\lambda_{|I^{(\lambda,J)}_x|+1}\right).
	\]
\end{prop}	
% Fix Indentation
\noindent
For instance, if $\lambda=(3,17,258,1500)$ and $x=(11,1600,2,300)$,
$x'=(16,5000,15,5000)$, then $I^\lambda_x=I^\lambda_{x'}=\{1,3\}$.
The values $x(i)$ of the  components $i\in I^\lambda_x$ are viewed as
\emph{small}, while the values $x(i)$ for $i\in [1,d]\smallsetminus I^\lambda_x$ are
\emph{large}. The gap between the values of large components and of
small components of $x$ is bigger
than $\lambda_{j+1}-\lambda_j$, where $j=|I^\lambda_x|$.
In our example, with $|I^\lambda_x|=2$, the gap is bigger than $\lambda_3-\lambda_2=258-17=241$.
\begin{figure}[t]	
\begin{tikzpicture}
    \begin{scope}
        \draw[black, thick] (0,0) -- (0,7);
        \draw[black, thick] (0,0) -- (3,0);
        \draw[black, thick] (3,0) -- (3,7);
        \node[anchor=south,rotate=90] at (0,3.5) {Number of tokens};

        \draw[dashed, black] (0.98,0) -- (0.98,7);
        \node[anchor=north] at (0.5,0) {$I$};
        \node[anchor=north] at (2,0) {$[1,d] \smallsetminus I$};
        
        \draw[densely dotted, black] (0,2) -- (3,2);
        \node[anchor=west] at (3,2) {$\lambda_{\sizeof{I}}$};
        
        \draw[densely dotted, black] (0,5.5) -- (3,5.5);
        \node[anchor=west] at (3,5.5) {$\lambda_{\sizeof{I}+1}$};
        \draw[densely dotted, black] (0,0.75) -- (3,0.75);
        \node[anchor=west] at (3,0.75) {$C\suf^A$};

        \drawconf{blue}{very thick}{
            1.5,1.4,1.1,0.9,1.9,
            7.0,6.5,5.5,5.8,6.1,
            5.9,5.7,5.8,5.9,6.4
        }{0.2}{0}
	    \node[anchor=west,text=blue] at (3,6.4) {$\bm{x}$};
    \end{scope}
    \begin{scope}[xshift=7cm,yshift=0cm]
        \draw[black, thick] (0,0) -- (0,7);
        \draw[black, thick] (0,0) -- (3,0);
        \draw[black, thick] (3,0) -- (3,7);
        \node[anchor=south,rotate=90] at (0,3.5) { Number of tokens};

        \draw[dashed, black] (0.98,0) -- (0.98,7);
        \node[anchor=north] at (0.5,0) {$I$};
        \node[anchor=north] at (2,0) {$[1,d] \smallsetminus I$};
        
        \draw[densely dotted, black] (0,2) -- (3.2,2);
        \node[anchor=west] at (3.2,2) {$\lambda_{\sizeof{I}}$};
        \draw[black, very thick] (3.2,1.9) -- (3.2,2.8);
        \draw[densely dotted, black] (0,2.8) -- (3.2,2.8);
	    \node[anchor=west] at (3.2,2.4)
	    {$x'(i)-(\lambda_{\sizeof{I}}-1)>0$};
	    \node[anchor=west] at (3.7,3.0)
	    {$i\in I$};
        
        \draw[densely dotted, black] (0,4.5) -- (3.2,4.5);
        \node[anchor=west] at (3.2,5.0)
	    {$\lambda_{\sizeof{I}+1}-x'(i)>0$};
	 \node[anchor=west] at (5.1,4.5)
	    {$i\notin I$};
        \draw[black, very thick] (3.2,4.5) -- (3.2,5.5);
        \draw[densely dotted, black] (0,5.5) -- (3.2,5.5);
        \node[anchor=west] at (3.2,5.5) {$\lambda_{\sizeof{I}+1}$};
        \draw[densely dotted, black] (0,0.75) -- (3,0.75);
        \node[anchor=west] at (3,0.75) {$C\suf^A$};

         \drawconf{red}{very thick}{
	    1.5,1.0,0.8,2.8,2.3,
            7.0,6.5,6.1,6.2,4.9,
            4.5,4.8,5.1,6.5,6.3
        }{0.2}{0}
	    \node[anchor=west,text=red] at (3,6.3) {$\bm{x'}$};
    \end{scope}
\end{tikzpicture}	
	\caption{Index set $I=I^\lambda_x$ (left);
	depiction of $\dist{\lambda}{I}{x'}$ (right).}\label{fig:extractor}
\end{figure}

Figure~\ref{fig:extractor} (left) depicts a marking $x\in\setN^d$
and highlights the index set $I=I^\lambda_x\subseteq[1,d]$. 
The marking $x$ is also an example of a marking 
above the level $C\suf^A$ (defined in~\eqref{eq:Csuf});
such markings are particularly important in our later reasoning.

Note that $I^\lambda_x=[1,d]$ if and only if $x(i)< \lambda_d$ for
every $i\in[1,d]$;
the condition  $x(i)\geq \lambda_{d+1}$ holds vacuously
for all $i\in\emptyset$.
If $J'=\{i\in[1,d]\mid x(i)\geq \lambda_d\}$, then 
$I^\lambda_x=I^{(\lambda,J)}_x$ for $J=[1,d]\smallsetminus J'$.
Specifically, if $J'=[1,d]$, then $I^\lambda_x=\emptyset$; the
condition
$x(i)<\lambda_{0}$ holds vacuously for all $i\in\emptyset$.
In general,  $I^\lambda_x=\emptyset$ if and only if for each
$\emptyset\neq I\subseteq[1,d]$, there exists $i\in I$ such that
$x(i)\geq \lambda_{|I|}$.

Later, we apply \emph{2-exp $d$-dim extractors} $\lambda$ to the $d$-dim nets $A$ under
consideration, meaning 
that 
\begin{equation*}
\lambda_d=f(\nnorm{A},d)\leq(2+\nnorm{A})^{2^{\text{poly}(d)}}
\end{equation*}
for some (implicit) 2-exp function $f(m,i)$ which is independent of $A$.
We use such extractors for which
the gaps $\lambda_{j+1}-\lambda_j$ between the large and small
components grow sufficiently with the index $j$.
However, before defining specific extractors, we introduce certain linear systems with
2-exp solutions.

\paragraph{Linear systems underlying the proof of Theorem~\ref{th:upper}.}
In Definition~\ref{def:crucialLS} we introduce a class of linear systems
on which our proof rests; Proposition~\ref{prop:Affsmall} shows that
these systems have 2-exp solutions.

The definition uses the notion of the \emph{distance of a vector to a
$\lambda$-case}, denoted $\dist{\lambda}{I}{x}$ for a vector
$x\in\setN^d$ and a $\lambda$-case $I$, where
$\emptyset\neq I\subseteq[1,d]$.
This distance is a nonnegative integer, and it is zero whenever
$I^\lambda_x=I$.
See Figure~\ref{fig:extractor} (right) for an illustration.

\begin{defi}[\textbf{Distance $\bm{\dist{\lambda}{I}{x}}$ of $\bm{x\in\setN^d}$ to the
	$\bm{\lambda}$-case $\bm{\emptyset\neq I\subseteq[1,d]}$}]
\hfill\\
Given a $d$-dim extractor $\lambda$, the \emph{distance of}
	$x\in\setN^d$ \emph{to} a \emph{$\lambda$-case} $I$, where
	$\emptyset\neq I\subseteq
[1,d]$, is 
	\[\dist{\lambda}{I}{x}=\max\left(\{0\}\cup
	\{x(i)-(\lambda_{|I|}-1)\mid i\in
I\}\cup\{\lambda_{|I|+1}-x(i)\mid i\in[1,d]\smallsetminus I\}\right).\]
\end{defi}	

% Fix Indentation
\noindent
The value $\dist{\lambda}{I}{x}$ is positive if
and only if there exists
$i\in [1,d]$ such that $i\in I$ and $x(i)\geq \lambda_{|I|}$, or
$i\in[1,d]\smallsetminus I$ and $x(i)<\lambda_{|I|+1}$.
For instance, if $\lambda=(3,17,258,1500)$ and $x=(11,1600,2,300)$,
then $\dist{\lambda}{\{1,3\}}{x}=0$, 
and
$\dist{\lambda}{\{1,2\}}{x}=\max(\{0\}\cup\{-5,1584\}\cup\{256,-42\})=1584$. 

\smallskip

The following definition of $(A,x_0,\lambda)$-linear systems aims at
transforming
 constraint~(\ref{eq:firstconstr}). 
For  \emph{reversible} $d$-dim nets $A$ and live 
markings $x_0$, 
our later choice of linear systems $S$
in this class will guarantee that 
every solution of $S$ presents
a nonempty set of at most $2^d$ \emph{live} markings of $A$ that are mutually \emph{virtually} reachable; in
fact, they are also mutually reachable since they all are above $C\vr^A$.
Hence, the non-reachability of dead markings in 
constraint~(\ref{eq:firstconstr}) is replaced with the positive
condition of virtual reachability of a specific subset of markings. 
The definition uses the 1-exp constant $C\shift^A$~(\ref{eq:Cshift}) and the 2-exp
constant $C\suf^A$~(\ref{eq:Csuf}) 
related to a $d$-dim
net $A$.

\begin{defi}[\textbf{$\bm{(A,x_0,\lambda)}$-linear systems for nets,
	markings, and 2-exp extractors}]\label{def:crucialLS}
Given a reversible $d$-dim net $A$ and a 2-exp extractor
	$\lambda=(\lambda_1,\lambda_2,\dots,\lambda_d)$,
	we define
the 2-exp bound	\begin{equation}\label{eq:BAffnets}
	B=B_{(A,\lambda)}=\max\{C\suf^A,\lambda_d+C\shift^A\}.
	\end{equation}
	Given a marking $x_0\in\setN^d\cap\upC{C\suf^A}$ (that is, 
	$\bigwedge_{i=1}^d x_0(i)\geq C\suf^A$), for each $\emptyset\neq I\subseteq[1,d]$ let 
	\[x^\close_I\in \setN^d\cap\upC{C\suf^A}\]
	be a fixed marking in $\Reach(x_0)$
	that is above $C\suf^A$ and minimizes the distance to
	the $\lambda$-case $I$, that is, $\dist{\lambda}{I}{x^\close_I}$ is
	minimal. 

Then the following 
linear system $S$ of dimension $d\cdot 2^d$, with
	a $d$-dim variable
	$\bvar{x}=(\bvar{x}_1,\bvar{x}_2,\dots,\bvar{x}_d)$ and $2^d-1$ $d$-dim variables 
	$\bvar{x}_I$, one for each $\emptyset\neq I\subseteq[1,d]$,
	is 
	\begin{center}
		an \emph{$(A,x_0,\lambda)$-linear system}:
	\end{center}		
\begin{equation}\label{eq:AxLS}
	(\bvar{x}\equiv_B x_0)
	\land
	\bigwedge_{\emptyset\neq I\subseteq[1,d]}\big((\bvar{x}_I\equiv_B x^\close_I)\land
	(\bvar{x}\virtual{*}\bvar{x}_I)\big),
\end{equation}
	where $\bvar{y}\equiv_B x$ is a shorthand for 

	\[\left(\bigwedge_{i\in[1,d], x(i)\leq B}
	(\bvar{y}_i=x(i))\right)\land 
	\left(\bigwedge_{i\in[1,d], x(i)> B} (\bvar{y}_i\geq
	B+1)\right).\]
\end{defi}	

\begin{prop}[\textbf{$\bm{(A,x_0,\lambda)}$-linear systems have 
	2-exp	solutions}]\label{prop:Affsmall}
\hfill\\
For every $(A,x_0,\lambda)$-linear system where $A$ has 
 dimension $d$,
there exists a solution 
	\begin{center}
	$x\in\setN^{d\cdot 2^d}$ such that 
	$\nnorm{x}\leq (2+\nnorm{A})^{2^{\text{poly}(d)}}$.
	\end{center}
	\end{prop}	
\begin{proof} 
Let $S$ be an $(A,x_0,\lambda)$-linear system where  $A$ has dimension $d$.
The system $S$ is
satisfiable, since the tuple $(x_0,(x^\close_I)_{\emptyset\neq I\subseteq [1,d]})$ is one
of its solutions.
	Due to Theorem~\ref{thm:smallsolutions},
there exists a solution
	$x\in\sem{S}$ such that
	\[\nnorm{x}\leq
		\mlcm{S} \cdot \left(1+ (\bar{d}+1)\,!\cdot (\bar{d}\cdot
		\nnorm{S}+1)^{\bar{d}}\right)\]
	where 	$\bar{d}=d\cdot 2^d$.
Recalling~(\ref{eq:AxLS})
and Corollary~\ref{cor:virtual2system}, we deduce that
	$\mlcm{S}\leq d\,!\cdot\nnorm{A}^d$, and $\nnorm{S}\leq
	1+B_{(A,\lambda)}\leq 
	(2+\nnorm{A})^{2^{\text{poly}(d)}}$. 
We thus obtain
\[\nnorm{x}\leq d\,!\cdot\nnorm{A}^d\cdot
 \left(1+ (d\cdot 2^d+1)\,!\cdot (d\cdot 2^d\cdot
		\nnorm{S}+1)^{d\cdot 2^d}\right),\]
which implies that 
	$\nnorm{x}\leq(2+\nnorm{A})^{2^{\text{poly}(d)}}$.
\end{proof}

% Fix Indentation
\noindent
Our goal now is to show certain \emph{lower bounds} for 
the increasing gaps
$\lambda_{j+1}-\lambda_j$ that ensure
that \emph{every solution} 
of an $(A,x_0,\lambda)$-linear system $S$
constitutes \emph{a set
of live markings} of $A$, under the condition that $x_0$ is a sufficiently
large \emph{live} marking.
Under this condition, the solution $(x_0,(x^\close_I)_{\emptyset\neq I\subseteq[1,d]})$ 
has the required property, since all markings $x^\close_I$ are reachable from
the live marking $x_0$, and are thus themselves live (by
Proposition~\ref{prop:liveunderreach}).
Once we establish that \emph{all solutions}
$(\bar{x}_0,(\bar{x}_I)_{\emptyset\neq I\subseteq[1,d]})$ of $S$ 
possess this property, we use the fact 
that the minimal solutions are guaranteed to be at most 2-exp by
Proposition~\ref{prop:Affsmall}.

This goal is achieved via Lemma~\ref{lem:finalbasicnets}, which builds on the quasi-dead marking property of \emph{reversible} nets established in Lemma~\ref{lem:reachquasi}.

\paragraph{Quasi-dead markings.}
We now show that, given a \emph{reversible} $d$-dim net $A$,
every nonlive marking $x_0\in\upC{\max\{C\suf^A,\nnorm{A}\cdot C\qd^A\}}$,
where $C\qd^A$ is a certain 2-exp constant,  can reach a dead marking via 
a  \emph{quasi-dead} marking $x\in\upC{C\suf^A}$.
By definition, quasi-dead markings must be above the level $C\suf^A$
and reach dead markings by executions
of length at most $C\qd^A$.
We use Figure~\ref{fig:markingstwo} for illustration.

\begin{defi}[\textbf{Quasi-dead markings, 2-exp function $\bm{f\bqd}$ and
constant	$\bm{C\bqd^A}$}]
\hfill\\	
For a 2-exp function $f\qd$ and a $d$-dim net $A$, a marking $x\in\setN^d$
	is \emph{$f\qd$-quasi-dead}, or simply \emph{quasi-dead} when $f\qd$
	is clear from context, if $x\in\upC{C\suf^A}$ 
and there exists an execution $x\step{\sigma}y$ such that $y$ is a dead
marking and $|\sigma|\leq
f\qd(\nnorm{A},d)=C\qd^A$.
\end{defi}

Lemma~\ref{lem:reachquasi} uses
Proposition~\ref{prop:simpletrick},
which is based on the following involved result~\cite[Theorem
2]{DBLP:conf/fsttcs/Leroux19}; 
 the result applies to general nets,
without assuming reversibility:
\begin{lemC}[\cite{DBLP:conf/fsttcs/Leroux19} {(\textbf{2-exp mutual
	reachability function $\bm{f\bmr}$, 2-exp constant
	$\bm{C\bmr^A}$})}]\label{lem:fmr}
	There exists a 2-exp function $f\mr$
	with the following property:
	For every $d$-dim net $A$, if $x\reach y$ and $y\reach x$, then
there exist action
	sequences $\sigma_1,\sigma_2$ such that 
	$x\step{\sigma_1}y\step{\sigma_2}x$ and
	\begin{equation*}
		|\sigma_1\sigma_2|\leq
		f\mr(\nnorm{A},d)\cdot\nnorm{x-y}=
		C\mr^A\cdot\nnorm{x-y}.
	\end{equation*}		
\end{lemC}

	\begin{prop}[\textbf{Down reachability $\bm{x\reach y}$ with
		$\bm{x\geq y}$}]\label{prop:simpletrick}
	\hfill\\
	Given a $d$-dim net $A$ and markings $x\geq y$ such that $x\reach y$, 
there exists a marking $x'\geq x$ satisfying the following conditions:
\begin{enumerate}[a)]	
	\item $\nnorm{x'-x}\leq C\mr^A\cdot \nnorm{x-y}$; 
		\item
			$x'\step{\sigma}y$ for some $\sigma$ 
			such that $|\sigma|\leq
			C\mr^A\cdot\nnorm{x-y}$;
		\item $x'(i)=x(i)$ if $x(i)=y(i)$
			(and $x'(i)\geq x(i)$ if $x(i)>y(i)$), for all
			$i\in[1,d]$.
\end{enumerate}
\end{prop}
\begin{proof}
Given a $d$-dim net $A$, and $x,y\in\setN^d$ such that $x\geq y$ and $x\reach y$,
	let $A'$ be 
the $d$-dim net obtained from $A$ by adding actions
	$\mathbf{e}_i$ for those $i\in[1,d]$ for which $x(i)>y(i)$, where $\mathbf{e}_i$
increments the
	$i$-th component by $1$.
	Formally,
	$(\mathbf{e}_i)_-=\mathbf{0}$ and $(\mathbf{e}_i)_+=e_i$,
	where $e_i$
is the $i$th unit vector satisfying
	$e_i(i)=1$ and $e_i(j)=0$ for 
	$j\neq i$.

Hence, there exist sequences $\sigma_1,\sigma_2$ of actions in $A'$
	satisfying
	\begin{equation}\label{eq:xmry}
	x\step{\sigma_1}y\step{\sigma_2}x\text{ in }A';
	\end{equation}
we can take $\sigma_1$ from $A$ due to the assumption $x\reach y$ in $A$,
whereas $\sigma_2$ can consist solely of the increment actions $\mathbf{e}_i$.
	By Lemma~\ref{lem:fmr}, there exist sequences
	$\sigma_1,\sigma_2$  of actions in $A'$
	that satisfy~\eqref{eq:xmry} and, moreover,
	\[|\sigma_1\sigma_2|\leq C\mr^A\cdot
\nnorm{x-y};\]
note that the actions $\mathbf{e}_i$ might occur
	in $\sigma_1$ as well. We safely use $C\mr^A=f\mr(\nnorm{A},d)$ instead of
	$C\mr^{A'}=f\mr(\nnorm{A'},d)$, since $\nnorm{A'}=\nnorm{A}$
	provided $\nnorm{A}\geq 1$; the claim is trivial if
	$\nnorm{A}=0$.

Any execution $x\step{\sigma_1}y$ in $A'$ can be clearly reorganized
	into the form 
	\[x\step{\sigma'}x'\step{\sigma}y,\] 
where 
the prefix $\sigma'$ is the projection of $\sigma_1$ onto
	the  set 
of increment actions $\mathbf{e}_i$, and
the suffix $\sigma$ is the projection of $\sigma_1$ onto
	the   set of original 
 actions in $A$.

	We have
	thus obtained $x'\geq x$ satisfying:
	\begin{enumerate}[a)]	
		\item $\nnorm{x'-x}\leq |\sigma'|\leq |\sigma_1|\leq C\mr^A\cdot \nnorm{x-y}$; 
		\item
			$x'\step{\sigma}y$ for some sequence $\sigma$ of actions in $A$
			such that $|\sigma|\leq |\sigma_1|\leq C\mr^A\cdot
			\nnorm{x-y}$;
	\item $x'(i)=x(i)$ if $x(i)=y(i)$ (since the action
		$\mathbf{e}_i$ is not in $A'$ for such $i\in[1,d]$).
			\qedhere
	\end{enumerate}
\end{proof}

\begin{lem}[\textbf{Large nonlive markings (virtually) reach  quasi-dead
	markings}]\label{lem:reachquasi}
	There exists a 2-exp function
$f\qd$ such that 
for every reversible $d$-dim net $A$ and 
	every nonlive marking
$x_0\in\upC{\max\{C\suf^A,\nnorm{A}\cdot C\qd^A\}}$, where
$C\qd^A=f\qd(\nnorm{A},d)$,
	there exists an $f\qd$-quasi-dead marking $x\in\Reach(x_0)$.
\end{lem}
\begin{proof}
Let $A$ be a reversible $d$-dim net and 
let	$x_0\in\upC{C\suf^A}$ be
	a nonlive marking of $A$.
	See Figure~\ref{fig:markingstwo} for illustration.
	The value $C\qd^A=f\qd(\nnorm{A},d)$ will arise from the
	following analysis.

	We fix a dead marking
	$y\in\Reach(x_0)$---it exists, since $x_0$ is nonlive---and a nonempty index set $I\subseteq [1,d]$ such that the
	restricted marking 
	$y\restr{I}\in[0,C\dead^A-1]^I$ is dead in the restricted net
	$A\restr{I}$ (recall~(\ref{eq:alldead})).
We have 
	\begin{equation}\label{eq:Iless}
		y(i)< x_0(i)\text{ for every }i\in I,
	\end{equation}		
because $y(i)< C\dead^A\leq C\suf^A\leq x_0(i)$ for every $i\in I$.

	Since  $x_0\reach y$, we obtain $y\vreach x_0$ by
the reversibility of $A$. We then consider a segmented
 virtual execution  
	\begin{equation}\label{eq:downup}
y=y_0\vreach y_1\vreach y_2\cdots\vreach y_k=x_0
	\end{equation}
	as in~(\ref{eq:virtsegment}).
	Hence, $\nnorm{y_{j+1}-y_j}\leq C\shift^A$ by~\eqref{eq:Cshift}, and 
for each $i\in[1,d]$ the following holds:
\begin{itemize}
\item If $y(i)<x_0(i)$, then
	$y(i)<y_1(i)<y_2(i)\cdots <y_{j_i}(i)=y_{j_i+1}(i)\cdots
	=y_k(i)=x_0(i)$;
\item	
If $y(i)\geq x_0(i)$, then $y_j(i)\geq x_0(i)$ for all
	$j\in[0,k]$.
	\end{itemize}
We have thus highlighted the strict increase in the case
	$y(i)<x_0(i)$, and the maintenance of the threshold $x_0(i)$ in
	the case $y(i)\geq x_0(i)$.

	Since $x_0(i)\geq C\suf^A$ for all $i\in[1,d]$, 
	there exists 	$j\leq C\suf^A$ such that 
	the marking $y_j$ in~\eqref{eq:downup}
	is above $C\suf^A$. By setting $x=y_j$, we can rewrite~(\ref{eq:downup}) as
	\begin{equation}\label{eq:xybound}
		y\vreach x\vreach x_0,\text{ where
		}x\in\upC{C\suf^A}\text{ and }\nnorm{x-y}\leq
		C\suf^A\cdot C\shift^A.
	\end{equation}
	We have thus obtained  a candidate for a quasi-dead marking $x\in\upC{C\suf^A}$, since 
	\begin{equation}\label{eq:candqd}
		x_0\reach x\reach y.
	\end{equation}
	Indeed, $x_0\reach y\vreach x$
	implies $x_0\vreach x$, hence  $x_0\reach x$ by
	Lemma~\ref{lem:fvr}, since 
	$x_0,x\in\upC{C\suf^A}$ and $C\suf^A=\max\{C\dead^A,C\vr^A\}\geq C\vr^A$. Due to
	reversibility, we also have $x\vreach x_0$, and thus  $x\reach
	x_0$, since 	$x,x_0\in\upC{C\vr^A}$.
Consequently, $x\reach y$ (because $x\reach x_0\reach y$).	
\begin{figure}[t]	
\begin{center}
\begin{tikzpicture}
    \begin{scope}
        \draw[black, thick] (0,0) -- (0,7);
        \draw[black, thick] (0,0) -- (8,0);
        \draw[black, thick] (8,0) -- (8,7);
        \node[anchor=south,rotate=90] at (0,3.5) {Number of tokens};

        \draw[dashed, black] (1.98,0) -- (1.98,7);
        \node[anchor=north] at (1,0) {$I$};
        \draw[dashed, black] (4.78,0) -- (4.78,7);
        \node[anchor=north] at (3.4,0) {$I' \smallsetminus I$};
        \node[anchor=north] at (6.5,0) {$[1,d] \smallsetminus I'$};
        
        \draw[densely dotted, black] (0,1.25) -- (8,1.25);
        \node[anchor=west] at (8,1.25) {$C\suf^A$};
        \draw[densely dotted, black] (0,4.5) -- (8,4.5);
	    \node[anchor=west] at (8,4.5) {$\nnorm{A}\cdot C\qd^A$};

        \drawconf{blue}{very thick}{
            6.9,4.5,6.1,5.5,6.5,
            6.0,6.1,5.7,5.0,4.8,
            5.0,5.4,5.9,6.22,6.16,
            4.9,4.5,4.7,5.1,5.0
        }{0.4}{0}
	    \node[anchor=west] at (8,5.0) {{\color{blue}$\bm{x_0}$} nonlive};

        \drawconf{applegreen}{very thick}{
            0.6,0.4,0.8,0.0,0.6,
            1.7,5.0,5.2,3.2,3.9,
            4.2,3.7,6.9,6.3,6.7,
            6.8,6.1,6.3,6.6,6.4
        }{0.4}{0}
	    \node[anchor=west] at (8,6.4)
	    {{\color{applegreen}$\bm{y}$} dead};

        \drawconf{red}{very thick}{
            1.9,1.5,1.8,1.7,1.3,
            2.3,5.8,5.65,3.95,4.6,
            4.95,4.3,6.35,6.26,6.2,
            6.2,5.2,5.9,5.8,6.0
        }{0.4}{0}
	    \node[anchor=west] at (8,6.0) {{\color{red}$\bm{x}$}
	    quasi-dead};
    \end{scope}
\end{tikzpicture}
\end{center}	
\caption{Illustration for the proof of
	Lemma~\ref{lem:reachquasi}.}\label{fig:markingstwo}
\end{figure}

% Fix Indentation
\noindent
	However,
	we have not bounded the length of an execution from $x$ to a
	dead marking.
	To handle this, we recall that our choice of 
	the virtual execution $y\vreach{x}$ in~\eqref{eq:xybound} obtained
	from~(\ref{eq:downup}) guarantees that
	for each $i\in[1,d]$ we have
\begin{center}
either $y(i)< x(i)\leq x_0(i)$ or $y(i)\geq x(i)\geq x_0(i)$. 
\end{center}
	As illustrated in Figure~\ref{fig:markingstwo},
we consider the set
\[I'=\{i\in[1,d]\mid y(i)< x(i)\}=\{i\in[1,d]\mid y(i)<
	x_0(i)\}\supseteq I;\] 
	the inclusion $I\subseteq I'$ follows from~\eqref{eq:Iless}.
	 Since
	$y\restr{I}$ is dead in $A\restr{I}$, the marking 
$y\restr{I'}$ is dead in the net $A\restr{I'}$ (by
Proposition~\ref{prop:restrdead}). 

Because $x\restr{I'}\geq y\restr{I'}$ and  $x\restr{I'}\reach
y\restr{I'}$ in  $A\restr{I'}$ (due to $x\reach y$ in  $A$),
we can apply Proposition~\ref{prop:simpletrick} to the net
$A\restr{I'}$. We obtain  $x'\geq
x\restr{I'}$ (where $x'\in\setN^{I'}$) satisfying:
\begin{enumerate}[a)]
	\item 
$\nnorm{x'-x\restr{I'}}\leq 
		C\mr^{A\restr{I'}}\cdot
		\nnorm{x\restr{I'}-y\restr{I'}}\leq C\mr^{A}\cdot
		\nnorm{x-y}\leq  C\mr^{A}\cdot C\suf^A\cdot C\shift^A$
		(using~\eqref{eq:xybound}); and
		\item
			$x'\step{\sigma}y\restr{I'}$ for some $\sigma$ 
			such that $|\sigma|\leq C\mr^{A\restr{I'}}\cdot
	\nnorm{x\restr{I'}-y\restr{I'}}\leq C\mr^{A}\cdot C\suf^A\cdot C\shift^A$.
	\end{enumerate}
(Condition c) in Proposition~\ref{prop:simpletrick} is irrelevant
here.)

In $A\restr{I'}$, we thus have $x\restr{I'}\reach y\restr{I'}$
and $x'\reach y\restr{I'}$. 
By reversibility of $A\restr{I'}$ (inherited from $A$, recall
Proposition~\ref{prop:restrpresrev}), we obtain 
$x\restr{I'}\vreach x'$. Since $x'\geq x\restr{I'}\in\upC{C\suf^A}$
and $C\suf^A\geq
C\vr^A\geq C\vr^{A\restr{I'}}$,
Lemma~\ref{lem:fvr} yields  
 $x\restr{I'}\step{\sigma'} x'$ for a sequence $\sigma'$ of actions in
 $A\restr{I'}$, and thus in $A$ by our conventions, such that
\begin{equation}\label{eq:boundsigma}
	|\sigma'|\leq f\vr(\nnorm{A\restr{I'}},|I'|)\cdot
\nnorm{x'-x\restr{I'}}\leq C\vr^A\cdot
C\mr^A\cdot C\suf^A\cdot C\shift^A
\end{equation}
(using a)). To summarize, we have obtained the execution 
\begin{equation}\label{eq:finalseq}
	x\restr{I'}\step{\sigma'}x'\step{\sigma}y\restr{I'}\text{ in }A\restr{I'},
\end{equation}
where $y\restr{I'}$ is a dead marking, $|\sigma'|$ is bounded
in~\eqref{eq:boundsigma}, and  $|\sigma|$ is bounded
in~b).
Hence,
 $|\sigma'\sigma|\leq
f\qd(\nnorm{A},d)$, if we set
\begin{equation}\label{eq:fqd}
	f\qd(\nnorm{A},d)=C\qd^A=(C\vr^A+1)\cdot C\mr^A\cdot
C\suf^A\cdot C\shift^A.
\end{equation}
	Recall that $x(i)\geq x_0(i)$ for all $i\in[1,d]\smallsetminus
	I'$. 
	Consequently, if the considered nonlive marking 
\begin{center}	$x_0$ is above the level
	$\nnorm{A}\cdot C\qd^A$,
\end{center}
	we have 
$x(i)\geq\nnorm{A}\cdot C\qd^A$ for all $i\in[1,d]\smallsetminus I'$. 
Since $|\sigma'\sigma|\leq C\qd^A$, the sequence $\sigma'\sigma$ can be
executed not only from $x\restr{I'}$ in $A\restr{I'}$
(recall~\eqref{eq:finalseq}) but
	also from $x$ in $A$. We obtain an execution
	$x\step{\sigma'\sigma}y'$ in $A$, where $y'$ is a dead marking
	because ${y'}\restr{I'}=y\restr{I'}$ (and thus
	${y'}\restr{I}=y\restr{I}$, where $y\restr{I}$ is a dead
	marking in $A\restr{I}$).
The candidate $x$ in~\eqref{eq:candqd} is thus indeed an $f\qd$-quasi-dead marking, for the
2-exp function $f\qd$ defined by~\eqref{eq:fqd}.
\end{proof}	

\paragraph{Live solutions of $\bm{(A,x_0,\lambda)}$-linear systems.}
Further, we fix a 2-exp function $f\qd$ as guaranteed by
Lemma~\ref{lem:reachquasi}.  This fixes the constant
$C\qd^A=f\qd(\nnorm{A},d)$, and thus the notion of quasi-dead marking,
for every $d$-dim net $A$.
The next lemma refers to Definition~\ref{def:crucialLS}.

\begin{lem}[\textbf{$\bm{(A,x_0,\lambda)}$-solutions are live, for
	sufficiently large live $\bm{x_0}$ 
	and $\bm{\lambda}$}]\label{lem:finalbasicnets}
\hfill\\
	Let $A$ be a reversible $d$-dim net, and let
	$\lambda=(\lambda_1,\lambda_2,\dots,\lambda_d)$ be an extractor
	satisfying
	\begin{itemize}
		\item
			$\lambda_1\geq
C\suf^A+\nnorm{A}\cdot C\qd^A$;
\item
	$\lambda_{i+1}\geq \lambda_1 + \nnorm{A}\cdot C\vr^A\cdot
			(\lambda_i-C\suf^A)$ for each
			$i\in[1,d{-}1]$.
	\end{itemize}
Let $x_0$ be a live marking of $A$ satisfying	
	\[x_0\in\upC{\max\{C\suf^A,\nnorm{A}\cdot
	C\qd^A\}}.\]
Let $S$ be an $(A,x_0,\lambda)$-linear system, and let
	$(\bar{x}_0,(\bar{x}_I)_{\emptyset\neq I\subseteq[1,d]})$ be a
	solution of $S$.  Then $\bar{x}_0$ is a live marking.
\end{lem}	
\begin{proof}
	Let $A,\lambda$, and $x_0$ be as in the statement.
	Let $S$ be an $(A,x_0,\lambda)$-linear
	system~\eqref{eq:AxLS}, namely
\begin{equation*}
	(\bvar{x}\equiv_B x_0)
	\land
	\bigwedge_{\emptyset\neq I\subseteq[1,d]}\big((\bvar{x}_I\equiv_B x^\close_I)\land
	(\bvar{x}\virtual{*}\bvar{x}_I)\big),
\end{equation*}
with $B=\max\{C\suf^A,\lambda_d+C\shift^A\}$.
Recall that $({x}_0,({x}_I^\close)_{\emptyset\neq I\subseteq[1,d]})$ is a
solution of $S$, and that it is a tuple of live markings above
$C\suf^A$ (they are above $C\suf^A$ by
Definition~\ref{def:crucialLS}, and live by
Proposition~\ref{prop:liveunderreach}, since $x_0$ is live and
${x}_I^\close\in\Reach(x_0)$).
Towards a contradiction, suppose that
	\begin{center}
$(\bar{x}_0,(\bar{x}_I)_{\emptyset\neq I\subseteq[1,d]})$ is a solution of $S$
where $\bar{x}_0$ is nonlive. 
	\end{center}
	Note that for each $\emptyset\neq I\subseteq[1,d]$, we have
	$\bar{x}_0\vreach \bar{x}_I$ and
	$\bar{x}_I\equiv_B
	x^\close_I$, that is, 
	\begin{itemize}
		\item			
			$\bar{x}_I(i)=x^\close_I(i)$ \ for all
			$i\in[1,d]$ such that $x^\close_I(i)\leq B$, and
\item
			$\bar{x}_I(i)>B$ \ for all
			$i\in[1,d]$ such that $x^\close_I(i)>B$.
	\end{itemize}			
Since $x^\close_I\in\upC{C\suf^A}$ by definition, it follows that 
	$\bar{x}_I\in\upC{C\suf^A}$ as well.
	Moreover, the sets of indices of small components 
	$I^\lambda_{\bar{x}_I}$ and $I^\lambda_{x^\close_I}$
	(as defined in	Proposition~\ref{prop:Ilambdax}) are
	identical,
	and the markings $\bar{x}_I$ and $x^\close_I$ coincide on
	these indices.

\smallskip

	Lemma~\ref{lem:reachquasi} implies the existence of
	\begin{center}
	a quasi-dead
		marking $x\in\upC{C\suf^A}$ 
		such that 
	$\bar{x}_0\reach x$; let
 $I=I^\lambda_{x}$. 
	\end{center}		
Since $x$ is a quasi-dead marking, we have $x\step{\sigma_1}y$ where
$y$ is a dead marking and
	$|\sigma_1|\leq C\qd^A$.  
	Note that $I\neq\emptyset$. Indeed, we have $y(i)< C\dead^A$
	for at least one $i\in[1,d]$ (recall~\eqref{eq:alldead}).
The case $I=I^\lambda_{x}=\emptyset$ would imply,	 for all
	$i\in[1,d]$,
	\[x(i)\geq
	\lambda_1\geq C\suf^A+\nnorm{A}\cdot
	C\qd^A\geq 
	C\dead^A+\nnorm{A}\cdot |\sigma_1|;\]
this would in turn imply $y\in\upC{C\dead^A}$---a contradiction.  
Consequently, $\bar{x}_{I}$ is well-defined.

	Since all markings $\bar{x}_0$, $\bar{x}_I, x$ are above $C\suf^A$
and $A$ is
	reversible, we have 
	\[\bar{x}_0\reach \bar{x}_I\reach \bar{x}_0\reach
	x\reach\bar{x}_0\text{, hence }\bar{x}_I\reach x\text{ in
	particular.}\]
Note that $\dist{\lambda}{I}{\bar{x}_I}=\dist{\lambda}{I}{x^\close_I}=0$
(the distance to the $\lambda$-case $I$;
recall Figure~\ref{fig:extractor}(right)).
	The equality
	$\dist{\lambda}{I}{\bar{x}_I}=\dist{\lambda}{I}{x^\close_I}$
holds due to $\bar{x}_I\equiv_{B}x^\close_I$. If we had 
	$\dist{\lambda}{I}{\bar{x}_I}>0$, then 
	the first small segment $\bar{x}_I\virtual{\sigma}x'$
	of the virtual execution
	$\bar{x}_I\vreach x$ in the form
	of~(\ref{eq:virtsegment})---for which
	$\nnorm{x'-\bar{x}_I}\leq C\shift^A$ and 
	$x'(i)$ is between the values $\bar{x}_I(i)$ and $x(i)$ for
	each
	$i\in[1,d]$---would decrease this distance 
(that is, $\dist{\lambda}{I}{x'}<\dist{\lambda}{I}{\bar{x}_I}$).
For the sequence $\sigma$ from $\bar{x}_I\virtual{\sigma}x'$, we obtain 
 $x^\close_I\virtual{\sigma}x''$, where  $x'\equiv_{\lambda_d}x''$
 (since  $\bar{x}_I\equiv_{B}x^\close_I$ and
$B=\lambda_d+C\shift^A$). Consequently, 
$x^\close_I\reach x''$ (as both $x^\close_I$ and $x''$ are above
$C\suf^A\geq C\vr^A$), 
and
$\dist{\lambda}{I}{x''}<\dist{\lambda}{I}{x^\close_I}$; this contradicts the choice
of  $x^\close_I$.
We thus have
\[\dist{\lambda}{I}{x}=\dist{\lambda}{I}{\bar{x}_I}=\dist{\lambda}{I}{x^\close_I}=0,\]
and all markings $x,\bar{x}_I,x^\close_I$ are above $C\suf^A$.
Hence, for each $i\in I$, the values $x(i)$ and
$\bar{x}_I(i)=x^\close_I(i)$ are in the interval
$[C\suf^A,\lambda_{|I|}-1]$.
The fact 
$\bar{x}_I\reach x$ implies that in the restricted net $A\restr{I}$ we
have 
$(\bar{x}_I)\restr{I}\step{\sigma_2}x\restr{I}$ where 
\[|\sigma_2|\leq f\vr(\nnorm{A},|I|)\cdot
\nnorm{x\restr{I}-(\bar{x}_I)\restr{I}}\leq C\vr^A\cdot
(\lambda_{|I|}-C\suf^A).\]
Combining with $x\step{\sigma_1}y$, where $|\sigma_1|\leq
C\qd^A$ and $y$ is dead, we obtain
\[(\bar{x}_{I})\restr{I}=(x^\close_{I})\restr{I}\step{\sigma_2}
x\restr{I}\step{\sigma_1}y\restr{I}\]
in $A\restr{I}$, where $|\sigma_2\sigma_1|\leq C\vr^{A}\cdot
(\lambda_{|I|}-C\suf^A)+C\qd^A$.
Since for each $i\in [1,d]\smallsetminus I$, we have
$x^\close_{I}(i)\geq\lambda_{|I|+1}\geq \nnorm{A}\cdot
|\sigma_2\sigma_1|$, it follows that 
\[x^\close_{I}\step{\sigma_2\sigma_1}y'\]
is an execution in $A$, where $y'\restr{I}=y\restr{I}$.
We complete the proof by showing that $y'$ is dead, which
contradicts the assumption that $x^\close_{I}$ is live (since 
$x_0\reach x^\close_{I}$ and $x_0$ is live).

Recall that $x\step{\sigma_1}y$ and $|\sigma_1|\leq C\qd^A$.
Since for each $i\in [1,d]\smallsetminus I$, we have 
\[x(i)\geq
\lambda_{|I|+1}\geq \lambda_1\geq C\dead^A+ \nnorm{A}\cdot C\qd^A,\]
it follows that 
$y(i)\geq C\dead^A$ for each  $i\in [1,d]\smallsetminus I$. This implies
that $y\restr{I}$ is dead in $A\restr{I}$, which in turn
implies that $y'$ is dead in $A$ (since
$y'\restr{I}=y\restr{I}$).
\end{proof}

\paragraph{Petri nets with states (PNSs).}
Lemma~\ref{lem:finalbasicnets}, which gives conditions under which 
solutions of $(A,x_0,\lambda)$-linear systems are live,
and Proposition~\ref{prop:Affsmall}, which guarantees 2-exp solutions of 
such systems, together establish Theorem~\ref{th:upper} 
for reversible nets $A$ that have a live marking $x_0$ above the
level $\max\{C\suf^A, \nnorm{A}\cdot C\qd^A\}$.

However, there exist structurally live conservative nets 
(hence, nets with live bottom SCCs)
that have no live marking above that level.
Indeed, there are nets in which the values of some 
components---indexed by a~set $\emptyset\neq I\subseteq[1,d]$---remain
small in every live marking, as demonstrated
by Example~\ref{exa:basiclive}.
In this case, it is natural to view the restrictions $x\restr{I}$ of
(live) markings $x$ as \emph{control states}; the components (that is,
Petri net places) outside $I$ are viewed as \emph{counters}.
Definition~\ref{def:PNS}
introduces \emph{Petri nets with
states} (PNSs), in a specific form tailored to our use.   
We will extend the previous results from nets to PNSs, with live
markings having sufficiently large counter values.

\begin{rem}
Petri nets with states can be viewed, in principle, 
as \emph{vector addition
	systems with states} (VASSs); a minor difference is 
	that transitions between states are
labelled with Petri net actions $a=(a_-,a_+)$ in PNSs, whereas in
	VASSs they are
	labelled merely with 
plain vectors encoding the effects $\Delta(a)$. 
	Our
	definition of PNSs is also related to the notion of
	\emph{rigid components} introduced in~\cite{DBLP:conf/stoc/Kosaraju82} in order to keep track 
	of the exact values of 
bounded components.
\end{rem}

% Fix Indentation
\noindent
Given a $d$-dim net $A$ and $I\subseteq[1,d]$,
we say that a \emph{nonempty finite} set $Q\subseteq\setN^I$ is
a~\emph{bSCC} of the restricted net $A\restr{I}$
if it corresponds to a bottom SCC 
in the
reachability graph of $A\restr{I}$; that is, if in $A\restr{I}$ we have 
$\Reach(x)=Q$ for all $x\in Q$.
Figure~\ref{fig:example} shows a 5-dim net $A_1$ and a restricted 4-dim
 net
$A_2=(A_1)\restr{[1,4]}$. Recalling Example~\ref{exa:basiclive}, we
can verify that $Q=\{(1,0,1,0),(0,1,2,0),(0,1,0,1),(1,0,0,1)\}$ is a
bSCC of $A_2$.

\begin{defi}[\textbf{PNS $\bm{G^A_{(I,Q)}}$, norm $\bm{\nnorm{G}}$,
	relations $\bm{\reach}$ and $\bm{\vreachtitle}$,
	reversible PNS}]\label{def:PNS}
	Given a $d$-dim net $A$, an index set $I\subseteq [1,d]$, and a (finite) bSCC 
$Q\subseteq\setN^I$, 
the \emph{Petri net with states (PNS)} $G^A_{(I,Q)}$ 
	is obtained from  $A$ by restricting \emph{markings}
	$\bar{x}\in\setN^d$ 
	to those satisfying $\bar{x}\restr{I}\in Q$. 
Setting  $J=[1,d]\smallsetminus I$,
	such a~marking  $\bar{x}\in\setN^d$
	is also written as 
	\begin{equation*}
		\textnormal{$(p,x)\in
	Q\times \setN^J$ where 
		$p=\bar{x}\restr{I}\in Q$ and
		$x=\bar{x}\restr{J}\in\setN^J$.}
	\end{equation*}
For $G=G^A_{(I,Q)}$, we say that $G$ is of \emph{dimension} $d$
(inherited from $A$), or that $G$ is a~\emph{$d$-dim PNS}.
We define the \emph{norm of} $G$ as
	\[\nnorm{G}=\max\{\nnorm{Q},\nnorm{A}\}.\]
The relations $\step{a}$,
	$\step{\sigma}$, $\reach$ in $G^A_{(I,Q)}$  coincide with the respective
	relations in $A$ restricted to markings in $Q\times \setN^J$.
	The \emph{virtual reachability relation} in  $G^A_{(I,Q)}$ is defined 
on the set $Q\times \setZ^J$ as follows:
	for action $a$ in $A$ we have 
	\begin{center}
	$(p,x)\virtual{a}(q,y)$ if $p\step{a\restr{I}}q$ in
	$A\restr{I}$ and $x\virtual{a\restr{J}}y$ in $A\restr{J}$.
	\end{center}
	This naturally induces the
	relations $(p,x)\virtual{\sigma}(q,y)$ for sequences
	$\sigma$ of
	actions in $A$, and the virtual reachability relation
	$(p,x)\vreach(q,y)$ on the set $Q\times \setZ^J$. 
Virtual reachability in a PNS $G^A_{(I,Q)}$ is thus based on
the standard reachability within $Q$ in $A\restr{I}$:
	\begin{center}
	$(p,x)\virtual{\sigma}(q,y)$ if and only if $p\step{\sigma}q$ in
		$A\restr{I}$ and $y-x =\Delta(\sigma)\restr{J}$.
	\end{center}
(By our
	convention, we may write 
	 $p\step{\sigma}q$ as a shorthand for
	  $p\step{\sigma\restr{I}}q$.)

	A \emph{PNS} $G^A_{(I,Q)}$ is \emph{reversible} if its virtual
	reachability relation
	$\vreach$ is symmetric; that is, $(p,x)\vreach(q,y)$ implies 
	$(q,y)\vreach(p,x)$.
\end{defi}

\begin{rem}
The dimension of $G=G^A_{(I,Q)}$, where $A$ is a $d$-dim net and
	$I\subseteq [1,d]$,
	could be naturally defined as $|J|$ for $J=[1,d]\smallsetminus
	I$. This would correspond to
the standard notion 
	of dimension in vector addition systems with states (VASSs). 
However, we define it as $d$ to emphasize that the underlying net $A$
remains the primary object of study.  

Note that the requirement that the state set $Q$ in $G=G^A_{(I,Q)}$ is a bSCC of
	$A\restr{I}$ guarantees that for any marking
	$\bar{x}=(p,x)\in Q\times \setN^J$, the reachability set $\Reach(\bar{x})$ in $A$ coincides with that in $G$.
\end{rem}

\paragraph{Extractors $\bm{\lambda}$ yielding reversible PNSs
$\bm{G^{(A,\lambda)}_{(I,Q)}}$ for
live bottom SCCs of nets $\bm{A}$.}

The following lemma is a simplified variant of Lemma~20 in
Section~6 of~\cite{DBLP:conf/fsttcs/Leroux19}. It concerns a general
SCC of a net; a bottom SCC 
is a special case. It allows us to separate, within an SCC, the
``control states''---components whose values are small in every marking
of the SCC---from the ``counters'', whose values are simultaneously
large in at least one marking.
The lemma uses the notion of \emph{$\nnorm{A}$-adapted
extractors}. For later convenience, we also introduce the notion of
\emph{strictly $\nnorm{A}$-adapted
extractors}.

\begin{defi}[\textbf{$\bm{\nnorm{A}}$-adapted and strictly $\bm{\nnorm{A}}$-adapted extractors
	$\bm{\lambda}$}]\label{def:Aadaptextract}
\hfill\\
	Given a $d$-dim extractor
	$\lambda=(\lambda_1,\lambda_2,\dots,\lambda_d)$, with
	$\lambda_0=1$ by our convention,
	and
	$m\in\setN_+$ (in particular, $m=\nnorm{A}$ for a $d$-dim net
	$A$), we say that $\lambda$ is \emph{$m$-adapted}
if  for each
	$i\in[0,d{-}1]$,
	\[\lambda_{i+1}\geq \lambda_i+m\cdot (\lambda_i)^i.\]
	 We say that $\lambda$ is  \emph{strictly
	$m$-adapted} if $\lambda_1\geq 1+m$ and
	for each
	$i\in[1,d{-}1]$,
	\begin{equation}\label{eq:genextr}
		\lambda_{i+1}\geq m\cdot (\lambda_i)^i+m\cdot
	i\cdot(\lambda_i)^i.
	\end{equation}
	(Note that every strictly $m$-adapted extractor is $m$-adapted.)
\end{defi}

\begin{lemC}[\cite{DBLP:conf/fsttcs/Leroux19} (\textbf{Small control-state projection with large
	counters})]\label{lem:sccfirst}
\hfill\\
	Let $X$ be an SCC in the reachability graph of a $d$-dim net $A$ 
 and
	let
	$\lambda=(\lambda_1,\lambda_2,\dots,\lambda_d)$ be a~$d$-dim
	extractor that is $\nnorm{A}$-adapted. 
	There exists a 
	set $I\subseteq[1,d]$
	satisfying the following conditions, where
	$J=[1,d]\smallsetminus I$:
	\begin{enumerate}[a)]	
\item
	$\nnorm{\rst{X}{I}}<\lambda_{|I|}$ (hence for all $x\in X$ and
			$i\in I$ we have $x(i)<\lambda_{|I|}$);			
\item
there exists			$x\in X$
such that for all $j\in J$ we have
		$x(j)\geq
		\lambda_{\sizeof{I}+1}-\nnorm{A}\cdot\sizeof{I}\cdot(\lambda_{\sizeof{I}})^{\sizeof{I}}$.
	\end{enumerate}
\end{lemC}

% Fix Indentation
\noindent
In the previous lemma, the mere existence of a respective set $I\subset[1,d]$
is sufficient for our aims; we do not require its uniqueness in
this case. This entails that we do not require the uniqueness of 
the PNS $G^{(A,\lambda)}_{(I,Q)}$ in the following definition.

\begin{defi}[\textbf{Extracting PNS
	$\bm{G^{(A,\lambda)}_{(I,Q)}}$ from a live bottom SCC of
	a net $\bm{A}$}]\label{def:extrPNS}
\hfill\\
Let a (finite or infinite) set $X\subseteq \setN^d$ be a live bottom SCC in the reachability graph of
	a $d$-dim net $A$. (The net $A$ is reversible by
	Lemma~\ref{lem:consstrbound}.)
	Let $\lambda$ be a strictly $\nnorm{A}$-adapted $d$-dim
extractor, and let $I\subseteq[1,d]$ be a set guaranteed by
Lemma~\ref{lem:sccfirst}. 
We then view the PNS $G=G^A_{(I,Q)}$ with  $Q=X\restr{I}$ as a PNS 
	$G^{(A,\lambda)}_{(I,Q)}$, \emph{extracted from the SCC $X$ via the extractor
	$\lambda$}.
\end{defi}

% Fix Indentation
\noindent
In the previous definition, we consider strictly $\nnorm{A}$-adapted 
extractors to guarantee that $Q=X\restr{I}$ is a bSCC of $A\restr{I}$
(as required in Definition~\ref{def:PNS}):

\begin{prop}[\textbf{PNSs $\bm{G^{(A,\lambda)}_{(I,Q)}}$ are
	well-defined, and reversible}]\label{prop:condproperPNS}
\hfill \\	
In every PNS ${G^{(A,\lambda)}_{(I,Q)}}$,
	the set ${Q}$ is indeed a
	bSCC of ${A\restr{I}}$. Moreover, ${G^{(A,\lambda)}_{(I,Q)}}$
	is reversible.
\end{prop}	
\begin{proof}
Let a PNS $G={G^{(A,\lambda)}_{(I,Q)}}$ be extracted from a live bottom
	SCC $X$ of a $d$-dim net $A$ as described
	in Definition~\ref{def:extrPNS}; this implies
that $A$ is reversible,	$Q=X\restr{I}$, and $\nnorm{Q}<\lambda_{|I|}$ (by
	condition a) in Lemma~\ref{lem:sccfirst}). Consequently, 
	\[|Q|\leq (1+\nnorm{Q})^{|I|}\leq (\lambda_{|I|})^{|I|}.\]

	If $I=\emptyset$, then $Q=\{()\}$ (the empty tuple is the
	unique
	control state), and we have $()\step{a}()$ for all actions in
	$A$. Hence, $Q$ is a trivial bSCC of $A\restr{I}$, and $G$ is
	reversible because $A$ is reversible.
	If $I=[1,d]$, then $A\restr{I}=A$ and $Q=X$.
	Here, the facts that $Q$ is a bSCC of $A\restr{I}$ and that $G$ is reversible follow from $X$ being a bottom SCC of $A$.
	We thus further assume $\emptyset\neq I\neq[1,d]$.
	
	To prove that  ${Q}=X\restr{I}$ is a bSCC of ${A\restr{I}}$,
we assume $p\step{a}p'$ 
	(that is, $p\step{a\restr{I}}p'$) 
	in $A\restr{I}$ with
	$p\in Q$, and show that $p'\in Q$; more precisely, we show
	that there exists a step 
	$(p,y)\step{a}(p',y')$ within $X$ for some
	$y,y'\in\setN^J$, where $J=[1,d]\smallsetminus I$.
	We use the fact that by condition b) in
	Lemma~\ref{lem:sccfirst}, there exists 
	$(q_0,x_0)\in X$ such that ($q_0\in Q$ and) for each $j\in J$,
	\[
	x_0(j)\geq
		\lambda_{\sizeof{I}+1}-\nnorm{A}\cdot\sizeof{I}\cdot(\lambda_{\sizeof{I}})^{\sizeof{I}}\geq
	\nnorm{A}\cdot (\lambda_{|I|})^{|I|}\geq \nnorm{A}\cdot |Q|.\]
(We use the property~\eqref{eq:genextr} of strict
	$\nnorm{A}$-adapted extractors.)
	Since $Q=X\restr{I}$, $X$ is an SCC, and $q,p\in Q$, there exists an execution 
	\[q_0\step{a_1}q_1\step{a_2}q_2\cdots \step{a_k}q_k=p\] in $A\restr{I}$ 
	with $k\leq |Q|-1$. The corresponding execution
	\[(q_0,x_0)\step{a_1}(q_1,x_1)\step{a_2}(q_2,x_2)\cdots
	\step{a_k}(q_k,x_k)=(p,y)\]
	 is contained within $X$, because $x_0$
	is above $\nnorm{A}\cdot |Q|\geq (k+1)\cdot\nnorm{A}$, and
	thus each $x_i$ is above
	$(k+1-i)\cdot\nnorm{A}\geq\nnorm{A}$.
The step
	$(p,y)\step{a}(p',y+\Delta(a)\restr{J})$ is thus within $X$ as
	well. 

	The proved fact that $p\step{a}p'$ with $p\in Q$ implies  a step $(p,y)\step{a}(p',y')$
	within $X$ shows not only 
	that $Q$ is a bSCC of $A\restr{I}$, but also  
that	${G^{(A,\lambda)}_{(I,Q)}}$ is reversible: 
	If
	$(p,x)\virtual{a}(p',x')$, that is, $p\step{a\restr{I}}p'$ and
	$x'-x=\Delta(a)\restr{J}$, then within the SCC $X$ we have
	$(p,y)\step{a}(p',y')$
	with $y'-y=x'-x=\Delta(a)\restr{J}$; this implies the
	existence of an execution
	$(p',y')\step{\sigma} (p,y)$ (within $X$), and thus
	$(p',x')\virtual{\sigma} (p,x)$.
\end{proof}	

\paragraph{Adapting the proof of Theorem~\ref{th:upper} from 
reversible nets to reversible PNSs}
The issue that a $d$-dim net $A$ with a live bottom SCC $X$ might not
possess a live marking 
above a sufficiently large level---specifically above
$\max\{C\suf^A,\nnorm{A}\cdot C\qd^A\}$ as stated in
Lemma~\ref{lem:finalbasicnets}---is, in principle,
resolved by extracting a PNS $G=G_{(I,Q)}^{(A,\lambda)}$ from $X$.
For a suitably chosen 2-exp extractor $\lambda$, the norm
$\nnorm{G}=\max\{\nnorm{Q},\nnorm{A}\}$ is (at most) 2-exp, and there
exists a live marking $(p_0,x_0)$ with $x_0$ above a sufficiently
large level (by part b) of Lemma~\ref{lem:sccfirst}).

In what follows, we therefore adapt the arguments from the previous proof
for reversible nets---which can be viewed as reversible PNSs with a
single state (the empty tuple)---to the full case of reversible PNSs.

\paragraph{(Virtual) reachability, 1-exp
$\bm{f\bvr}$ and $\bm{C\bvr^G}$, 2-exp
$\bm{C\bsuf^G=\max\{C\bdead^A,C\bvr^G\}}$.}
We now recall a variant of~\cite[Lemma~5]{DBLP:conf/fsttcs/Leroux19} in a more
general form than in Lemma~\ref{lem:fvr}:

\begin{lem}[\textbf{Virtual and standard reachability in
reversible PNSs, $\bm{f\bvr}$ and $\bm{C\bvr^G}$}]\label{lem:pnsreachfirst}
	There exists a 1-exp function $f\vr$ such that 
for every reversible $d$-dim PNS $G$, the following conditions hold,
	where	
$C\vr^G=f\vr(\nnorm{G},d)$:
	\begin{enumerate}
		\item
			If $(p,x)\vreach (q,y)$, then
			$(p,x)\virtual{\sigma}(q,y)$
for some $\sigma$ satisfying
			$|\sigma|\leq
					C\vr^G\cdot \nnorm{y-x}$.
\item
	If $(p,x)\vreach (q,y)$
			and  $x,y\in\upC{C\vr^G}$, 
			then $(p,x)\step{\sigma}(q,y)$
	with
		$|\sigma|\leq
		C\vr^G\cdot \nnorm{y-x}$.
	\end{enumerate}

\end{lem}
\begin{rem}
The cited result in~\cite{DBLP:conf/fsttcs/Leroux19} concerns
a broader class 
of reversible PNSs than those considered here and is technically more
detailed.
As in other parts of this paper, we prioritize a simplified presentation 
that suffices for our needs.
\end{rem}

% Fix Indentation
\noindent
Dead markings of a PNS $G=G_{(I,Q)}^A$ are the dead markings $y$ of the
net $A$ satisfying $y\restr{I}\in Q$. Therefore, we do not introduce
a separate constant
$C\dead^G$, but simply use $C\dead^A$.  The sufficiently large 2-exp
constant from~\eqref{eq:Csuf} now becomes
\begin{equation}\label{eq:CsufG}
C\suf^G=\max\{C\dead^A,C\vr^G\}.
\end{equation}

\paragraph{Virtual reachability in reversible PNSs via simple cycles
in control-state units.}
Control-state units of PNSs can be naturally viewed as graphs, allowing us
to use standard graph-theoretic terminology:

\begin{defi}[\textbf{Arc-labelled multigraph $\bm{(Q,E)}$ corresponding to
	a PNS $\bm{G_{(I,Q)}^A}$}]
	Given a $d$-dim PNS $G=G_{(I,Q)}^A$, by $(Q,E)$ we denote 
the arc-labelled multigraph 
over the set of actions $A$, 	where 
$p\step{a}q$ is an arc in $E$ if $p\step{a\restr{I}}q$ holds in
the net $A\restr{I}$.
\\
By a \emph{path} in $(Q,E)$, we mean a sequence
	$p_0\step{a_1}p_1\cdots \step{a_k}p_k$; it is a \emph{cycle}
	if $p_k=p_0$, and a \emph{simple
cycle} if, moreover, $0\leq i<j\leq k-1$ implies $p_i\neq p_j$. 
By the
\emph{effect} of a cycle $p\step{\sigma}p$, we mean
	the vector $\Delta(\sigma)\restr{J}\in\setZ^J$, where
	$J=[1,d]\smallsetminus I$
(noting that $\Delta(\sigma)\restr{I}=\mathbf{0}$).
\end{defi}	

% Fix Indentation
\noindent
The following proposition motivates  focusing on
 ``cyclic virtual reachability''.

\begin{prop}[\textbf{Virtual reachability in PNSs is based on
	control-unit cycles}]\label{prop:quickreachcycle}
Given a reversible $d$-dim PNS $G=G_{(I,Q)}^A$, if $(p,x)\vreach (q,y)$, then
there exists an action sequence $\sigma$ with $|\sigma|\leq
	(|Q|-1)$---which implies $\nnorm{\Delta(\sigma)}\leq
	(|Q|-1)\cdot\nnorm{A}$---such that
	\[(p,x)\virtual{\sigma}(q,x')\vreach(q,y),\]
where $x'=x+\Delta(\sigma)\restr{J}$ for $J=[1,d]\smallsetminus I$.
\end{prop}
\begin{proof}
If $(p,x)\vreach (q,y)$ in a reversible PNS $G=G_{(I,Q)}^A$,
we may simply consider a
	shortest path $p\step{\sigma}q$ in the corresponding graph $(Q,E)$;
we thus obtain $(p,x)\virtual{\sigma}(q,x')$ with
	$|\sigma|\leq |Q|-1$.
By reversibility of $G$, the relation $\vreach$ is symmetric, which
	implies  $(q,x')\vreach(p,x)$; together with the assumption $(p,x) \vreach
(q,y)$, this yields $(q,x') \vreach (q,y)$. 
\end{proof}

% Fix Indentation
\noindent
We note that simple cycles constitute the building blocks of
cyclic virtual executions:

\begin{prop}[\textbf{Zero-effect all-state cycles, and simple cycles as
	basic blocks}]\label{prop:cycles}
	Given a reversible $d$-dim PNS $G=G_{(I,Q)}^A$, the following
	conditions hold:
	\begin{enumerate}
		\item	There exists a cycle
	$p_0\step{\sigma_0}p_0$ in the corresponding graph $(Q,E)$
	that visits all states in $Q$ and has zero effect; that
			is, it satisfies 
			not only
	$\Delta(\sigma_0)\restr{I}=\mathbf{0}$ but also
	$\Delta(\sigma_0)\restr{J}=\mathbf{0}$ for
			$J=[1,d]\smallsetminus I$.
\item 
	For every state $p_0\in Q$, the effects of cycles
			$p_0\step{\sigma}p_0$
			are precisely those vectors in $\setZ^J$ that
			can be expressed as sums of the effects of
			simple cycles in $(Q,E)$.
	\end{enumerate}
	\end{prop}
\begin{proof}
	(1) By definition, $Q$ is a bottom SCC in $A\restr{I}$.
	Consequently, there is a cycle $p\step{\sigma}p$ in the graph
	$(Q,E)$ that visits all states $q\in Q$; the cycle corresponds to a virtual
	execution
	$(p,\mathbf{0})\virtual{\sigma}(p,\Delta(\sigma)\restr{J})$.
	By the symmetry of the relation $\vreach$, there must also
	exist a virtual execution
	$(p,\Delta(\sigma)\restr{J})\virtual{\sigma'}(p,\mathbf{0})$
	for some sequence $\sigma'$. Therefore,
	$p\step{\sigma\sigma'}p$ is the
	claimed zero-effect cycle visiting all states.

	(2) Let $\mathcal{E}\subseteq \setZ^J$ be the set
	of effects of simple cycles in the graph $(Q,E)$.
Any cycle $p_0\step{\sigma}p_0$ in 
	$(Q,E)$ is either a~simple cycle---in which case
	$\Delta(\sigma)\restr{J}\in\mathcal{E}$---or it can be
	decomposed as 
	$p_0\step{\sigma_1}q\step{\sigma_2}q\step{\sigma_3}p_0$, where
	$q\step{\sigma_2}q$ is a simple cycle. In the latter case, the
	effect of the cycle $p_0\step{\sigma_1\sigma_3}p_0$ belongs to
	$\mathcal{E}^*$ by the inductive hypothesis; since
	$\Delta(\sigma_2)\restr{J}\in\mathcal{E}$, it follows
	that $\Delta(\sigma_1\sigma_2\sigma_3)\restr{J}$ belongs to
	$\mathcal{E}^*$ as well. 

	On the other hand, if $z_0\in \mathcal{E}$---due to 
	 a simple
	cycle $q\step{\sigma}q$ with $\Delta(\sigma)\restr{J}=z_0$---
	then for any $p_0\in Q$, we may consider a cycle
	$p_0\step{\sigma_1}q\step{\sigma}q\step{\sigma_2}p_0$, where 
	$p_0\step{\sigma_1\sigma_2}p_0$ is a zero-effect cycle visiting
	all states in $Q$. Consequently, the effect of the cycle
	$p_0\step{\sigma_1\sigma\sigma_2}p_0$ is $z_0$.
	This implies that for any
	$z\in\mathcal{E}^*$, there exists a cycle 
	$p_0\step{\sigma}p_0$ with the effect $\Delta(\sigma)\restr{J}=z$.
\end{proof}

\paragraph{Cyclic virtual reachability in a PNS $\bm{G}$ via a
``simple-cycle'' net $\bm{\ASC^G}$.}
Recall that a $d$-dim net $A$ is a finite set of actions
 $a=(a_-,a_+)\in\setN^d\times\setN^d$, with
$\Delta(a)=(a_+-a_-)\in\setZ^d$. For the virtual reachability relation $\vreach$,
only the set $A\D=\{\Delta(a)\mid a\in A\}$ is relevant, since
$x\vreach y$ holds
if and only if $(y-x)\in A\D^*$. 
In the following definition, for $z\in \setZ^J$ we
use the notation $z_+$, $z_-$ where, for each $i\in J$,
\begin{center}
$z_+(i)=\max\{z(i),0\}$ and $z_-(i)=\max\{-z(i),0\}$.
\end{center}
	For instance, $(2,-3,0)_+=(2,0,0)$ and  $(2,-3,0)_-=(0,3,0)$.
Thus $z=z_+-z_-$, and each $z\in\setZ^J$ can be implicitly viewed as
a net action $z=(z_+,z_-)$ with $\Delta(z)=z$.

\begin{defi}[\textbf{Simple-cycle \bm{$|J|$}-dim net \bm{$\ASC^G$} associated with
	a PNS \bm{$G=G^A_{(I,Q)}$}}]\label{def:ASC}
	Given a $d$-dim PNS $G=G^A_{(I,Q)}$ with the corresponding
	graph $(Q,E)$
and $J=[1,d]\smallsetminus I$, 
	the \emph{simple-cycle net} $\ASC^G$
is the $|J|$-dim net 
	\begin{equation*}
		\ASC^G=(\ASC^G)\D=\{z\in\setZ^J\mid z \text{ is the effect of a simple
		cycle in }(Q,E)\}.
	\end{equation*}

\end{defi}
% Fix Indentation
\noindent
We note that, given a $d$-dim PNS $G=G^A_{(I,Q)}$, we have
\begin{equation}\label{eq:normASC}
\nnorm{\ASC^G}\leq |Q|\cdot\nnorm{A}\leq
	\nnorm{A}\cdot(1+\nnorm{Q})^{|I|}.
\end{equation}
The following fact is a corollary of
Proposition~\ref{prop:cycles}(2).

	\begin{cor}[\textbf{Reducing cyclic virtual reachability in
		$\bm{G}$ 
		to the net 
		$\bm{\ASC^G}$}]\label{cor:cycles}
\hfill\\
		Given a reversible $d$-dim PNS $G=G^A_{(I,Q)}$, for any $p\in Q$ we
		have 
		\begin{center}
		$(p,x)\vreach (p,y)$ in $G$ if and only if $x\vreach y$
		in $\ASC^G$.
		\end{center}			
	\end{cor}

\paragraph{Extending $\bm{(A,x_0,\lambda)}$-linear systems to PNSs.}
We now extend the notion of $(A,x_0,\lambda)$-linear systems
from Definition~\ref{def:crucialLS} to PNSs.

\begin{defi}[\textbf{$\bm{(G,(p_0,x_0),\lambda)}$-linear systems for
	reversible PNSs}]\label{def:crucialLSPNS}
\hfill\\
	Given a reversible $d$-dim PNS	$G=G^A_{(I,Q)}$, where
	$J=[1,d]\smallsetminus I$, and a $|J|$-dim extractor
	$\lambda=(\lambda_1,\lambda_2,\dots,\lambda_{|J|})$,
	we define
the bound	\begin{equation}\label{eq:BPNS}
	B=B_{(G,\lambda)}=\max\{C\suf^G,\lambda_{|J|}+C\shift^{\ASC^G}\}.
	\end{equation}
	Given a marking $(p_0,x_0)\in Q\times \setN^J$ 
	with $x_0\in\upC{C\suf^G}$,
	for each $\emptyset\neq J'\subseteq J$ let 
	\[x^\close_{J'}\in \setN^J\cap\upC{C\suf^G}\]
	be a fixed vector such that
	$(p_0,x_0)\reach(p_0,x^\close_{J'})$ and  
	$\dist{\lambda}{J'}{x^\close_{J'}}$ is
	minimal. 
Then the following 
	linear system $S$ of dimension $|J|\cdot 2^{|J|}$, with
	a $|J|$-dim variable
	$\bvar{x}=(\bvar{x}_1,\bvar{x}_2,\dots,\bvar{x}_{|J|})$ and
	$2^{|J|}-1$ $|J|$-dim variables 
	$\bvar{x}_{J'}$, one for each $\emptyset\neq J'\subseteq J$,
	is 
	\begin{center}
		a \emph{$(G,(p_0,x_0),\lambda)$-linear system}:
	\end{center}		
\begin{equation}\label{eq:GxLS}
	(\bvar{x}\equiv_B x_0)
	\land
	\bigwedge_{\emptyset\neq J'\subseteq
	J}\big((\bvar{x}_{J'}\equiv_B x^\close_{J'})\land
	(\bvar{x}\virtual{*}\bvar{x}_{J'})\big),
\end{equation}
	where $\bvar{x}\virtual{*}\bvar{x}_{J'}$ refers to the virtual
	reachability in $\ASC^G$. 
\end{defi}

% Fix Indentation
\noindent
Note that $(x,x')$ is a solution of the expression
$\bvar{x}\virtual{*}\bvar{x}_{J'}$ referring to $\ASC^G$ if and
only if $(p_0,x)\vreach(p_0,x')$ holds in $G$ (due to Corollary~\ref{cor:cycles}).

Our aim is to show that if $(p_0,x_0)$ is a live marking of
$G^A_{(I,Q)}$, where $x_0$ is above a sufficiently large level
relative to $\nnorm{G}=\max\{\nnorm{Q},\nnorm{A}\}$, then  $(p_0,\bar{x}_0)$ is live for every
solution $(\bar{x}_0,(\bar{x}_{J'})_{\emptyset\neq J'\subseteq J})$
of a $(G,(p_0,x_0),\lambda)$-linear system, for a suitable 2-exp
extractor $\lambda$.
As previously, we use the notion of quasi-dead markings to this aim.

\begin{defi}[\textbf{Quasi-dead markings in PNSs}]
\hfill\\	
	For a 2-exp function $f\qd$ and a $d$-dim PNS $G=G^A_{(I,Q)}$,
	where $J=[1,d]\smallsetminus I$,
	a marking $(p_0,x)\in Q\times \setN^J$
	is \emph{$f\qd$-quasi-dead} in $G$, or simply
	\emph{quasi-dead} in $G$ when $f\qd$
	is clear from context, if $x\in\upC{C\suf^G}$ 
	and there exists an execution $(p_0,x)\step{\sigma}(q,y)$ such
	that $(q,y)$ is a dead
marking and $|\sigma|\leq
f\qd(\nnorm{G},d)=C\qd^G$.
\end{defi}

\begin{lem}[\textbf{Nonlive $\bm{(p_0,x_0)}$ with large $\bm{x_0}$
	reaches a  quasi-dead
	marking $\bm{(p_0,x)}$}]\label{lem:reachquasiPNS}
	There exists a 2-exp function
	$f\qd$ such that 
	for every reversible $d$-dim PNS $G=G^A_{(I,Q)}$ and 
	every nonlive marking $(p_0,x_0)\in Q\times \setN^J$, where
	$J=[1,d]\smallsetminus I$ and
	\begin{equation}\label{eq:condxzero}
x_0\in\upC{\left(2\cdot(|Q|-1)\cdot\nnorm{A}+
		\max\{C\suf^G,\nnorm{A}\cdot C\qd^G\}\right)},
	\end{equation}
		there exists an $f\qd$-quasi-dead marking
	$(p_0,x)\in\Reach((p_0,x_0))$.
\end{lem}
\begin{proof}
	Let $G=G^A_{(I,Q)}$ be a reversible $d$-dim PNS, where
	$J=[1,d]\smallsetminus I$, and let $(p_0,x_0)$
be a nonlive marking with $x_0\in\setN^J$ satisfying	
	\begin{equation}\label{eq:xzerobegin}
		x_0\in\upC{\left(2\cdot(|Q|-1)\cdot\nnorm{A}+C\suf^G\right)}.
	\end{equation}
	The value $C\qd^G=f\qd(\nnorm{G},d)$ will arise from the
	following analysis.

Recall that $Q$ is a
	bSCC in $A\restr{I}$.
If there exists $p\in Q$ that is nonlive in $A\restr{I}$---that is, there exists an action
	$a\in A$ that is not enabled at any $q\in Q$ in
	$A\restr{I}$---then all $q\in Q$ are dead markings in
	$A\restr{I}$ (by our terminology);
consequently, $(p_0,x_0)$ is dead in the net $A$, and thus in the PNS $G$ as well.
	We thus further assume that $Q$ is a live bottom SCC in
	$A\restr{I}$.

	We now fix a dead marking $(q,y)$ such that $(p_0,x_0)\reach
	(q,y)$. 
Let $\emptyset\neq J'\subseteq J$ be such that 
	$y(i)< C\dead^A$ for all $i\in J'$ and 
	\begin{equation}\label{eq:Jprime}
		\textnormal{$(q,y\restr{J'})$ is
		dead in the net $A\restr{(I\cup J')}$.}
	\end{equation}
	By reversibility of $G$ and the
	assumption~\eqref{eq:xzerobegin},
 we may fix an
	execution 
\begin{equation*}
(p_0,x_0)\step{\sigma_0}(q,y')\reach (q,y)
\end{equation*}
where  $p_0\step{\sigma_0}q$ is a shortest path from $p_0$ to $q$ in the
	graph $(Q,E)$ associated with $G$. 
	(Recall Proposition~\ref{prop:quickreachcycle},
	Lemma~\ref{lem:pnsreachfirst}, and the fact that $C\suf^G\geq
	C\vr^G$ by~\eqref{eq:CsufG}.)
Since $|\sigma_0|\leq |Q|-1$, we obtain that, for all $i\in J$,
	\begin{equation}\label{eq:yprimefirst}
		y'(i)\geq x_0(i)-(|Q|-1)\cdot\nnorm{A},
	\end{equation}
and thus
	\begin{equation}\label{eq:yprime}
		y'\in\upC{\left((|Q|-1)\cdot\nnorm{A}+C\suf^G\right)}.
	\end{equation}
Since $y'\vreach y$ in the
	$|J|$-dim ``simple-cycle'' net $\ASC^G$ associated with $G$
	(by Corollary~\ref{cor:cycles}), 
due to the reversibility of $G$---implying the reversibility of
	$\ASC^G$---we may consider a segmented
 virtual execution  
	\begin{equation}\label{eq:downupPNS}
y=y_0\vreach y_1\vreach y_2\cdots\vreach y_k=y'
	\end{equation}
	as in~(\ref{eq:virtsegment}), for the net $\ASC^G$.	
	Consequently, for all $j\in[0,k-1]$ we have
	\begin{equation}\label{eq:CshiftASC}
	\nnorm{y_{j+1}-y_j}\leq
		C\shift^{\ASC^G}=f\shift(\nnorm{\ASC^G},|J|).
\end{equation}
Moreover, 
 the following holds for each $i\in J$:
\begin{itemize}
\item If $y(i)<y'(i)$, then
	$y(i)<y_1(i)<y_2(i)\cdots <y_{j_i}(i)=y_{j_i+1}(i)\cdots
	=y_k(i)=y'(i)$;
\item	
If $y(i)\geq y'(i)$, then $y_j(i)\geq y'(i)$ for all
	$j\in[0,k]$.
	\end{itemize}
By~\eqref{eq:yprime}, 
	there exists $j\leq C\suf^G+(|Q|-1)\cdot\nnorm{A}$ such that 
	the marking $y_j$ in~\eqref{eq:downupPNS}
	is above the level $C\suf^G+(|Q|-1)\cdot\nnorm{A}$. Setting $x'=y_j$, we
	obtain
 \[(q,y)\vreach (q,x')\step{\sigma'}(p_0,x),\]
where
$q\step{\sigma'}p_0$ is a shortest path from $q$ to $p_0$ in the graph
$(Q,E)$. Consequently, $|\sigma'|\leq |Q|-1$, and $x',x$ are both above
$C\suf^G$, because
\begin{equation}\label{eq:xprimex}
	\nnorm{x'-x}\leq (|Q|-1)\cdot\nnorm{A}.
\end{equation}
Since $(p_0,x_0)\reach(q,y)\vreach(q,x')\step{\sigma'}(p_0,x)$
and all $x_0,x',x$ are above $C\suf^G\geq C\vr^G$,
 we obtain
\begin{equation*}
(p_0,x_0)\reach(q,x')\reach(p_0,x)\reach(p_0,x_0),
\end{equation*}
where the last relation follows from reversibility.
Since $(p_0,x_0)\reach(q,y)$, we obtain
\[(p_0,x_0)\reach(p_0,x)\step{\sigma_1}(q,x')\step{\sigma_2}(q,y)\]
for some action sequences
$\sigma_1,\sigma_2$.
The marking $(p_0,x)$ is a candidate for a quasi-dead marking with
the control state $p_0$.
We can bound the length of $\sigma_1$, using
Lemma~\ref{lem:pnsreachfirst} and~\eqref{eq:xprimex}: 
\begin{equation}\label{eq:sigmaone}
	|\sigma_1|\leq C\vr^G\cdot \nnorm{x'-x}\leq
C\vr^G\cdot(|Q|-1)\cdot\nnorm{A}.
\end{equation}
However, we have no bound on $|\sigma_2|$. To handle this, we proceed
as in the proof of Lemma~\ref{lem:reachquasi}.
We set 
\begin{equation}\label{eq:Jprimeprime}
	J''=\{i\in J\mid y(i)< y'(i)\}=\{i\in J\mid y(i)<x'(i)\},
\end{equation}
	noting that $J'\subseteq J''\subseteq J$; recall~\eqref{eq:Jprime} and
	the fact that 
		 $y(i)<C\dead^A\leq C\suf^G\leq y'(i)$ for
	all $i\in J'$.
Since the marking  $(q,y\restr{J'})$ is dead in the net  $A\restr{(I\cup
J')}$, the marking
$(q,y\restr{J''})$ is dead in the net $A\restr{(I\cup J'')}$.

Since $(q,x'\restr{J''})\reach(q,y\restr{J''})$ in  
 $A\restr{(I\cup J'')}$, and  $(q,x'\restr{J''})\geq
 (q,y\restr{J''})$,
 we may apply
 Proposition~\ref{prop:simpletrick}. Consequently, there exists 
 $x''\in \setN^{J''}$ such that
 $x''\geq x'\restr{J''}$ 
and $(q,x'')\step{\sigma}(q,y\restr{J''})$ for some sequence $\sigma$,
where both $\nnorm{x''-x'\restr{J''}}$ and $|\sigma|$ are bounded by
\begin{equation}\label{eq:boundfmr} 
f\mr(\nnorm{A},d)\cdot \nnorm{x'-y}\leq C\mr^A\cdot
	(C\suf^G+(|Q|-1)\cdot\nnorm{A})\cdot C\shift^{\ASC^G}.
\end{equation}
Using reversibility, we obtain $(q,x'\restr{J''})\vreach (q,x'')$
in the net $A'=A\restr{(I\cup J'')}$, as well as in the PNS
$G'=G\restr{(I\cup J'')}=G^{A'}_{(I,Q)}$. Since $x'\restr{J''}$ and
$x''$ are above $C\suf^G\geq C\vr^G\geq C\vr^{G'}$, there exists an
action sequence $\rho$ satisfying
$(q,x'\restr{J''})\step{\rho}(q,x'')$ and 

\begin{equation}\label{eq:boundrho}
	|\rho|\leq C\vr^G\cdot \nnorm{x''-x'\restr{J''}}\leq
C\vr^G\cdot f\mr(\nnorm{A},d)\cdot \nnorm{x'-y}.
\end{equation}
We have thus obtained that, in the net $A\restr{(I\cup J'')}$,
\[(p_0,x\restr{J''})\step{\sigma_1}(q,x'\restr{J''})\step{\rho}(q,x'')\step{\sigma}(q,y\restr{J''}).\]
Inspecting the bound on the length $|\sigma_1\rho\sigma|$ provided
by~\eqref{eq:sigmaone}, \eqref{eq:boundrho}, 
\eqref{eq:boundfmr}, \eqref{eq:CshiftASC},
and \eqref{eq:normASC}, we deduce the existence 
 of a 2-exp function $f\qd$ such that
\[(p_0,x\restr{J''})\step{\sigma_1\rho\sigma}(q,y\restr{J''})\text{ with }
|\sigma_1\rho\sigma|\leq f\qd(\nnorm{G},d).\]
We have $(p_0,x)\step{\sigma_1}(q,x')$ in $G$; however, the sequence $\rho\sigma$ might not be executable from $(q,x')$ in
$G$.
Recalling~\eqref{eq:Jprimeprime}, \eqref{eq:yprimefirst}, and the definition 
 $x'=y_j$ for a respective $y_j$ from~\eqref{eq:downupPNS}, we note
 that,
for all $i\in J\smallsetminus J''$,
\[x'(i)\geq y'(i)\geq x_0(i)-(|Q|-1)\cdot\nnorm{A}.\]
Consequently, if $x_0$ (satisfying~\eqref{eq:xzerobegin}) is above the level
$(|Q|-1)\cdot\nnorm{A}+\nnorm{A}\cdot C\qd^G$, then we obtain 
\[(p_0,x)\step{\sigma_1}(q,x')\step{\rho\sigma}(q,y'')\]
in $G$, where ${y''}\restr{J''}=y\restr{J''}$. Since
$(q,y\restr{J''})$ is dead in $A\restr{(I\cup J'')}$, the marking
$(q,y'')$ is dead in $A$ (and thus in $G$). Consequently, $(p_0,x)$ is an $f\qd$-quasi-dead marking in $G$.
\end{proof}	

\paragraph{Guaranteed existence of live markings $\bm{(p_0,x_0)}$ with large
counter values $\bm{x_0}$.}
We now show that from a live bottom SCC $X$ of a $d$-dim net $A$, we can always extract a
PNS $G=G_{(I,Q)}^A$ where $\nnorm{G}=\max\{\nnorm{Q},\nnorm{A}\}$ is
bounded by a 2-exp function and which possesses a live
marking $(p_0,x_0)$ such that $x_0$ satisfies
condition~\eqref{eq:condxzero}. To this end,
we fix a 2-exp function $f\qd$ provided by Lemma~\ref{lem:reachquasiPNS}.
By $f\suf$, we mean a 2-exp function satisfying
$f\suf(\nnorm{G},d)=C\suf^G=\max\{C\dead^{A},C\vr^G\}$ for every $d$-dim
PNS $G_{(I,Q)}^A$ (which coincides with $A$ if $I=\emptyset$).

\begin{defi}[\textbf{Feasible PNS $\bm{G^{(A,\lambda')}_{(I,Q)}}$ extracted
	from a live bottom SCC $\bm{X}$}]\label{def:feasiblePNS}
	Let a partial function $\bar{\lambda}\colon (\setN_+)^3\to
	\setN_+$ be defined inductively as
	follows for all triples $(m,d,i)\in (\setN_+)^3$ with
	$i\in[1,d]$;
we write $\bar{\lambda}^{(m,d)}_i$ to denote $\bar{\lambda}(m,d,i)$: 
	\begin{itemize}
		\item			
			$\bar{\lambda}^{(m,d)}_1=1+\max\{f\suf(m,d),m\cdot
			f\qd(m,d)\}$;
\item
	$\bar{\lambda}^{(m,d)}_{i+1}=\left[2\cdot((\bar{\lambda}^{(m,d)}_i)^i-1)\cdot
			m+ \max\{f\suf(\bar{\lambda}^{(m,d)}_i,d), m\cdot
			f\qd(\bar{\lambda}^{(m,d)}_i,d)\}\right]+m\cdot
			i\cdot(\bar{\lambda}^{(m,d)}_i)^i.$
	\end{itemize}
For any live bottom SCC $X$ in the reachability graph of
	a $d$-dim net $A$, 
	a $d$-dim PNS $G=G_{(I,Q)}^A$ is \emph{feasible} if 
	$G=G_{(I,Q)}^{(A,\lambda')}$, where
	$\lambda'=(\lambda'_1,\lambda'_2,\dots,\lambda'_d)$ is the extractor satisfying 
	$\lambda'_i=\bar{\lambda}^{(\nnorm{A},d)}_i$ for all
	$i\in[1,d]$. 
	\\
	(Recall Definition~\ref{def:extrPNS} and note that $\lambda'$
	is strictly $\nnorm{A}$-adapted.)
\end{defi}

% Fix Indentation
\noindent
Observe that the function $\bar{f}(m,d)=\bar{\lambda}^{(m,d)}_d$ is a
2-exp function ($\bar{f}(m,d)\leq (2+m)^{2^{\text{poly}(d)}}$).

\begin{prop}[\textbf{Feasible PNSs possess live markings with
	large counter values}]\label{prop:feasiblelive}
	Let $X$ be a live bottom SCC of
	a $d$-dim net $A$, 
and let $G=G_{(I,Q)}^{(A,\lambda')}$ be a feasible PNS extracted from
	$X$.
Then there exists  a (live)	marking
	$(p_0,x_0)\in X$ with $p_0\in Q=X\restr{I}$ and
	\[x_0\in\upC{\left(
2\cdot(|Q|-1)\cdot\nnorm{A}+
		\max\{C\suf^G,\nnorm{A}\cdot C\qd^G\}\right)}.\]
\end{prop}	
\begin{proof}
Let $A,X,G=G_{(I,Q)}^{(A,\lambda')}$ be as in the statement.

		If $I=[1,d]$, then $Q=X$, and the claim holds
		trivially.
	If $I=\emptyset$, then the PNS $G$ coincides with the net $A$
	(in this case, $Q=\{()\}$ and thus $|Q|=1$), and, by
	definition of $G_{(I,Q)}^{(A,\lambda')}$, 
	there exists $x_0\in X$ such that, for all
	$i\in[1,d]$,
	\[x_0(i)\geq 	\lambda'_1\geq
	1+\max\{f\suf(\nnorm{A},d),\nnorm{A}\cdot
			f\qd(\nnorm{A},d)\}\geq
 \max\{C\suf^A,\nnorm{A}\cdot
	C\qd^A\}.\]
	We therefore further assume $\emptyset\neq I\neq [1,d]$.
By definition of $G=G_{(I,Q)}^{(A,\lambda')}$,
we have $\nnorm{Q}<\lambda'_{|I|}$, and there exists $(p_0,x_0)\in X$
	where, for every
	$i\in J=[1,d]\smallsetminus I$, 
\[x_0(i)\geq\lambda'_{|I|+1}-\nnorm{A}\cdot
	|I|\cdot(\lambda'_{|I|})^{|I|}.\]
	The definition of $\lambda'$ (in
	Definition~\ref{def:feasiblePNS}) implies that,  for every
	$i\in J=[1,d]\smallsetminus I$, 
	\[x_0(i)\geq
	2\cdot((\lambda'_{|I|})^{|I|}-1)\cdot\nnorm{A}+
		\max\{f\suf(\lambda'_{|I|},d),\nnorm{A}\cdot
		f\qd(\lambda'_{|I|},d)\}.\]
	Since $(\lambda'_{|I|})^{|I|}-1\geq (1+\nnorm{Q})^{|I|}-1\geq
	|Q|-1$ and $\lambda'_{|I|}\geq
	\max\{\nnorm{Q},\nnorm{A}\}=\nnorm{G}$,
	we obtain,  for every
	$i\in J=[1,d]\smallsetminus I$, 
	\[x_0(i)\geq
	2\cdot(|Q|-1)\cdot\nnorm{A}+
		\max\{f\suf(\nnorm{G},d),\nnorm{A}\cdot
		f\qd(\nnorm{G},d)\},\]
that is,
	\[x_0(i)\geq
2\cdot(|Q|-1)\cdot\nnorm{A}+
		\max\{C\suf^G,\nnorm{A}\cdot C\qd^G\}.\qedhere\]
\end{proof}

\paragraph{Live solutions of $\bm{(G,(p_0,x_0),\lambda)}$-linear systems.}
Recall Definition~\ref{def:crucialLSPNS} of $(G,(p_0,x_0),\lambda)$-linear systems.
We now show that, given a feasible PNS $G=G^{(A,\lambda')}_{(I,Q)}$,
if $(p_0,x_0)$ is a live marking of $G$ guaranteed by
Proposition~\ref{prop:feasiblelive}
and $S$ is a $(G,(p_0,x_0),\lambda)$-linear system for an extractor
$\lambda$ analogous to that in Lemma~\ref{lem:finalbasicnets},
then $(p_0,\bar{x}_0)$ is a live marking of $G$ 
for every solution
$(\bar{x}_0,(\bar{x}_{J'})_{\emptyset\neq J'\subseteq J})$
of $S$.
The proof is analogous to that of Lemma~\ref{lem:finalbasicnets} and
is therefore presented here in a more concise manner.

\begin{lem}[\textbf{$\bm{(G,(p_0,x_0),\lambda)}$-solutions are live,
	if $\bm{G}$ is feasible and 
	 $\bm{x_0,\lambda}$ are large}]\label{lem:finalPNS}
	Let $G=G_{(I,Q)}^{(A,\lambda')}$ be a feasible PNS, extracted
	from a live bottom SCC $X$ of
	a $d$-dim net $A$.
Let
	$\lambda=(\lambda_1,\lambda_2,\dots,\lambda_{|J|})$,
	where $J=[1,d]\smallsetminus I$, be an extractor
	satisfying:
	\begin{itemize}
		\item
		$\lambda_1\geq
C\suf^G+\nnorm{A}\cdot C\qd^G$;
\item
	$\lambda_{i+1}\geq \lambda_1 + \nnorm{A}\cdot C\vr^{G}\cdot
			(\lambda_i-C\suf^G)$ for each
			$i\in[1,|J|{-}1]$.
	\end{itemize}
	Let $(p_0,x_0)$ be a live marking of $G$ where $x_0\in\setN^J$
	satisfies 
	\[x_0\in\upC{\left(
2\cdot(|Q|-1)\cdot\nnorm{A}+
		\max\{C\suf^G,\nnorm{A}\cdot C\qd^G\}\right)}.\]
Let $S$ be a $(G,(p_0,x_0),\lambda)$-linear system,
and let $(\bar{x}_0,(\bar{x}_{J'})_{\emptyset\neq
	J'\subseteq J})$ be a solution of $S$.
Then
 $(p_0,\bar{x}_0)$ is a live marking. 
\end{lem}	
\begin{proof}
Let $A,G=G_{(I,Q)}^{(A,\lambda')},\lambda$, and $(p_0,x_0)$ be as in
	the statement.
Let $A'=\ASC^G$ be the ``simple-cycle'' $|J|$-dim net associated with
	$G$, where $J=[1,d]\smallsetminus I$.

	Let $S$ be a $(G,(p_0,x_0),\lambda)$-linear
	system~\eqref{eq:GxLS}, namely
\begin{equation*}
	(\bvar{x}\equiv_B x_0)
	\land
	\bigwedge_{\emptyset\neq J'\subseteq
	J}\big((\bvar{x}_{J'}\equiv_B x^\close_{J'})\land
	(\bvar{x}\virtual{*}\bvar{x}_{J'})\big),
\end{equation*}
	where $B=\max\{C\suf^G,\lambda_{|J|}+C\shift^{A'}\}$
and
	$\bvar{x}\virtual{*}\bvar{x}_{J'}$ refers to the virtual
	reachability in $A'$. 
	Recall that $(p_0,x_0)\reach (p_0,x^\close_{J'})$ for each
	$\emptyset\neq J'\subseteq J$; this implies $x_0\vreach
	x^\close_{J'}$ in $A'$.

Towards a contradiction, suppose that
	\begin{center}
		$(\bar{x}_0,(\bar{x}_{J'})_{\emptyset\neq J'\subseteq
		J})$ is a solution of $S$
		where $(p_0,\bar{x}_0)$ is nonlive. 
	\end{center}
	The sets of indices of $\lambda$-small components 
	$I^{\lambda}_{\bar{x}_{J'}}\subseteq J$ and
	$I^{\lambda}_{x^\close_{J'}}\subseteq J$
	(as defined in	Proposition~\ref{prop:Ilambdax}) are
	identical,
	and the vectors $\bar{x}_{J'}$ and $x^\close_{J'}$---viewed as
	markings of the net $A'=\ASC^G$---coincide on
	these indices.
	By definition, all vectors $x_0,x^\close_{J'}$, and therefore also 
	$\bar{x}_0, \bar{x}_{J'}$, are above the level $C\suf^G$.
	Lemma~\ref{lem:reachquasiPNS} implies the existence of
	\begin{center}
	a quasi-dead
		marking $(p_0,x)\in\Reach((p_0,\bar{x}_0))$ with
		$x\in\upC{C\suf^G}$; let
		$I'=I^{\lambda}_{x}\subseteq J$. 
	\end{center}		
	Since $(p_0,x)$ is a quasi-dead marking, we have
	$(p_0,x)\step{\sigma_1}(q,y)$ where
	$(q,y)$ is a dead marking and
	$|\sigma_1|\leq C\qd^G$.  
Note that $I'\neq\emptyset$.
	Indeed, we have $y(i)< C\dead^A$
	for at least one $i\in J$, since otherwise $q$ would be dead
	in $A\restr{I}$, which would imply that $(p_0,x_0)$ is
	nonlive (because
	$(p_0,x_0)\step{\sigma_1}(q,x_0+\Delta(\sigma_1)\restr{J})$). 
	The case $I'=I^{\lambda}_{x}=\emptyset$ would imply
that $x$ is above the level $\lambda_1\geq C\suf^G+\nnorm{A}\cdot
	C\qd^G\geq C\suf^G+\nnorm{A}\cdot |\sigma_1|$, in which case $y$ would be above $C\suf^G\geq
	C\dead^A$---a contradiction. Consequently, 
	$\bar{x}_{I'}$ is well-defined.

	Since $(p_0,\bar{x}_0)\reach (p_0,x)$, we have
	$\bar{x}_0\vreach x$ in $A'$. By reversibility of $A'$
	(both $A$ and $G$ are reversible), it follows that
	$\bar{x}_{I'}\vreach x$ in $A'$ (since $\bar{x}_0\vreach
	\bar{x}_{I'}$ and $\bar{x}_0\vreach x$); this implies
	$(p_0,\bar{x}_{I'})\reach (p_0,x)$ in $G$ (and in $A$), because
	both $\bar{x}_{I'}$ and $x$ are above $C\suf^G$.

The equalities 
\[\dist{\lambda}{I'}{x}=\dist{\lambda}{I'}{\bar{x}_{I'}}=\dist{\lambda}{I'}{x^\close_{I'}}=0\] 
are established analogously to Lemma~\ref{lem:finalbasicnets},
by considering the distance of $\bar{x}_{I'}$ to the $\lambda$-case $I'$
	and a segmented virtual execution $\bar{x}_{I'}\vreach x$ in
	$A'=\ASC^G$.

For each $i\in I'$, the values $x(i)$ and
$\bar{x}_{I'}(i)=x^\close_{I'}(i)$ are in the interval
$[C\suf^G,\lambda_{|I'|}-1]$; therefore, 
$\nnorm{x\restr{I'}-(\bar{x}_{I'})\restr{I'}}\leq \lambda_{|I'|}-C\suf^G$.
The fact 
$(p_0,\bar{x}_{I'})\reach (p_0,x)$ implies that
there exists an action sequence $\sigma_2$ such that 
$(p_0,(\bar{x}_{I'})\restr{I'})\step{\sigma_2} (p_0,x\restr{I'})$ in
the restricted PNS $G\restr{I\cup I'}$ (that is, in $A\restr{I\cup I'}$)
and
\[|\sigma_2|\leq f\vr(\nnorm{G},|I\cup I'|)\cdot
\nnorm{x\restr{I'}-(\bar{x}_{I'})\restr{I'}}\leq C\vr^{G}\cdot
(\lambda_{|I'|}-C\suf^G).\]
Combining with $(p_0,x)\step{\sigma_1}(q,y)$, where $|\sigma_1|\leq
C\qd^G$ and $(q,y)$ is dead, we obtain
\[(p_0,(\bar{x}_{I'})\restr{I'})=(p_0,(x^\close_{I'})\restr{I'})\step{\sigma_2}
(p_0,x\restr{I'})\step{\sigma_1}(q,y\restr{I'})\]
in $G\restr{I\cup I'}$, where $|\sigma_2\sigma_1|\leq C\vr^{G}\cdot
(\lambda_{|I'|}-C\suf^G)+C\qd^G$.
Since for each $i\in J\smallsetminus I'$, we have
$x^\close_{I'}(i)\geq\lambda_{|I'|+1}\geq \nnorm{A}\cdot
|\sigma_2\sigma_1|$, it follows that 
\[(p_0,x^\close_{I'})\step{\sigma_2\sigma_1}(q,y')\]
is an execution in $G$, where $y'\restr{I'}=y\restr{I'}$.
We complete the proof by showing that $(q,y')$ is dead, which
contradicts the assumption that $(p_0,x^\close_{I'})$ is live (since 
$(p_0,x_0)\reach(p_0,x^\close_{I'})$ and $(p_0,x_0)$ is live).

Recall that $(p_0,x)\step{\sigma_1}(q,y)$ and $|\sigma_1|\leq C\qd^G$.
Since for each $i\in J\smallsetminus I'$, we have $x(i)\geq
\lambda_{|I'|+1}\geq \lambda_1\geq C\dead^A+ \nnorm{A}\cdot C\qd^G$, it follows that 
$y(i)\geq C\dead^A$ for each  $i\in J\smallsetminus I'$. This implies
that $(q,y\restr{I'})$ is dead in $A\restr{I\cup I'}$, which in turn
implies that $(q,y')$ is dead in $A$ (since
$y'\restr{I'}=y\restr{I'}$).
\end{proof}

\paragraph{Proof of Theorem~\ref{th:upper}.} 

Given a $d$-dim
net $A$ possessing a live bottom SCC $X\subseteq\setN^d$, we consider 
 a feasible PNS  $G=G_{(I,Q)}^{(A,\lambda')}$ as defined in
Definition~\ref{def:feasiblePNS}. We set $J=[1,d]\smallsetminus I$.

Since $\nnorm{Q}<\lambda'_{|I|}$, it
follows that 
\[\nnorm{G}=\max\{\nnorm{Q},\nnorm{A}\}\leq
(2+\nnorm{A})^{2^{\text{poly}(d)}}.\]
The 2-exp bound follows from the definition of $\lambda'$ and
the closure properties of 1-exp and 2-exp functions discussed at the
beginning of this section.
Moreover, by Proposition~\ref{prop:feasiblelive}, there exists a live	marking
	$(p_0,x_0)\in X$ with $p_0\in Q=X\restr{I}$ and
	\[x_0\in\upC{\left(
2\cdot(|Q|-1)\cdot\nnorm{A}+
		\max\{C\suf^G,\nnorm{A}\cdot C\qd^G\}\right)}.\]
Recall that $|Q|\leq (1+\nnorm{Q})^{|I|}$ (where $|I|\leq d$) and
that $f\suf$ and $f\qd$ are 2-exp functions.
Lemma~\ref{lem:finalPNS} implies that every solution
of a $(G,(p_0,x_0),\lambda)$-linear system $S$, 
where $\lambda$ is defined
by 
\[\lambda_1=
C\suf^G+\nnorm{A}\cdot C\qd^G, \text{ and } 
\lambda_{i+1}= \lambda_1 + \nnorm{A}\cdot C\vr^{G}\cdot
			(\lambda_i-C\suf^G)\text{ for each }
			i\in[1,|J|{-}1],\]
			yields  
a live marking $(p_0,\bar{x}_0)$.
The dimension of $S$ is at most $d\cdot 2^d$, and
\[\max\{\mlcm{S},\nnorm{S}\}\leq (2+\nnorm{A})^{2^{\text{poly}(d)}}.\]
Indeed, the virtual reachability in $S$ refers to the ``simple-cycle
net''  $\ASC^G$; consequently,
Corollary~\ref{cor:virtual2system} implies that 
\[\mlcm{S}\leq |J|\,!\cdot\nnorm{\ASC^G}\leq d\,!\cdot
|Q|\cdot\nnorm{A}\leq  d\,!\cdot (1+\nnorm{Q})^d \cdot\nnorm{A},\]
where $\nnorm{Q}\leq (2+\nnorm{A})^{2^{\text{poly}(d)}}$.
Since
$B=B_{(G,\lambda)}=\max\{C\suf^G,\lambda_{|J|}+C\shift^{\ASC^G}\}$,
where
\begin{equation*}
\lambda_{|J|}\leq (2+\nnorm{A})^{2^{\text{poly}(|J|)}}\leq (2+\nnorm{A})^{2^{\text{poly}(d)}}
	\text{ and }f\shift(\nnorm{\ASC^G},|J|)\leq f\shift(|Q|\cdot \nnorm{A},d),
\end{equation*}
	we also obtain $\nnorm{S}\leq  (2+\nnorm{A})^{2^{\text{poly}(d)}}$.

Analogously to Proposition~\ref{prop:Affsmall}, we use
Theorem~\ref{thm:smallsolutions} to deduce that
there exists a solution of $S$ that yields $(p_0,\bar{x}_0)$ (which is
a live marking of $A$) such that
\[\nnorm{(p_0,\bar{x}_0)}\leq
(2+\nnorm{A})^{2^{\text{poly}(d)}}.\tag*{\qed}\]

\subsection{Dead markings}\label{subsec:dead}
\hfill\\

In this section, we provide further details regarding the
characterisation~\eqref{eq:alldead} of dead markings.
Recall that in our terminology, a marking $y\in\setN^d$ of a $d$-dim net $A$
is dead if there exists an action $a\in A$ that is dead at $y$, that is,
$a$ is disabled at every marking $y'\in\Reach(y)$. The following
proposition is a straightforward observation.
\begin{prop}[\textbf{If a restriction of a marking is dead, then the
	marking is dead}]\label{prop:restrdead}
	Given a $d$-dim net $A$ and $I\subseteq [1,d]$, if 
 $y\restr{I}$ is dead in $A\restr{I}$, then $y$ is dead in $A$.
\end{prop}
\begin{proof}
The claim follows from the fact that, for every $y\in\setN^d$, the set
$\{y'\restr{I}\mid y\reach y'$ in $A\}$ is a subset of the set
$\{y'\in\setN^I\mid y\restr{I}\reach y'$ in $A\restr{I}\}$.
\end{proof}

% Fix Indentation
\noindent
We now adapt Rackoff's result for coverability
from~\cite{DBLP:journals/tcs/Rackoff78}.
\begin{lem}[\textbf{Deadness is determined by components with
	``small'' values}]\label{lem:rackoffdead}
\hfill\\
	There exists a 2-exp function $f\dead$ with the following property:
\\
For every
$d$-dim net $A$, a marking $y\in\setN^d$ is dead in $A$ if and only if 
 $y\restr{\textsc{s}_y}$ is dead in the restricted net
	$A\restr{\textsc{s}_y}$,
	where $\textsc{s}_y=\{i\in [1,d]\mid
y(i)<f\dead(\nnorm{A},d)\}$.
\end{lem}
\begin{proof}
Note that an action $a=(a_-,a_+)$ in a net $A$
	is not dead
	at a marking $x$ if and only if $x$ can cover $a_-$, that is, 
	if and only if there exists $x'$ such that $x\reach x'\geq a_-$.
	Along the lines of~\cite{DBLP:journals/tcs/Rackoff78}
	(Lemmas~3.3. and~3.4), we define a function
	$g:\setN\times\setN\to\setN$ as follows:
\begin{itemize}
	\item			
 $g(m,0)=0$;
\item
	for $i\geq 1$,	$g(m,i)= (m + m\cdot g(m,i{-}1))^{i}+g(m,i{-}1)$.
\end{itemize}		

	% Fix Indentation
	\noindent
	By induction on $|I|$, it is straightforward to verify that 
	for a $d$-dim net $A$, 
an action $a=(a_-,a_+)\in A$, an index set $I\subseteq [1,d]$, and a
	marking
	$x\in\setN^I$ of $A\restr{I}$, the following holds: if  $x$ can cover
	$(a_-)\restr{I}$ (in the net $A\restr{I}$), then 
there exist an action sequence $\sigma$ and a marking $x'\in\setN^I$ such that 
	\begin{equation*}
		x\step{\sigma}x'\geq (a_-)\restr{I}\text{ and }|\sigma|\leq
	g(\nnorm{A},|I|). 
	\end{equation*}
The claim is trivial for $I=\emptyset$.	In the case $I\neq\emptyset$,
we observe that 
if a shortest execution  $x\reach x'\geq (a_-)\restr{I}$ in
	$A\restr{I}$
	 has a prefix
	reaching the value $n=\nnorm{A} + \nnorm{A}\cdot
	g(\nnorm{A},|I|{-}1)$ or larger in some
	component, then the length of the suffix is bounded
	by $g(\nnorm{A},|I|{-}1)$ by the inductive hypothesis, and the
	length of the prefix is bounded by $n^{|I|}$.

This implies that for any $d$-dim net $A$ and $x\in\setN^d$, 
the values
	$x(i)$, $i\in[1,d]$, for which  
	$x(i)\geq \nnorm{A}+\nnorm{A}\cdot g(\nnorm{A},d)$ are
	irrelevant to 
	the question of whether $x$ is dead. 
As the required 2-exp function, we can thus take
	\[f\dead(m,i)=m+m\cdot g(m,i).\qedhere\] 
\end{proof}	

\begin{rem}
We note that the recent
paper~\cite{kunnemann_et_al:LIPIcs.ICALP.2023.131} has improved Rackoff's
bound in~\cite{DBLP:journals/tcs/Rackoff78}, but such fine-tuned
results are not needed at our level of analysis.
\end{rem}

\subsection{Virtual reachability in reversible 
nets}\label{subsec:virtrevers}
\hfill\\

In this section, we formalize the previous informal claim regarding small-segment
virtual executions~(\ref{eq:virtsegment})
that are ``directed'' from a marking $x$ to a marking $y$.

We recall the notation $A\D=\{\Delta(a)\mid a\in
A\}$ and the fact that
the virtual reachability relation $\vreach$ in any
\emph{reversible} $d$-dim net $A$ is
symmetric: $x\virtual{*}y$ implies $y\virtual{*}x$.
This follows from the fact that $x\virtual{*}y$ in $A$ is equivalent to
$(y-x)\in A\D^*$, and that 
$A\D^*$  is a subgroup of
$(\setZ^d,+)$ if $A$ is reversible.

For any $d\in\Nat$, we define the function 
\begin{equation*}
\sign:\setZ^d\to\{-1,0,1\}^d
\end{equation*}
as follows:
 for $x\in\setZ^d$ and $i\in[1,d]$, we set
\begin{itemize}
	\item		
 $\sign(x)(i)=-1$ if $x(i)<0$;
\item
$\sign(x)(i)=0$ if $x(i)=0$;
\item
		$\sign(x)(i)=1$ if $x(i)>0$.
\end{itemize}
		We define the following partial order $\sleq$ on $\setZ^d$:
\begin{equation*}
	\textnormal{$x\sleq y$\ if $\sign(x)=\sign(y)$ and $|x(i)|\leq
	|y(i)|$ for all $i\in [1,d]$.} 
\end{equation*}
For instance, $(0,5,-3)\sleq (0,11,-6)$, but 
 $(0,5,-3)\not\sleq (1,11,-6)$, since
 $\sign((0,5,-3))=(0,1,-1)\neq\sign((1,11,-6))=(1,1,-1)$.
Note that $\mathbf{0}\sleq x$ implies $x=\mathbf{0}$.
For $X\subseteq \setZ^d$, 
\[\minsleq{X}\]
denotes the set of 
minimal elements of $X$ w.r.t.\ $\sleq$\,; it is finite
since
$\sleq$ is a well-partial-order.

\begin{prop}[\textbf{``Directed'' decomposition of elements in group $\bm{L}$ by
	$\bm{\minsleq{L}}$}]\label{prop:onestepsubgroup}
	If $L\subseteq \setZ^d$ is a group and
	$y\in L$, then there exists $z\in\minsleq{L}$ such that
	$z\sleq y$, $y=z+(y-z)$,   
	and $(y-z)\in L$. Moreover, $\nnorm{y-z}<\nnorm{y}$
 if $y\neq\mathbf{0}$.
\end{prop}	
\begin{proof}
Let $L\subseteq \setZ^d$ be a group, and
let $y\in L$.
	Since $\sleq$ is a well-partial-order, there exists $z\in\minsleq{L}$ such that $z\sleq
	y$. We have $(y-z)\in L$ because $L$ is a group.
To show the strict decrease in norm, consider a component $i\in[1,d]$:
	\begin{itemize}
		\item
			If $y(i)=0$, then $z(i)=0$, since
			$\sign(y)=\sign(z)$; therefore, $y(i)-z(i)=0$.
\item
If $y(i)>0$, then $0<z(i)\leq y(i)$, 
which implies $0\leq y(i)-z(i)<y(i)$; therefore,
$|y(i)-z(i)|<|y(i)|$.
\item
If $y(i)<0$, then $y(i)\leq z(i)<0$, 
			which implies $y(i)<y(i)-z(i)\leq 0$;
			therefore, 
$|y(i)-z(i)|<|y(i)|$.
\end{itemize}
Consequently, $\nnorm{y-z}<\nnorm{y}$ if
	$y\neq\mathbf{0}$.
\end{proof}

\begin{cor}[\textbf{Segmenting virtual executions of reversible nets}]\label{cor:subgroup}
\hfill\\
	Given a reversible net $A$ (for which $A\D^*$ is a group), if
	$x\virtual{*}y$, then 
	there exist $z_1,z_2,\dots,z_k$ in $\minsleq{A\D^*}$
such that
\begin{equation}\label{eq:decompseq}
x\virtual{*}(x+z_1)\virtual{*}(x+z_1+z_2)\cdots\virtual{*}(x+z_1+z_2\cdots
	+z_k)=y
\end{equation}	
and for each $j\in[1,k]$ it holds that
 $z_j\sleq y-(x+z_1+z_2\cdots +z_{j-1})$.
\end{cor}
\begin{proof}
This is a direct consequence of Proposition~\ref{prop:onestepsubgroup},
since $x\virtual{*}y$ is equivalent to $(y-x)\in A\D^*$. 
By applying the proposition repeatedly, we construct the sequence:
in the first step, we choose
$z_1\in\minsleq{A\D^*}$ such that 
$z_1\sleq (y-x)$. Thus $y-x=z_1+(y-x-z_1)$, which implies
$y-x-z_1=y-(x+z_1)\in A\D^*$; that is,
$(x+z_1)\vreach y$. Moreover, if  $x\neq y$, then
$\nnorm{y-(x+z_1)}<\nnorm{y-x}$.
\end{proof}

% Fix Indentation
\noindent
Recall that we used
$C\shift^A=f\shift(\nnorm{A},d)$ to bound the norms of segments in the virtual
executions~(\ref{eq:virtsegment}) of a $d$-dim net $A$. In principle,
by~\eqref{eq:Cshift}
 we claimed  
that $\nnorm{\minsleq{A\D^*}}\leq f\shift(\nnorm{A},d)$, where
$f\shift$ is a 1-exp function. The following lemma establishes this
result; it is slightly more general, since it shows that 
	\[\nnorm{\minsleq{A^*}}\leq (2+\nnorm{A})^{\text{poly}(d)}\] 
for every finite set $A\subseteq \setZ^d$ (where $A^*$ is not necessarily a group).

\begin{lem}[\textbf{For every finite set $\bm{A\subseteq \setZ^d}$,	
the norm $\bm{\nnorm{\minsleq{A^*}}}$ is 1-exp}]\label{lem:Lminbound}
\hfill\\
For every finite set $A\subseteq \setZ^d$, it holds that
	\[\norm{}{\minsleq{A^*}}\leq  (3+\nnorm{A}\cdot |A|)^d\leq
	(3+\nnorm{A}\cdot (1+2\cdot\nnorm{A})^d)^d.\]
\end{lem}
\begin{proof}
By decomposing the set $A^*$ according to the function $\sign$
 (recall that $x\sleq y$ requires $\sign(x)=\sign(y)$, so vectors in
 different sign classes are incomparable), we obtain:
		\[\minsleq{A^*}=\bigcup_{s\in \{-1,0,1\}^d}\minsleq{\{y\in A^*
	\mid \sign(y)=s\}}.\]
	Thus, it suffices to provide the required bound on
	$\nnorm{\minsleq{\{y\in A^* \mid \sign(y)=s\}}}$ 
	for an arbitrary vector  $s\in\{-1,0,1\}^d$, which is henceforth fixed.
Observe that by flipping the sign of the $i$th component of each
	vector in $A$ for all $i\in[1,d]$ such that $s(i)=-1$, we may
	assume without loss of generality that $s\in \{0,1\}^d$. We
	set  \[I=\{i\in[1,d]\mid s(i)=0\}.\]
	For $k=|A|$, we henceforth view $A$ as a $d\times k$ matrix
	(with $d$ rows and $k$ columns), where
the columns correspond to the elements of $A$.
	Consequently, we now aim to provide a bound on the
	norm $\nnorm{\min{}_{\leq}(Y)}$, where $\leq$
	denotes the
	standard component-wise order and the set $Y\subseteq \setN^d$
	is defined as follows:
	\[Y=\{Av \mid v\in\setN^k
	\land \bigwedge_{i\in I}(Av)(i)=0\land
	\bigwedge_{i\in[1,d]\smallsetminus I}(Av)(i)\geq 1\}.\]
We proceed similarly to the proof
	of~\cite[Corollary~1]{DBLP:conf/rta/Pottier91}, which deals with
systems of linear inequalities.
Let $B$ be the $d \times (k+d+1)$ matrix
\[B=\left(\begin{array}{c|c|c} A & -\mathrm{Id}_d & -s\end{array}\right),\]
that is, the $k$ columns of $A$, followed by the columns of the negated
$d\times d$ identity matrix, followed by the single column $-s$.
We set 
	\[M=\{x\in\setN^{k+d+1}\mid
	Bx=\mathbf{0}\}=\{(v,z,t)\in\setN^k\times\setN^d\times\setN\mid
	Av-z-ts=\mathbf{0}\},\]
where the three blocks of $B$ act on $v$, $z$, and $t$, respectively.
	We then observe that 
	\begin{equation}\label{eq:YviaM}
		Y=\{Av\mid \exists z\in\setN^d: (v,z,1)\in M\land
	\bigwedge_{i\in I}z(i)=0\}.
	\end{equation}
 Lemma~\ref{lem:smallsolution} implies that $M=X^*$, 
		where $X=\min{}_{\leq}(M\smallsetminus\{\mathbf{0}\})$ and,
		since $\rank{B}\leq d$ and $\norm{1}{B}\leq
		k\cdot\norm{}{A}+2$,
	\begin{equation}\label{eq:XboundforM}
		\norm{1}{X}\leq (3+k\cdot\norm{}{A})^d.
	\end{equation}
Finally, we fix an arbitrary  $y\in \min{}_{\leq}(Y)$ and bound its
	norm $\nnorm{y}$.
	By~\eqref{eq:YviaM}, we may fix $v$ and $z$ such that $y=Av$,
	$(v,z,1)\in M$,
and $z(i)=0$ for all $i\in I$.
	Since $M=X^*$, the tuple $(v,z,1)$ is a sum of elements of $X$;
	their last components are nonnegative and sum to $1$, so exactly
	one summand $(v',z',1)$ has last component $1$, and it satisfies
	$(v',z')\leq (v,z)$ and
	$\norm{1}{(v',z',1)}\leq (3+k\cdot\norm{}{A})^d$
	by~\eqref{eq:XboundforM}.
	As $z'\leq z$ and $z(i)=0$ for all $i\in I$, we have $z'(i)=0$ for
	all $i\in I$, and hence $Av'=s+z'$ belongs to $Y$
	by~\eqref{eq:YviaM}.
	Because 
	\[Av'=s+z'\leq s+z=Av=y\] and $y\in \min{}_{\leq}(Y)$, it
	follows that $z'=z$. 
	Consequently, discarding the last component of $(v',z',1)$,
	\[\nnorm{z}\leq\norm{1}{z}=\norm{1}{z'}\leq (3+k\cdot\norm{}{A})^d-1.\]
	Since $Av-z=y-z=s\in\{0,1\}^d$,
	we conclude that $\nnorm{y}\leq\nnorm{z}+1\leq
	(3+k\cdot\norm{}{A})^d$.
\end{proof}

\section{Bounds on Minimal Solutions of Linear Systems}\label{sec:smallsolutions}

In this section, we prove Theorem~\ref{thm:smallsolutions}, which we
repeat here for convenience.

\medskip
\noindent
	\textbf{Theorem~\ref{thm:smallsolutions}} (\textbf{Bounds on solutions of linear
	systems}).
\hfill\\
	\emph{
Every satisfiable $d$-dim linear system $S$ 
 has a solution $x\in\setZ^d$ such that
	\[\nnorm{x}\leq
		\mlcm{S} \cdot \left(1+ (d+1)\,!\cdot (d\cdot
		\nnorm{S}+1)^{d}\right).\]
}
\medskip

We first prove the theorem for the case where the linear system $S$
	does not contain any divisibility constraints. We use the
	following result, which is a weaker form of the theorem from~\cite{Gathen1978}.

	\begin{lemC}[\cite{Gathen1978} (\textbf{Bounds on solutions to
		inequalities 
		via subdeterminants})]\label{lem:Gathen}
Let $C$ and $c$ be $n\times d$ and $n\times 1$ integer matrices, respectively,
and let $r$ be the rank of $C$. Let $m$ be an
	upper bound on the absolute values of the $(r-1)\times (r-1)$
	and $r\times r$ subdeterminants of the $n\times (d+1)$
	augmented matrix
	$(C, c)$. If $Cx\geq c$ has an integer solution, then it has
	a solution $x\in \setZ^d$ such that $|x(i)|\leq (d+1)\cdot m$ for
	all $i\in [1,d]$.
\end{lemC}

\begin{lem}[\textbf{Systems without divisibility constraints}]
\label{lem:nodivsmallsolutions}
\hfill\\
Every satisfiable $d$-dim linear system $S$ without divisibility constraints has a solution $x \in \mathbb{Z}^d$ such that 
\[
    \nnorm{x} \leq (d+1)\,!\cdot(\nnorm{S}+1)^d.
\]
\end{lem}
\begin{proof}
Let $S$ be a satisfiable $d$-dim linear system, expressed as a
	propositional formula whose atoms are equality and inequality
	constraints (with no divisibility constraints).
Each equality constraint
	$\scalar{\alpha}{\bvar{x}}=b$ in $S$ can be replaced by the
	conjunction
	$\scalar{\alpha}{\bvar{x}}\geq b\,\wedge\,
	\scalar{-\alpha}{\bvar{x}}\geq -b$ without affecting
	$\sem{S}$ or $\nnorm{S}$. 

	We may thus assume,
	without loss of generality, that the atoms in $S$ consist only
	of inequality constraints.
By
expressing $S$ in disjunctive normal form, we obtain
	linear systems $S_1,S_2,\ldots,S_k$, each given as a
	conjunction of inequalities and negations of
	inequalities, where
	\[\sem{S}=\bigcup_{j=1}^k\sem{S_j}\text{, and }
	\nnorm{S_j}\leq \nnorm{S}\text{ for all }j\in[1,k].\]
Since $\sem{S}\neq\emptyset$, it follows that $\sem{S_j}\neq\emptyset$ for some $j\in[1,k]$.
It therefore suffices to establish the claim for the case where  
$S$ is a~conjunction of inequality constraints and their negations; we
	will assume this henceforth.

Let $S'$ be the linear system obtained from $S$ by replacing
	each subformula $\neg(\scalar{\alpha}{\bvar{x}}\geq b)$ with
	the
	inequality constraint $\scalar{-\alpha}{\bvar{x}}\geq -b+1$.
Then
	\begin{equation}\label{eq:semSprime}\sem{S'}=\sem{S}=\{x\in\setZ^d\mid
	Cx\geq c\},\end{equation}
where $C$ and $c$ are some $n\times d$ and $n\times 1$ integer
	matrices, respectively, with
	\[
		\max\{\nnorm{C},\nnorm{c}\}=\nnorm{S'}\leq\nnorm{S}+1.\]
        Let $r$ be the rank of $C$, and let $m$ be an upper bound on
	the absolute values of the $(r-1)\times (r-1)$ and $r\times r$
	subdeterminants of the $n\times (d+1)$ augmented matrix $(C, c)$.
Since  $r\leq d$, it follows that
	\[m\leq d\,!\cdot (\nnorm{S}+1)^d.\]
	Applying Lemma~\ref{lem:Gathen} to~\eqref{eq:semSprime}, we
conclude that $S'$ admits a solution $x$ such that 
	\[\nnorm{x}\leq (d+1)\cdot m\leq (d+1)\,!\cdot
	(\nnorm{S}+1)^d.\qedhere\]
\end{proof}

\noindent
\textbf{Proof of Theorem~\ref{thm:smallsolutions}}. 
Let $S$ be a satisfiable $d$-dim linear system, with a~variable vector
$\bvar{x}$ ranging over $\setZ^d$. We set
$\ell=\mlcm{S}$ and fix a solution $\sol$ of $S$.
Let 
\begin{center}
$b\in \{0,1,\dots,\ell{-}1\}^d$ be the vector of remainders of
	$\sol$ modulo $\ell$;
\end{center}
specifically, we define $b(i)=(\sol(i)\bmod\ell)$ for all $i\in[1,d]$. 

We aim to construct a $d$-dim linear system $S_b$, with a variable
vector
$\bvar{y}$ and without divisibility constraints, such that 
\begin{equation}\label{eq:semSb}
\sem{S_b}=\{y\in \setZ^d\mid b+\ell y\in\sem{S}\}.
\end{equation}
We construct $S_b$ from $S$ by replacing each constraint as follows:
  \begin{itemize}
  \item $\scalar{\alpha}{\bvar{x}}=c$ is replaced by 
	  $\scalar{\alpha}{\bvar{y}}=\frac{c-\scalar{\alpha}{b}}{\ell}$
		  if $\ell$ divides $c-\scalar{\alpha}{b}$, and by
		  \emph{false} otherwise;
  \item $\scalar{\alpha}{\bvar{x}}\geq c$ is replaced by
	  $\scalar{\alpha}{\bvar{y}}\geq  \left
		  \lceil{\frac{c-\scalar{\alpha}{b}}{\ell}}\right
		  \rceil$;
  \item $\scalar{\alpha}{\bvar{x}}\equiv c\,(\bmod\, m)$ is replaced
	  by \emph{true} if
		  $\scalar{\alpha}{b}\equiv c\,(\bmod\, m)$ and by
		  \emph{false}
		  otherwise.
  \end{itemize}

% Fix Indentation
\noindent
It is straightforward to verify that $\sem{S_b}$ satisfies the 
condition~\eqref{eq:semSb}, by noting that $y\in\sem{\text{new
constraint}}$ if and only if
$x\in\sem{\text{old constraint}}$ for $x=b+\ell y$
(using the fact that $\ell$ is a multiple of each $m$ appearing in the
divisibility constraints).
Moreover, $\sem{S_b}\neq\emptyset$ because the vector $y$ defined by 
$y(i)=\frac{\sol(i)-b(i)}{\ell}$ for each $i\in[1,d]$ is a solution.

Note that if $\scalar{\alpha}{\bvar{x}}\sim c$ is a constraint of
$S$, where $\sim$ is the equality or the inequality, then
the values $|\alpha(i)|$ for $i\in[1,d]$ and $|c|$ are bounded by
$\nnorm{S}$. It follows that
$|\scalar{\alpha}{b}|\leq d\cdot\nnorm{S}\cdot(\ell-1)$, and 
 $|\frac{c-\scalar{\alpha}{b}}{\ell}|\leq
d\cdot\nnorm{S}$. We have thus shown that 
\[\nnorm{S_b}\leq d\cdot\nnorm{S}.\] 
Since $S_b$ is satisfiable and contains no divisibility constraints, 
Lemma~\ref{lem:nodivsmallsolutions} implies that $S_b$ has a solution
$y\in\setZ^d$ such that 
\[\nnorm{y}\leq (d+1)\,!\cdot(\nnorm{S_b}+1)^d.\]
Hence $x=b+\ell y$
is a solution of $S$, satisfying
\[\nnorm{x}\leq (\ell-1) +\ell\cdot (d+1)\,!\cdot(\nnorm{S_b}+1)^d\leq
	\ell\cdot\left(1+ (d+1)\,!\cdot(d\cdot \nnorm{S}+1)^d\right).\tag*{\qed}\]

\section{Bounds on Sizes of Linear Systems for Groups}\label{sec:virtual}
In this section, we prove Theorem~\ref{thm:virtual2system}, which we
repeat here for convenience.

\medskip
\noindent
	\textbf{Theorem~\ref{thm:virtual2system}} (\textbf{Groups expressed by linear systems of bounded size}).
\hfill\\
	\emph{Let $L$ be the group spanned by a finite set $X\subseteq
	\setZ^d$. There exists a~$d$-dim linear system $S$ such that
	\begin{center}
	$\sem{S}=L$ and $\max\{\nnorm{S},\mlcm{S}\}\leq d\,!\cdot \norm{}{X}^d $.
	\end{center}
		}
\medskip

\noindent
Recall that the group $L$ spanned by a finite set $X\subseteq \setZ^d$ is 
the set of integer linear combinations of elements of $X$. The
elements of $X$ can be viewed as the columns of 
an integer $d\times k$ matrix $C$, where $k=|X|$; in this case,
$\nnorm{X}=\nnorm{C}$. 

To prove the theorem, we consider an integer $d\times k$ matrix $C$
and 
denote by $L_C$ the
 group (also called a \emph{lattice}) spanned by the columns
of $C$; that is,
\begin{equation*}
L_C=\{x\in\setZ^d\mid \exists y\in \setZ^k: Cy=x\}.
\end{equation*}
We aim to construct a linear system $S$ such that $\sem{S}=L_C$ and
\begin{equation}\label{eq:boundSLC}
	\max\{\nnorm{S},\mlcm{S}\}\leq d\,!\cdot \nnorm{C}^d.
\end{equation}	
\paragraph{Case $\bm{\text{rank}(C)=d}$.}
We first consider the case where $\rank{C} = d$; that is, we assume that the rows of $C$ are linearly independent. From~\cite[Chapter 4]{schrijver1986theory}, we recall that there exists a square $d \times d$ non-singular integer matrix $H$ such that $L_H = L_C$. We can take $H$ to be the first $d$ columns of the Hermite normal form of $C$ (the remaining $k-d$ columns being zero vectors). It follows that 
\[
    L_C = L_H= \{x\in\setZ^d\mid \exists y\in \setZ^d: Hy=x\}=\{x \in \setZ^d \mid H^{-1}x \in \setZ^d\};
\]
note that $H^{-1}x \in \setZ^d$ implies that $x = H(H^{-1}x) \in \setZ^d$.

Although the entries of $H^{-1}$ may be non-integers, the matrix
$\det(H) \cdot H^{-1}$ is an integer matrix, since $H^{-1} =
\frac{1}{\det(H)}\cdot \adj(H)$, where $\adj(H)$ is the adjugate of
$H$ (and thus an integer matrix). In particular, $H^{-1}(\det(H) \cdot
z)$ is an integer vector for any $z \in \setZ^d$. This implies that
for any $x \in \setZ^d$, 
\[
	\text{the vector } H^{-1}x \text{ is an integer vector if and
	only if } H^{-1}(x \bmod \ell) \text{ is an integer vector,}
\]
where $\ell = |\det(H)|$ and the modulo operation is applied component-wise.

Defining $B = \{b \in \{0, 1, \dots, \ell{-}1\}^d \mid H^{-1}b \in \mathbb{Z}^d\}$, we obtain $L_C = \sem{S}$, where 
\[
    S = \bigvee_{b \in B} \left( \bigwedge_{i=1}^d \bvar{x}_i \equiv
    b(i) \!\!\!\! \pmod \ell \right);
\]
here $\bvar{x} = (\bvar{x}_1,\bvar{x}_2, \dots, \bvar{x}_d)$ is the
variable vector of $S$ ranging over $\setZ^d$.

To ensure the bound~(\ref{eq:boundSLC}),
it suffices to show that $|\det(H)|\leq d\,!\cdot\nnorm{C}^d$.
Consider any $d\times d$ matrix $M$ composed of $d$
linearly independent columns of $C$; hence $L_M$ is a subgroup of
$L_C=L_H$.
From~\cite[Chapter 4]{schrijver1986theory}, it follows that $\det(H)$
divides $\det(M)$, which implies that $|\det(H)|\leq |\det(M)|$.
Since $|\det(M)|\leq d\,!\cdot\nnorm{M}^d$ and
$\nnorm{M}\leq \nnorm{C}$, we have completed the case for
$\rank{C}=d$.

\paragraph{Case $\bm{\text{rank}(C)=r<d}$.}
We now consider a $d \times k$ integer matrix $C$ with $\rank{C} = r < d$. Without loss of generality, we assume (possibly after reordering the rows of $C$) that $C$ is of the form
\[
	C = \begin{bmatrix} \bar{C} \\ C' \end{bmatrix},
\]
where $\bar{C}$ is an $r \times k$ matrix with $\rank{\bar{C}} = r$
(that is, $\bar{C}$ is a row basis of $C$), and each of the $d-r$ rows
of $C'$ is a rational linear combination of the $r$ rows of $\bar{C}$.
Hence, there exists a unique rational $(d-r) \times r$ matrix $T$ such
that $C' = T\bar{C}$.

We now examine the transformation of $\bar{C}$ into its Hermite normal
form in more detail. This is achieved by a sequence of elementary
column operations on $\bar{C}$, comprising: 
\begin{enumerate*}[(i)]
    \item multiplying a column by $-1$, 
    \item swapping two columns, and 
    \item adding an integer multiple of one column to another.
\end{enumerate*}
These operations correspond to multiplying $\bar{C}$ from the right by a unimodular matrix $U$ (an integer $k \times k$ matrix with $|\det(U)| = 1$). Multiplying $C$ by $U$, we obtain
\[
	CU = \begin{bmatrix} \bar{C} \\ C' \end{bmatrix} U = 
		\begin{bmatrix} \bar{H} & 0 \\ H' & 0 \end{bmatrix} = [H \mid 0],
\]
where $\bar{H}$ is a non-singular $r \times r$ integer matrix such
that $L_{\bar{H}}
= L_{\bar{C}}$, and $H' = T\bar{H}$. The latter holds because $C' = T\bar{C}$ and elementary column operations preserve linear relations between rows.

Since elementary column operations preserve the spanned group, we have
\[
    L_{C} = L_{H} = \{x \in \setZ^d \mid \exists y \in
    \setZ^r : \bar{H}y = x\restr{[1,r]} \land H'y = x\restr{[r+1,d]}\}.
\]
By rewriting the conjunction as $y = \bar{H}^{-1}x\restr{[1,r]}\land
H'\bar{H}^{-1}x\restr{[1,r]} = x\restr{[r+1,d]}$, and noting that
$H'\bar{H}^{-1} = T$, we obtain
\begin{equation}\label{eq:LbarC}
	L_{C}=\{x\in\setZ^d\mid x\restr{[1,r]}\in L_{\bar{C}}\land
x\restr{[r+1,d]}=T x\restr{[1,r]}\}.
\end{equation}

% Fix Indentation
\noindent
We note that the rational $(d{-}r)\times r$ matrix $T$ is determined
by any submatrix $M$ consisting of 
$r$ linearly independent columns of $C$.
Letting
$\bar{M}=M\restr{[1,r]}$
and $M'=M\restr{[r+1,d]}$, we have $M'=T\bar{M}$ (as $C'=T\bar{C}$), which
implies that the $r$ rows of $\bar{M}$ are linearly independent;
hence, $\bar{M}$
is non-singular.
 Using Cramer's rule,
we deduce that all entries of $T$ are of the form
$\frac{\det(M'')}{\det(\bar{M})}$, with the same denominator
$\det(\bar{M})$ and
varying integer numerators. The absolute values of the denominator and all numerators are
bounded by $r\,!\cdot \nnorm{M}^r$, where $r<d$ and $\nnorm{M}\leq \nnorm{C}$.

The condition $x\restr{[r+1,d]}=T x\restr{[1,r]}$ can be expressed
as a conjunction of homogeneous linear constraints for $i\in[r{+}1,d]$ of the form 
\[
	\det(\bar{M}) \cdot \bvar{x}_i - \sum_{j=1}^r \det(M''_{ij}) \cdot
    \bvar{x}_j = 0,
\]
where the absolute values of all coefficients are bounded by
$r\,! \cdot \nnorm{C}^r$. 
Since $r < d$, the group $L_{C}$ in~\eqref{eq:LbarC} can indeed
be expressed by a linear system $S$ satisfying~\eqref{eq:boundSLC}.
Here, we apply the results from the case $\rank{C} = d$ to the
submatrix $\bar{C}$ with $\rank{\bar{C}} = r$ (noting that
$\nnorm{\bar{C}} \leq \nnorm{C}$). This completes the proof of Theorem~\ref{thm:virtual2system}.

\section{\EXPSPACE-hardness 
(a counterpart of Theorem~\ref{th:upper})}\label{sec:hardness}
In this section, we prove the following theorem for a subclass
of $1$-conservative nets.
A~net $A$ is a
\emph{population protocol net} (a \emph{PP-net} for short) if each action
$(\myvec{a_-},\myvec{a_+})$ in $A$ satisfies either
$\normone{\myvec{a_-}} = \normone{\myvec{a_+}} = 1$ or
$\normone{\myvec{a_-}} = \normone{\myvec{a_+}} = 2$.
A $d$-dim \emph{net} $A$ is \emph{ordinary} if $A\subseteq \{0,1\}^d\times\{0,1\}^d$. 

\begin{thm}[\textbf{EXPSPACE-hardness for a simple subclass of
	$\bm{1}$-conservative nets}]\label{thm:lowerbound}
	The structural liveness problem is
    \EXPSPACE-hard for ordinary reversible PP-nets.
\end{thm}

\begin{rem}
It is evident that \ppnets are indeed a subclass of
$1$-conservative nets; they model the \emph{population
protocols} introduced
in~\cite{DBLP:journals/dc/AngluinADFP06}. In these protocols,
an arbitrary number of pairwise indistinguishable
finite-state agents interact in pairs. 
Here, a marking is a function that assigns to each state $q$ the
number of agents currently in that state.
When two
agents in states $q_1$ and $q_2$ interact, they
change their states to $q_3$ and $q_4$, 
according to a~transition function; the states $q_1,q_2,q_3,q_4$ are
not necessarily pairwise distinct. In \cite{DBLP:journals/fuin/JancarV22},
 the structural liveness was studied  for \emph{immediate observation
nets}. This is a subclass of
PP-nets modelling \emph{immediate observation protocols} introduced in
\cite{DBLP:journals/dc/AngluinAER07}, 
and for their generalised variants; these classes of nets
were introduced and studied
in~\cite{DBLP:conf/apn/EsparzaRW19,DBLP:conf/rp/RaskinW20,DBLP:conf/concur/RaskinWE20,DBLP:phd/dnb/Weil-Kennedy23}.
The structural liveness problem for these classes was shown to be 
\PSPACE-complete \cite{DBLP:journals/fuin/JancarV22};
Theorem~\ref{thm:lowerbound} thus demonstrates that the problem for the class
of PP-nets is substantially harder, even when restricted to ordinary
and (structurally) reversible PP-nets.
\end{rem}

\paragraph{A useful \EXPSPACE-hard problem \Cover.}
Our proof of Theorem~\ref{thm:lowerbound} draws on ideas from
\cite{DBLP:journals/acta/JancarP19}. First, we recall the
well-known result presented in \cite{mayr1982complexity}
showing that the uniform word problem for commutative
semigroups 
is \EXPSPACE-complete. An instance of this
problem consists of a~finite alphabet $\Sigma$, a~finite set
of equations $u\equiv v$ for $u,v\in\Sigma^*$ (implicitly
comprising $ab\equiv ba$ for all $a,b\in\Sigma$), and
$u_0,v_0\in\Sigma^*$; the question is whether $u_0\equiv v_0$.
The crux of the high complexity lies in the fact that the
commutative semigroup defined by $a\equiv b^{2^{2^n}}$,
which can be written in space $O(2^n)$ when using binary
notation for exponents, can be embedded in a commutative
semigroup of size $O(n)$, even when using unary notation for
exponents. In fact, even the (weaker) \emph{coverability}
version in which $u_0,v_0\in\Sigma$ and we ask whether
$u_0\equiv v_0w$ for some $w\in\Sigma^*$ is
\EXPSPACE-complete. 
Moreover, we can assume that 
in each given equation $u\equiv v$ neither $u$ nor $v$
contains more than one occurrence of any $a \in \Sigma$.
(The crux is that $aau'\equiv v$ can be replaced with  
$a_1a_2u' \equiv v$, and $a_1 \equiv
a$, $a_2 \equiv a$.)

It is straightforward to formulate the aforementioned \EXPSPACE-complete
coverability problem in terms of Petri nets.
We say that a \emph{net} $A$ is
\emph{strongly reversible} if $A=A^{-1}$
(hence if $(a_-,a_+) \in A$ implies $(a_+,a_-) \in A$).
Given $d\in\setN$, for each $i\in[1,d]$, we denote by $\mathbf{e}_i$ the
unit vector from $\{0,1\}^d$ satisfying 
$\mathbf{e}_i(i)=1$ and $\mathbf{e}_{i}(j)=0$ for all
$j\neq i$.
Recall that, in our context, the term \emph{places} is a synonym for \emph{vector components}.

\begin{prop}[\textbf{Problem \Cover for ordinary strongly reversible
	nets}]\label{prop:problemcover}
\hfill\\	
	The following problem \Cover is \EXPSPACE-hard:
\\
Instance: An ordinary strongly reversible $d$-dim net $A$, and two places
	$\pinit,\pcov\in[1,d]$.
\\
	Question: Does there exist $y\in\setN^d$ such that $\mathbf{e}_\pinit\reach
	(\mathbf{e}_\pcov+y)$~?
\end{prop}
\begin{proof}
An instance of the aforementioned coverability problem for
	commutative semigroups, with equations $u\equiv v$ over
	$\Sigma$ and distinguished elements
	$u_0,v_0\in\Sigma$,
	can be viewed as an instance of \Cover: We consider a
	bijection $\textsc{b}$ from $\Sigma$
onto $\{1,2,\dots,|\Sigma|\}$, and the
$|\Sigma|$-dim net where each equation $u\equiv v$ gives rise to two
	actions $(\chi_u,\chi_v)$ and $(\chi_v,\chi_u)$; for the
	characteristic vector
	$\chi_w\in \{0,1\}^{|\Sigma|}$ we have $\chi_w(i)=1$ iff
	$\textsc{b}^{-1}(i)$ occurs in $w$. Finally, we set
	$\pinit=\textsc{b}(u_0)$ and $\pcov=\textsc{b}(v_0)$.
\end{proof}	

\paragraph{Reducing \Cover to \SCover for $\bm{1}$-conservative nets.}
We prove Theorem~\ref{thm:lowerbound} by showing a polynomial
reduction from the problem \textsc{Cover} to
the structural liveness problem for ordinary reversible \ppnets. For convenience, we use two intermediate problems, called \SCover
and \PPSCover. The letter \textsc{S} refers to \emph{store places}, 
which enable a~simple transformation of \Cover to a~corresponding
problem for $1$-conservative nets. 

\begin{quote}
    \textsc{Problem} \SCover
    \smallskip
\begin{description}
        \item[\emph{Instance}]
An ordinary strongly reversible $1$-conservative $d$-dim net $A$, and
three places $\pinit,\pcov,\pstore\in[1,d]$.
 \item[\emph{Question}]
	Do there exist $k\in\setN$ and $y\in\setN^d$ such that 
		\[(\mathbf{e}_\pinit+k\cdot \mathbf{e}_\pstore)\reach
		(\mathbf{e}_\pcov+y)~?\]
 \end{description}
 \end{quote}
The problem \PPSCover is the subproblem of \SCover in which the given
net $A$ is additionally required to be a PP-net.

\begin{prop}[\textbf{EXPSPACE-hard \Cover reduces to \SCover}]
\hfill\\
	The \Cover problem (see Proposition~\ref{prop:problemcover}) is polynomially reducible to \SCover.
\end{prop}
\begin{proof}
Consider an instance of \Cover, that is,
an ordinary strongly reversible $d$-dim net $A$ and
$\pinit,\pcov\in[1,d]$.
We extend $A$ to a $2d$-dim net $A'$, where the components $d+1, d+2, \dots, 2d$ are referred to as the \emph{store places}. These places allow for the removal of any potential imbalance that occurs when an action $a$ produces more (or fewer) tokens than it consumes. Specifically, $A' \subseteq \{0,1\}^{2d} \times \{0,1\}^{2d}$ is constructed as follows:
	\begin{itemize}
	\item		
Each action $a=(a_-,a_+)\in\{0,1\}^d\times\{0,1\}^d$ in $A$ gives rise 
 to 
\\
the action $a'=(a'_-,a'_+)\in\{0,1\}^{2d}\times\{0,1\}^{2d}$ in $A'$,
where
		\begin{itemize}
			\item
				$a'_-(i)=a_-(i)$ and $a'_+(i)=a_+(i)$
				for all $i\in[1,d]$;
\item
if $k=\normone{a_-} - \normone{a_+}\geq 0$, then
\begin{itemize}
			 \item
				 $a'_-(i)=0$ for all
				 $i\in[d+1,2d]$,
	\item				$a'_+(i)=1$ for all
				 $i\in[d+1,d+k]$,
 \item
		$a'_+(i)=0$ for all
				 $i\in[d+k+1,2d]$;
\end{itemize}
\item
if $k=\normone{a_+} - \normone{a_-}>0$, then
				\begin{itemize}
\item
			$a'_-(i)=1$ for all $i\in[d+1,d+k]$, 
		\item		 $a'_-(i)=0$ for all
				 $i\in[d+k+1,2d]$,
\item
	$a'_+(i)=0$ for all $i\in[d+1,2d]$.
				\end{itemize}
				\end{itemize}
		(Hence, $\normone{a'_-}=\normone{a'_+}$.)
	
	\item
Moreover, $A'$ contains actions $(\mathbf{e}_{d+1},\mathbf{e}_{d+d})$ and 
		$(\mathbf{e}_{d+d},\mathbf{e}_{d+1})$, and   
		for each $i\in[1,d-1]$ it contains actions
		$(\mathbf{e}_{d+i},\mathbf{e}_{d+i+1})$ and 
		$(\mathbf{e}_{d+i+1},\mathbf{e}_{d+i})$. 
		(Hence, any token in a store place can move freely to any other store place.)
\end{itemize}
Since $A$ is ordinary and strongly reversible, it is
 clear that $A'$ is also ordinary and strongly reversible; moreover,
	$A'$ is $1$-conservative. 
	It is easily verified that $(A,\pinit,\pcov)$ is a positive instance of \Cover if and
	only if  $(A',\pinit,\pcov,d+1)$ is a positive instance of
	\SCover. 
Indeed, if $\pcov$ is coverable from $\pinit$ by an execution of $A$, then
it suffices to provide an adequate number of tokens in the store place
	$d+1$ to allow the execution to be mimicked by $A'$.
	Conversely, the projection of any execution of $A'$ onto the
	first $d$ components constitutes a valid execution of $A$.
\end{proof}

\paragraph{Reducing \SCover to \PPSCover.}
We now aim to reduce the \SCover problem to its subproblem \PPSCover,
which requires the input net to be a PP-net.  
This reduction is based on the following construction, which is an instance of
the standard approach of simulating an action by a sequence of simpler
actions, controlled by additional \emph{control places}. 

For convenience, an action $a=(a_-,a_+)$ of an ordinary
$1$-conservative $d$-dim net $A$ with
$\normone{a_-}=\normone{a_+}=k$ is also written as 
\begin{equation*}
\pptrans{a}{i_1,i_2,\dots,i_k}{i'_1,i'_2,\dots,i'_k},
	\text{ or simply }(i_1,i_2,\dots,i_k)\rightarrow(i'_1,i'_2,\dots,i'_k),
\end{equation*}
where $\{i_j\mid j\in[1,k]\}=\{i\mid a_-(i)=1\}$ and
 $\{i'_j\mid j\in[1,k]\}=\{i'\mid a_+(i')=1\}$.
When $k=1$, we write $\pptransone{a}{i_1}{i'_1}$.
Furthermore, recall that for a $d$-dim net $A$, the set
$P=\{1,2,\dots,d\}$ is viewed as the set of \emph{places} of $A$.

\begin{prop}[\textbf{$\bm{1}$-conservative nets simulated by PP-nets}]\label{prop:constoPP}
\hfill\\	
There exists a polynomial-time construction that transforms
any ordinary strongly reversible $1$-conservative net $A$ with a
set of places $P$ 
into an ordinary strongly reversible PP-net $A'$ with a
	set of places $P'\supseteq P$ and a distinguished place
	$\prun\in P'\smallsetminus P$ satisfying the following two
	conditions:
\begin{enumerate}
	\item		
For all markings $x,y$ of $A$:
		$x \step{*} y$ for $A$ if and only if $x' \step{*} y'$ for
		$A'$, where $x'$ and $y'$ are obtained from $x$ and $y$,
		respectively, by
		adding one token to the place $\prun$ (and no tokens to the
		places in $P'\smallsetminus (P\cup\{\prun\})$).
    \item
For each marking $x$ of $A'$:
		if $x\restr{P'\smallsetminus P}=\mathbf{0}$, then no
		action is enabled at $x$; if 
		$\normone{x\restr{P'\smallsetminus P}}=1$, then  
		$\normone{y\restr{P'\smallsetminus P}}=1$ for all
		$y\in\Reach(x)$.
\end{enumerate}
\end{prop}
\begin{figure}
\begin{center}
\begin{tikzpicture}[node distance=1.5cm,>=stealth',bend angle=45,auto]
    \tikzstyle{place}=[circle,thick,draw=black!75,fill=blue!20,minimum size=4mm]
    \tikzstyle{transition}=[rectangle,thick,draw=black!75,
    fill=black!20,minimum size=3mm]
    \tikzstyle{dots}=[circle,thick,draw=none,minimum size=6mm]

    \tikzstyle{every label}=[black]

    \begin{scope}
        \node [place] (pi1) [label=left:$i_1$] {};
        \node [place] (pi2) [label=left:$i_2$,right of=pi1] {};
        \node [place] (pi3) [label=left:$i_3$,right of=pi2] {};

        \node [transition] (t) [below of=pi2] {$a$}
        edge [pre] (pi1)
        edge [pre] (pi2)
        edge [pre] (pi3);

        \node [place] (po2) [label=left:$i'_2$,below of=t] {}
        edge [pre] (t);
        \node [place] (po1) [label=left:$i'_1$,left of=po2] {}
        edge [pre] (t);
        \node [place] (po3) [label=left:$i'_3$,right of=po2] {}
        edge [pre] (t);
    \end{scope}
    \begin{scope}[xshift = 5cm,yshift = -1.5cm]
        \node [place] (prun) [label=left:$\prun$] {};
        \node [transition] (t1) [above right of=prun] {$a_1$}
        edge [pre] (prun);
        \node [transition] (t1r) [below right of=prun] {$a_1^R$}
        edge [post] (prun);
        \node [place] (p1) [label=below:$p^a_1$,below right of=t1] {}
        edge [pre] (t1)
        edge [post] (t1r);
        \node [place] (po1) [label=left:$i'_1$,below right of=t1r] {}
        edge [pre] (t1)
        edge [post] (t1r);
        \node [place] (pi1) [label=left:$i_1$,above right of=t1] {}
        edge [post] (t1)
        edge [pre] (t1r);

        \node [transition] (t2) [above right of=p1] {$a_2$}
        edge [pre] (p1);
        \node [transition] (t2r) [below right of=p1] {$a_2^R$}
        edge [post] (p1);
        \node [place] (p2) [label=below:$p^a_2$,below right of=t2] {}
        edge [pre] (t2)
        edge [post] (t2r);
        \node [place] (po2) [label=left:$i'_2$,below right of=t2r] {}
        edge [pre] (t2)
        edge [post] (t2r);
        \node [place] (pi2) [label=left:$i_2$,above right of=t2] {}
        edge [post] (t2)
        edge [pre] (t2r);

        \node [transition] (t3) [above right of=p2] {$a_3$}
        edge [pre] (p2);
        \node [transition] (t3r) [below right of=p2] {$a_3^R$}
        edge [post] (p2);
        \node [place] (po3) [label=left:$i'_3$,below right of=t3r] {}
        edge [pre] (t3)
        edge [post] (t3r);
        \node [place] (pi3) [label=left:$i_3$,above right of=t3] {}
        edge [post] (t3)
        edge [pre] (t3r);

        \draw[shorten >=0.3mm,->] (t3) to[in=0,out=90] ($(3cm,3cm)$)
         to[in=90,out=180] (prun);
        \draw[shorten >=0.3mm,->] (prun) to[in=180,out=270] ($(3cm,-3cm)$)
         to[in=270,out=0] (t3r);
    \end{scope}
\end{tikzpicture}
\caption{Transformation of the pair
$(\myvec{a_-},\myvec{a_+})$, $(\myvec{a_+},\myvec{a_-})$  where $\normone{\myvec{a_-}}
    = \normone{\myvec{a_+}} = 3$.}
\label{fig:contopp1}
\end{center}
\end{figure}

\begin{proof}
Let $A$ be an ordinary strongly reversible
    $1$-conservative net with a set of places $P$.
To extend $A$ to the required PP-net $A'$,
	we introduce a new control place $\prun$, 
together with further auxiliary places as specified by the following construction of the actions of $A'$.

For each pair of mutually reversed actions
	$a=(\myvec{a_-},\myvec{a_+})$ and
	$a^R=(\myvec{a_+},\myvec{a_-})$  in $A$, written as 
	\begin{center}
$\pptrans{a}{i_1,i_2,\dots,i_k}{i'_1,i'_2,\dots,i'_k}$ and
$\pptrans{a^R}{i'_1,i'_2,\dots,i'_k}{i_1,i_2,\dots,i_k}$,
	\end{center}
		we proceed as follows (see
	Figure~\ref{fig:contopp1}):
	\begin{enumerate}[(a)]	
        \item   If $k=1$, and we have 
		$\pptransone{a}{i_1}{i'_1}$ and 
		    $\pptransone{a^R}{i'_1}{i_1}$, then $A'$ contains
		    the (mutually reversed) actions 
		    \begin{center}
		    $\pptrans{a_1}{i_1,\prun}{i'_1,\prun}$ and
        $\pptrans{a_1^R}{i'_1,\prun}{i_1,\prun}$.
		    \end{center}			    
	    \item
		If $k > 1$ then
$A'$ contains fresh auxiliary places $p^a_1,p^a_2,\dots,p^a_{k-1}$, and 
 the following actions:
			\begin{enumerate}[(i)]
            \item 
            $\pptrans{a_1}{i_1,\prun}{i'_1,p^a_1}$ and 
            $\pptrans{a_1^R}{i'_1,p^a_1}{i_1,\prun}$;
            \item
            $\pptrans{a_j}{i_j,p^a_{j-1}}{i'_j,p^a_j}$ and
             $\pptrans{a_j^R}{i'_j,p^a_j}{i_j,p^a_{j-1}}$,
			for all $\ininter{j}{2}{k{-}1}$, 

            \item 
            $\pptrans{a_k}{i_k,p^a_{k-1}}{i'_k,\prun}$ and
            $\pptrans{a_k^R}{i'_k,\prun}{i_k,p^a_{k-1}}$.
        \end{enumerate}
    \end{enumerate}
The net $A'$ is clearly an ordinary strongly reversible PP-net; 
	the definition of its actions ensures that condition $(2)$ holds.
A step $x\step{a}y$ of $A$ is simulated
by the execution $x'\step{a_1\cdots a_k}y'$ of $A'$, 
	while $y\step{a^R}x$
	is simulated by $y'\step{a^R_k\cdots a^R_1}x'$, 
	where $x'$ and $y'$ are obtained from $x$ and $y$ by placing a token
	in $\prun$. Condition $(1)$ thus holds as well.
\end{proof}

\begin{prop}[\textbf{\SCover reduces to \PPSCover}]
\hfill\\
	The problem \SCover is polynomially reducible to \PPSCover.
\end{prop}
\begin{proof}
	Let $(A,\pinit, \pcov, \pstore)$ be an instance of \SCover,
where $A$ is an ordinary strongly reversible $1$-conservative net with
	a set of places $P$. 
Let $A'$ be the
	PP-net obtained from $A$ via Proposition~\ref{prop:constoPP},
with a set of places $P'\supseteq P$, where $P'\smallsetminus P$
	contains
	 the distinguished place $\prun$.
We construct $A''$ from $A'$ by adding two new places $\pinit'$ and
	$\pcov'$;
each action $a=(a_-,a_+)\in A'$ is extended to an
	action in $A''$ by setting $a_-(i)=a_+(i)=0$ for
	$i\in\{\pinit',\pcov'\}$.

Moreover, let $A''$ also contain
	the following 
	actions, each together with its reverse: 
	\begin{center}
	$(\pinit',\pstore)\rightarrow (\pinit,\prun)$, 
$(\pcov,\prun)\rightarrow(\pcov',\pstore)$,

	$(\pinit,\prun)\rightarrow(\pinit',\pstore)$,
	$(\pcov',\pstore)\rightarrow(\pcov,\prun)$. 
	\end{center}
Thus, $A''$ is an ordinary strongly reversible PP-net. We conclude the proof by verifying that 
	$(A,\pinit, \pcov, \pstore)$
	is a positive instance of $\SCover$ if{f}
	$(A'',\pinit', \pcov', \pstore)$ is a positive instance of \PPSCover.

	$(\Rightarrow)$:
Let Exec be an execution of $A$ demonstrating
$(\mathbf{e}_\pinit+k\cdot \mathbf{e}_\pstore)\reach
		(\mathbf{e}_\pcov+y)$.
By Proposition~\ref{prop:constoPP}(1), there exists a corresponding
execution  Exec$'$ of the PP-net $A'$ where both the initial marking
	$\mathbf{e}_{\pinit} + k \cdot
	\mathbf{e}_{\pstore}$ and the target marking
	$\mathbf{e}_{\pcov} + y$ are augmented by a token in
	$\prun$.
	Consequently, if $A''$ starts at $\mathbf{e}_{\pinit'}+(k{+}1)\cdot
	\mathbf{e}_\pstore$, it can perform the action
	$(\pinit',\pstore)\rightarrow (\pinit,\prun)$, simulate the
	execution Exec$'$ of $A'$, and conclude by performing
	$(\pcov,\prun)\rightarrow(\pcov',\pstore)$.

	$(\Leftarrow)$:
Let Exec be an execution of $A''$ demonstrating
	$(\mathbf{e}_{\pinit'}+k\cdot \mathbf{e}_\pstore)\reach
		(\mathbf{e}_{\pcov'}+y)$.
		By Proposition~\ref{prop:constoPP}(2), we deduce that Exec maintains
		exactly one token within the set of places
 $\{\pinit',\pcov'\}\cup (P'\smallsetminus P)$. Furthermore, whenever 
		the token resides in $\pinit'$, the only enabled action
		is $(\pinit',\pstore)\rightarrow (\pinit,\prun)$.
		Consequently, Exec can be (re)constructed so that it starts with
		$(\pinit',\pstore)\rightarrow (\pinit,\prun)$,
		simulates
		an execution Exec$'$ of $A'$, and concludes with
		$(\pcov,\prun)\rightarrow(\pcov',\pstore)$.
By Proposition~\ref{prop:constoPP}(1), the existence of Exec$'$ implies 
that there is an execution of $A$ starting from
$(\mathbf{e}_{\pinit}+(k{-}1)\cdot \mathbf{e}_\pstore)$ that reaches a
marking with a token in $\pcov$.
\end{proof}

% Fix Indentation
\noindent
The proof of Theorem~\ref{thm:lowerbound} is thus completed by the following lemma.
Note that this reduction does not preserve \emph{strong}
reversibility.

\begin{lem}[\textbf{\PPSCover reduces to structural liveness}]\label{lem:PPSCtoSLP}
\hfill\\
	The \PPSCover problem is polynomially reducible to the structural
	liveness problem for ordinary reversible PP-nets.
\end{lem}
\begin{figure}[t]
\begin{center}
\begin{tikzpicture}[node distance=1.7cm,>=stealth',bend angle=10,auto]
    \tikzstyle{place}=[circle,thick,draw=black!75,fill=blue!20,minimum size=4mm]
    \tikzstyle{transition}=[rectangle,thick,draw=black!75,
    fill=black!20,minimum size=3mm]
    \tikzstyle{dots}=[circle,thick,draw=none,minimum size=4mm]

    \tikzstyle{every label}=[black]

    \begin{scope}[xshift=5cm,yshift=3cm]
        \node [place] (pI) [label=right:$\pinit$] {};
        \node [place] (pC) [label=right:$\pcov$,below of=pI] {};
    \end{scope}
    \begin{scope}[yshift=3cm]
        \node [place] (Z2) [label=left:$\pdec$] {};
        \node [place] (Z2tmp) [label=left:$\pdec'$,left of=Z2] {};
        \node [transition] (tZ1) [below of=Z2] {$a_1$}
        edge [post] (Z2);
        \node [place] (Z1) [label=left:$\pinc$,below of=tZ1] {}
        edge [post] (tZ1);
        \node [transition] (tZ) [above of=Z2] {$a_3$}
        edge [pre] (Z2tmp);
        \draw[shorten >=0.3mm,shorten <=0.3mm,<->] (tZ) to (Z2);
        \node [transition] (tZ2) [left=0.5cm of tZ] {$a_2$}
        edge [pre] (Z2)
        edge [post] (Z2tmp);
        \node [transition] (tZ3) [left=0.5cm of tZ2] {$a^R_2$}
        edge [pre] (Z2tmp)
        edge [post] (Z2);
    \end{scope}
    \begin{scope}[xshift=3cm,yshift=2cm]
        \node [place] (S1) [label=left:$\pstore$] {}
        edge [pre] (tZ);
    \end{scope}
    \begin{scope}[xshift=2.5cm,yshift=0.5cm]
        \node [transition] (tC) [] {$a_\pcov$}
        edge [pre] (pC)
        edge [pre] (S1)
        edge [post] (Z1)
        edge [post] (Z2);
    \end{scope}
    \begin{scope}[xshift=4.5cm,yshift=-0.25cm]
        \node [transition] (tIC) [] {$a_\lara{\pinc,\pcov}$}
        edge [pre] (S1)
        edge [post] (pC);
        \draw[shorten >=0.3mm,shorten <=0.3mm,<->] (tIC) to (Z1);
    \end{scope}
    \begin{scope}[xshift=1.5cm,yshift=5cm]
        \node [transition] (tI) [] {$a_\pinit$}
        edge [pre] (Z2)
        edge [post] (pI);
        \node [transition] (tRI) [right of=tI] {$a_\pinit^R$}
        edge [pre] (pI)
        edge [post] (Z2);
    \end{scope}
\end{tikzpicture}
\caption{Construction in the proof of
Lemma~\ref{lem:PPSCtoSLP}.}\label{fig:covtoslp}
\end{center}
\end{figure}

\begin{proof}
	Let $(A,\pinit, \pcov, \pstore)$ be an instance of \PPSCover. Thus, $A$
is an ordinary strongly reversible $d$-dim PP-net, with the set $P$ of
	$d$ places, three of which are 
 $\pinit, \pcov, \pstore$.
	We extend $A$ to an ordinary, reversible (though not
	strongly reversible) PP-net $A'$ such that 
	$(A,\pinit, \pcov, \pstore)$ is a~positive instance of \PPSCover if and
only if $A'$ is structurally live.

The set $P'$ of places of $A'$ is constructed from $P$ by adding 
	fresh places  $\pinc,\pdec,\pdec'$; hence $A'$ is a
	$(d{+}3)$-dim net,  
	and each $a=(a_-,a_+)\in A$ is extended to become an
	action in $A'$, by setting $a_-(i)=a_+(i)=0$ for
	all $i\in\{\pinc,\pdec,\pdec'\}$.
The net $A'$ further contains
the following actions,
partially depicted in Figure~\ref{fig:covtoslp}:
    \begin{enumerate}
        \item 
        $\pptrans{a_\pcov}{\pcov,\pstore}{\pinc,\pdec}$;
\item for each $p \in P \smallsetminus \{\pstore\}$,	    
	\begin{center}
	$\pptrans{a_\lara{\pinc,p}}{\pinc,\pstore}{\pinc,p}$
    and
	$\pptrans{a_\lara{\pdec,p}}{\pdec,p}{\pdec,\pstore}$;
	\end{center}		    
\item        $\pptransone{a_1}{\pinc}{\pdec}$,
		$\pptransone{a_2}{\pdec}{\pdec'}$;
        $\pptransone{a^R_2}{\pdec'}{\pdec}$,
\item
	$\pptrans{a_3}{\pdec,\pdec'}{\pdec,\pstore}$;
        \item 
        $\pptransone{a_\pinit}{\pdec}{\pinit}$
        and
        $\pptransone{a_\pinit^R}{\pinit}{\pdec}$.
    \end{enumerate}
Since $A$ is an ordinary strongly reversible PP-net,
it is evident that $A'$ is an ordinary PP-net. We now show that $A'$
	is also (structurally) reversible:
Since $a^R_\pinit$ is the reverse of $a_\pinit$, 
	$a^R_2$ is the reverse of $a_2$, and
	$\Delta(a_\lara{\pinc,p})=-\Delta(a_\lara{\pdec,p})$ (for each $p\in
	P\smallsetminus\{\pstore\}$),
it suffices to show a sequence $\sigma$ of actions of $A'$ such that 
	$\Delta(a_\pcov\,a_1\,a_3\,\sigma)=\mathbf{0}$, i.e.,
$\Delta(\sigma)(\pdec)=-2$, $\Delta(\sigma)(\pdec')=1$,
	$\Delta(\sigma)(\pcov)=1$, and 
	$\Delta(\sigma)(p)=0$
	for all other places $p$.
We can take 
	$\sigma=a_2\, a_2\, a_3\, a_\lara{\pinc,\pcov}$.

We call a \emph{marking} $x$ of $A'$ \emph{initial} if
$x(\pinit)=1$ and $x(p)=0$ for all places $p$ outside the set
$\{\pinit,\pstore\}$. We further identify the following ``initial'' \emph{property} \textsc{IP}: 
for every marking
	$x$ of $A'$ with $x(p)\geq 1$ for some
	$p\in P'\smallsetminus P=\{\pinc,\pdec,\pdec'\}$ there is  an initial marking $x'$
	such that $x\reach x'$ (which entails
	$\normone{x'}=\normone{x}$ since $A'$ is conservative, which
	in turn makes this initial marking unique).
	 Indeed, it suffices to use the
	actions $a_1,a_2,a^R_2,a_3$, $a_\lara{\pdec,p}$ (for $p\in
	P\smallsetminus\{\pstore\}$), and $a_\pinit$ in an appropriate sequence.

We now verify that  
 $(A,\pinit,\pcov,\pstore)$ is a positive instance of \PPSCover if and
 only if $A'$ is structurally live.

$(\Leftarrow)$:	
		Let $x_0$ be a live marking of
		$A'$. Since $a_\pcov$ is not dead at $x_0$, by \textsc{IP}
we deduce that	there is an initial marking $x_1$ such that $x_0\reach
		x_1$; $x_1$ is necessarily live. 
		Let $x_1\step{\sigma}x_2$ be a shortest execution from
		$x_1$ enabling $a_\pcov$; hence $x_2(\pcov)\geq 1$.
Since shortest, $\sigma$ cannot contain any action from the set 
		$\{a_\pcov,a_1\}\cup \{a_\lara{\pinc,p}\mid p\in
		P\smallsetminus\{\pstore\}\}$.
Let $\ell$ be the number of occurrences of actions
		from the set
		$\{a^R_\pinit,a_2,a^R_2,a_3,a_\pinit\}\cup 
		\{a_\lara{\pdec,p}\mid p\in
		P\smallsetminus\{\pstore\}\}$ in $\sigma$.
Let $\sigma'$ arise from $\sigma$ by omitting all these occurrences,
		and let $x'_1$ arise from $x_1$ by
		increasing the number of tokens in $\pstore$ by
		$\ell$ (to be generous). Then we clearly have
		$x'_1\step{\sigma'}x'_2$, where $x'_2(\pcov)\geq 1$;
		since
		$(x'_1)\restr{P}\step{(\sigma')\restr{P}}(x'_2)\restr{P}$
		is an execution of $A$, we get that 
 $(A,\pinit,\pcov,\pstore)$ is a
		positive instance of \PPSCover.

		$(\Rightarrow)$:
		Let $k,y$ be such that 
		$(\mathbf{e}_\pinit+k\cdot \mathbf{e}_\pstore)\reach
	(\mathbf{e}_\pcov+y)$ in $A$ and, moreover, $k\geq |P|$.  We  show that 
		the initial marking $x_0$ of $A'$ for which	
		$x_0(\pstore)=k+1$ (and	$x_0(\pinit)=1$) is
		live.

Note that no action is dead at $x_0$: we can enable and
		execute
		$a_\pcov$ that puts a token in both $\pinc$ and $\pdec$, and by
		using the actions $a_\lara{\pinc,p}$ and
		$a_\lara{\pdec,p}$ we can then enable any action, except
		$a^R_2$ and $a_3$; but executing the enabled $a_2$
		enables $a^R_2$, and $a_1$ then enables $a_3$.

To complete the proof,
		it suffices to show that $x_0$ is a
		home-marking, that is, $x_0\reach x$ implies
$x\reach x_0$. For the sake of contradiction,
suppose that 
	$x_0\reach x_1\step{a}x_2$ is an execution where $x_1\reach x_0$
		and $x_2\not\reach x_0$. This implies that $a$ has no reverse
		action, and it is thus an action among
		$a_\pcov,a_1,a_3,a_\lara{\pinc,p},a_\lara{\pdec,p}$. But
		then $x_2(\pdec)\geq 1$ or  $x_2(\pinc)\geq 1$,
and by \textsc{IP} we get $x_2\reach
		x_3$ for an initial marking $x_3$. Since
		$\normone{x_3}=\normone{x_0}$, we get 
 $x_3=x_0$, which contradicts $x_2\not\reach x_0$.
\end{proof}

\bibliographystyle{alphaurl}
\bibliography{bibliography}

\end{document}